\documentclass[aps,prb,twocolumn,showpacs,preprintnumbers,amsmath,amssymb,floatfix,unsortedaddress,10pt]{revtex4-1}

\usepackage{graphicx,xcolor,hyperref,multirow,nicefrac}

\begin{document}

\title{Diffusion quantum Monte Carlo study of excitonic complexes in
  two-dimensional transition-metal dichalcogenides}

\author{E.\ Mostaani}

\affiliation{Department of Physics, Lancaster University, Lancaster
  LA1 4YB, United Kingdom}

\affiliation{Cambridge Graphene Centre, Engineering Department,
  University of Cambridge, 9 J.\ J.\ Thomson Avenue, Cambridge CB3
  0FA, United Kingdom}

\author{M.\ Szyniszewski}

\affiliation{National Graphene Institute, University of Manchester,
  Booth Street East, Manchester M13 9PL, United Kingdom}

\affiliation{Department of Physics, Lancaster University, Lancaster
  LA1 4YB, United Kingdom}

\author{C.\ H.\ Price}

\affiliation{Department of Physics, Lancaster University, Lancaster
  LA1 4YB, United Kingdom}

\author{R.\ Maezono}

\affiliation{Japan Advanced Institute of Science and Technology,
  School of Information Science, Asahidai 1-1, Nomi, Ishikawa
  923-1292, Japan}

\author{M.\ Danovich}

\affiliation{National Graphene Institute, University of Manchester,
  Booth Street East, Manchester M13 9PL, United Kingdom}

\author{R.\ J.\ Hunt}

\affiliation{Department of Physics, Lancaster University, Lancaster
  LA1 4YB, United Kingdom}

\author{N.\ D.\ Drummond}

\affiliation{Department of Physics, Lancaster University, Lancaster
  LA1 4YB, United Kingdom}

\author{V.\ I.\ Fal'ko}

\affiliation{National Graphene Institute, University of Manchester,
  Booth Street East, Manchester M13 9PL, United Kingdom}

\date{\today}

\begin{abstract}
Excitonic effects play a particularly important role in the
optoelectronic behavior of two-dimensional semiconductors.  To
facilitate the interpretation of experimental photoabsorption and
photoluminescence spectra we provide (i) statistically exact diffusion
quantum Monte Carlo binding-energy data for a Mott-Wannier model of
(donor/acceptor-bound) excitons, trions, and biexcitons in
two-dimensional semiconductors in which charges interact via the
Keldysh potential, (ii) contact pair-distribution functions to allow a
perturbative description of contact interactions between charge
carriers, and (iii) an analysis and classification of the different
types of bright trion and biexciton that can be seen in single-layer
molybdenum and tungsten dichalcogenides.  We investigate the stability
of biexcitons in which two charge carriers are indistinguishable,
finding that they are only bound when the indistinguishable particles
are several times heavier than the distinguishable ones.
Donor/acceptor-bound biexcitons have similar binding energies to the
experimentally measured biexciton binding energies. We predict the
relative positions of all stable free and bound excitonic complexes of
distinguishable charge carriers in the photoluminescence spectra of
WSe$_2$ and MoSe$_2$.
\end{abstract}


\maketitle


\section{Introduction}

The last decade has witnessed a remarkable surge of interest in the
properties of truly two-dimensional (2D), atomically thin
semiconductors.  These include monolayer transition-metal
dichalcogenides (TMDCs) such as MoS$_2$, MoSe$_2$, WS$_2$, and
WSe$_2$, which acquire a direct-gap character in hexagonal monolayer
form.\cite{Mak_2010,Splendiani_2010,Kormanyos_2015,Ross_2014} The
direct gap and strong optical absorption of TMDCs suggest a range of
potential optoelectronic applications, e.g., in photodetectors,
photovoltaics, and light-emitting diodes.  A particularly interesting
aspect of monolayer TMDCs is the strong excitonic effects present in
their photoabsorption and photoluminescence
spectra,\cite{Ramasubramaniam_2012,Cheiwchanchamnangij_2012,Qiu_2013}
including nonhydrogenic Rydberg spectra\cite{Chernikov_2014,Ye_2014}
and lines ascribed to trions (charged
excitons)\cite{Ross_2013,Mak_2013,Lin_2014} and biexcitons (bound
pairs of excitons).\cite{Mai_2014,Shang_2015,You_2015,Plechinger_2015}
The nonhydrogenic nature of the excitonic energy spectrum is due to
lateral polarization effects in 2D crystals, which modify the form of
the Coulomb interaction between charge carriers.  Mott-Wannier models
of 2D trions and biexcitons have been studied using quantum Monte
Carlo (QMC)
methods,\cite{Bressanini_1998,Tan_2005,Ganchev_2015,Velizhanin_2015,Mayers_2015,Spink_2015,Kylanpaa_2015,Chaves_2016,Szyniszewski_2017},
variational methods\cite{Berkelbach_2013,Zhang_2015,Kidd_2016}, and
hyperspherical harmonics approaches,\cite{Kezerashvili_2016} and
interpolation formulas linking the 2D-screened and $1/r$ Coulomb
interaction regimes have been proposed.  Here we extend these studies
to provide numerically exact binding-energy data for all nonlocal
screening strengths, including an analysis of limiting behavior, and
we classify the types of trion and biexciton that can be observed in
different TMDCs.  We also investigate donor- and acceptor-bound
charge-carrier complexes in TMDCs, such as donor-bound biexcitons and
quintons, which have not to our knowledge been studied before.

The rest of the article is structured as follows.  In
Sec.\ \ref{sec:bright_dark} we describe the band structures of
molybdenum and tungsten dichalcogenides and analyze the nature of the
trions and biexcitons in these materials; furthermore, we perform a
group theoretical analysis of exciton properties.  In
Sec.\ \ref{sec:excitonic_properties} we explain the Keldysh form of
the screened Coulomb interaction between charges in 2D semiconductors,
describe the ways in which charge-carrier complexes are expected to
dissociate and recombine, and explain the importance of the contact
pair distribution function (PDF)\@. In Sec.\ \ref{sec:methodology} we
describe our computational methodology for solving the Mott-Wannier
model of charge-carrier complexes. We present our numerical results
for the binding energies and PDFs of the different complexes in
Sec.\ \ref{sec:results}.  Finally, we draw our conclusions in
Sec.\ \ref{sec:conclusions}.

\section{Bright and dark biexcitons and trions in molybdenum and tungsten
dichalcogenides \label{sec:bright_dark}}

\subsection{Classification of trions and biexcitons
\label{sec:classification_T_XX}}

\begin{figure}[!htbp]
\begin{center}
\includegraphics[clip,width=0.45\textwidth]{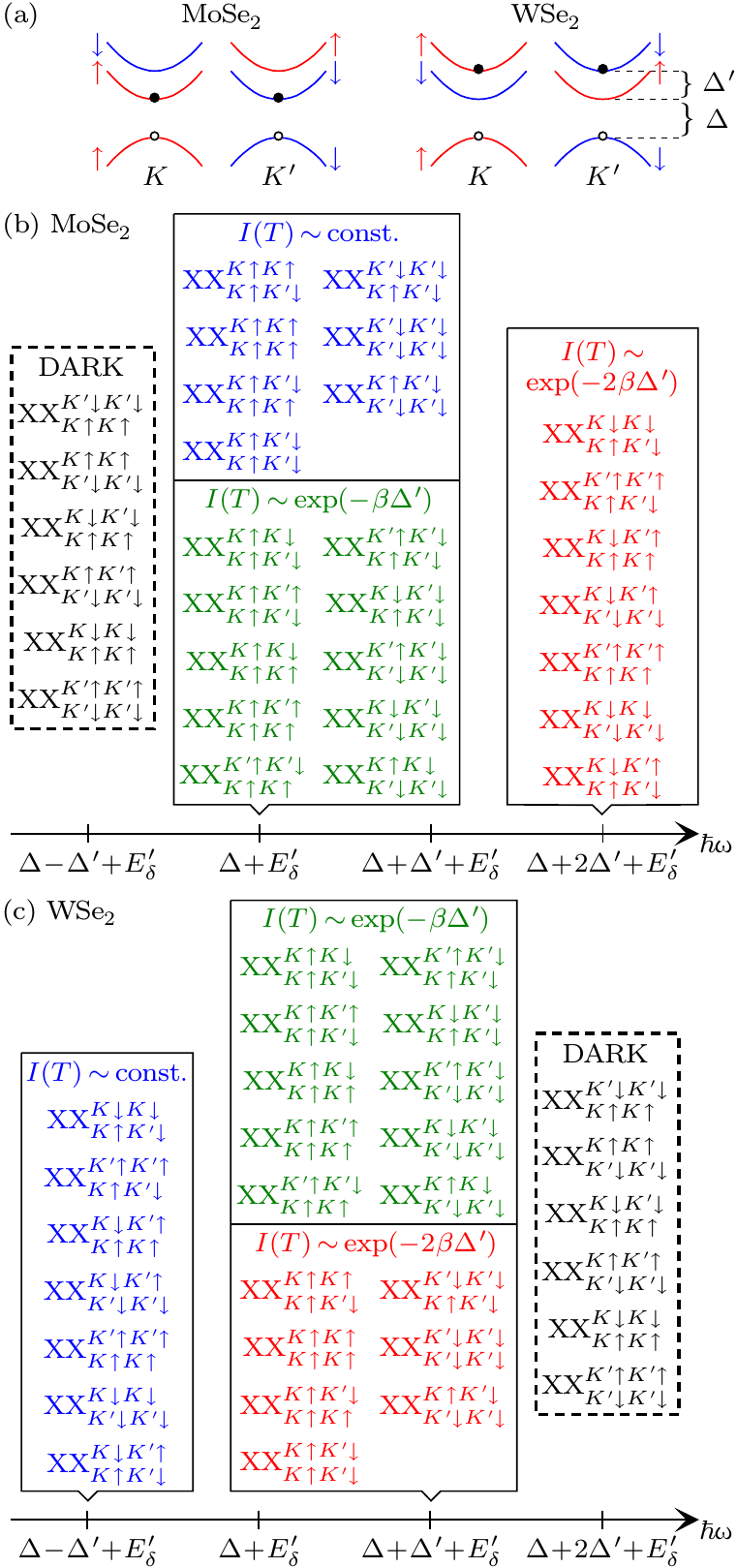}
\end{center}
\caption{(Color online) (a) Spin-split valence and conduction bands
  for MoSe$_2$ (left) and WSe$_2$ and WS$_2$ (right). We only show the
  spin-splitting of the conduction band; the spin-splitting of the
  valence band is much larger, so that no holes in the lower
  spin-split valence band are expected at room
  temperature.\cite{Kormanyos_2015} (b) and (c) Classification of
  biexciton recombination processes in molybdenum and tungsten
  dichalcogenides, respectively. $\Delta$ is the band gap, while
  $\Delta^\prime$ is the spin-splitting of the conduction band.
  $E_\delta^\prime \equiv E_{\rm XX} - E_{\rm X}$ is the difference
  between the total energies $E_{\rm XX}$ and $E_{\rm X}$ of a
  biexciton and an exciton. $\hbar \omega$ indicates the photon
  energies at which peaks in photoluminescence spectra are expected to
  appear. XX$_{k_3 \sigma_3 k_4 \sigma_4}^{k_1 \sigma_1 k_2 \sigma_2}$
  denotes a biexciton consisting of conduction-band electrons in
  valleys $k_1$ and $k_2$ with spins $\sigma_1$ and $\sigma_2$ and
  valence-band holes in valleys $k_3$ and $k_4$ with spins $\sigma_3$
  and $\sigma_4$. For example, the biexcitons shown in (a) are both
  denoted by XX$_{K \downarrow K' \uparrow}^{K \downarrow K'
    \uparrow}$.
 \label{fig:classification}}
\end{figure}

In monolayer molybdenum and tungsten dichalcogenides the
conduction-band minimum and valence-band maximum occur at the $K$ and
$K'$ points of the hexagonal Brillouin zone. Spin-orbit coupling
induces a significant splitting of both the valence band and the
conduction band at $K$ and $K'$.  In molybdenum diselenides, the
valence-band maximum has the same spin as the conduction-band minimum
within each valley, while in tungsten dichalcogenides such states have
opposite spins.\cite{Kormanyos_2015} Figure
\ref{fig:classification}(a) presents examples of the ways in which
biexcitons can be formed in molybdenum and tungsten
dichalcogenides. The spin-splitting of the valence band (0.15--0.5 eV)
is sufficiently large that no holes in the lower spin-split valence
band are expected at room temperature; however, the spin-splitting of
the conduction band ($\Delta'=3$--50 meV) is small enough that
electrons can be found in the upper spin-split conduction band at room
temperature.\cite{Kormanyos_2015}

\begin{figure}[!htbp]
\begin{center}
\includegraphics[clip,width=0.45\textwidth]{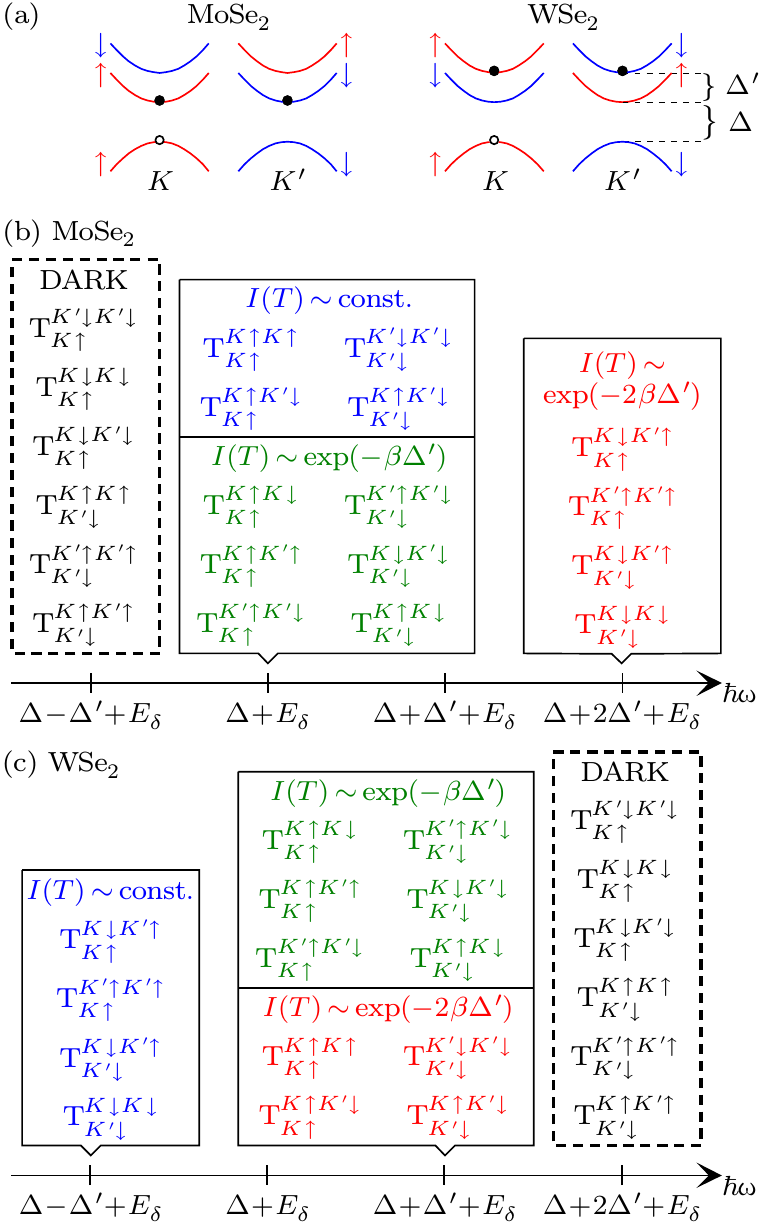}
\end{center}
\caption{(Color online) As Fig.\ \ref{fig:classification}, but for
  negative trions in molybdenum and tungsten
  dichalcogenides. $E_\delta \equiv E_{\rm X^-}$ is the total energy
  $E_{\rm X^-}$ of a negative trion. T$_{k_3 \sigma_3}^{k_1 \sigma_1
    k_2 \sigma_2}$ denotes a trion consisting of conduction-band
  electrons in valleys $k_1$ and $k_2$ with spins $\sigma_1$ and
  $\sigma_2$ and a valence-band hole in valley $k_3$ with spin
  $\sigma_3$.
 \label{fig:classification-trion}}
\end{figure}

An exciton, biexciton or trion is said to be either \textit{dark} or
\textit{semidark} when the recombination of an electron and hole is
forbidden by spin and momentum conservation; otherwise the complex is
said to be \textit{bright}. Semidark complexes are those in which
recombination can in fact take place due to intervalley scattering
with an accompanying energy shift. The precise photon energies depend
on whether the electrons occupy the higher- or lower-energy spin-split
bands in the initial and final states.  Furthermore, the intensity of
a spectral line depends on the thermal occupancy of the initial state.
Figures \ref{fig:classification}(b) and \ref{fig:classification}(c)
present a classification of biexcitons in molybdenum and tungsten
dichalcogenides with respect to the recombination energy and the
intensity of the emitted photons. This intensity has the following
temperature dependence:
\begin{widetext}
\begin{equation}
  I (T) \sim \left\{ \begin{array}{ll} \color{blue}{\text{const}.} &
    \text{for no electrons in the upper spin-split conduction band}
    \\ {\color[HTML]{008000}e^{- \Delta' / (k_{\rm B}T)}} & \text{for
      one electron in the upper spin-split conduction band}
    \\ \color{red}{e^{- 2 \Delta' / (k_{\rm B}T)}} & \text{for two
      electrons in the upper spin-split conduction band}
  \end{array} \right. ,
\end{equation}
\end{widetext}
where $\Delta'$ is the spin-orbit-induced splitting of the conduction
band, $k_{\rm B}$ is Boltzmann's constant, and $T$ is the
temperature. A similar classification can be made for trions: see
Fig.\ \ref{fig:classification-trion}. In a photoluminescence
experiment, we expect to see energies attributed to different kinds of
biexcitons and trions and emission lines of varying intensity, as
explained in Sec.\ \ref{sec:binding_energies}.

The opposite spin-splittings of the conduction and valence bands in
tungsten dichalcogenides results in the ground-state trions and
biexcitons being dark, with the two electrons residing in opposite
valleys. These dark complexes are coupled through an intervalley
electron-electron scattering to their excited bright counterparts with
both electrons residing in the upper spin-split conduction band. This
coupling gives a finite oscillator strength to the dark ground states
that is proportional to $[\mu_{\rm bd}/(2\Delta')]^2$, where $\mu_{\rm
  bd}$ is the coupling matrix element between dark and bright
states. As a result, the expected photoluminescence spectrum contains
two additional lines resulting from the recombination of these
``semidark'' trions and biexcitons, at an energy shifted downwards by
$2\Delta'$ relative to the bright complexes, and having a
temperature-independent intensity.

\subsection{Group theoretical analysis of excitons \label{sec:group_theory}}

Exciton wave functions can be classified according to the irreducible
representation (irrep) of the point-group symmetry of the TMDC
crystal, $D_{3h}$. As the states in the two valleys are degenerate,
one can treat the two valleys simultaneously by using the extended
group $D_{3h}''=D_{3h}+tD_{3h}+t^2D_{3h}$, where $t$ denotes
translation by a lattice vector. The character table of the extended
group is given in Table \ref{table:d3htt}.

The total exciton wave function $X$ is given in general by the product
of three components: the spatial envelope function $\Phi$, the Bloch
or lattice wave functions of the electron and hole $U_k$, and the spin
part $\chi$:
\begin{equation}
X = \Phi({\bf r}_{\rm e}, {\bf r}_{\rm h})\otimes U_k({\bf r}_{\rm e},
{\bf r}_{\rm h}) \otimes \chi(s_{\rm e}, s_{\rm h}).
\end{equation}
The representations of the wave functions by irreps consist of the
direct product of the individual irreps corresponding to the three
components: $\Gamma_X = \Gamma_{\Phi}\otimes \Gamma_U\otimes
\Gamma_{\chi}$. The tightly bound ground-state excitons are
characterized by a maximally symmetrized envelope function
corresponding to the identity irrep $\Gamma_{\Phi}=A_1^+$. Therefore
the representations of the exciton states are determined by the irreps
of the lattice and spin parts.

The conduction- and valence-band Bloch states transform according to
the 2D irreps $E_1'$ and $E_2'$, respectively. Using the product
table, Table \ref{table:prod_d3htt}, the lattice part of the exciton
wave function transforms as
\begin{equation}
E'^+_1\otimes E'^+_2 = E^+\oplus E'^+_3,
\end{equation}
where the 2D irrep $E^+$ corresponds to intravalley excitons in the
$K$ and $K'$ valleys, and $E'^+_3$ corresponds to intervalley
excitons, which are dark due to momentum conservation. In the
following, we will consider the $E^+$ intravalley excitons only.

The exciton spin part consists of two spin-$\nicefrac{1}{2}$ particles
corresponding to the spinor 2D irrep $D_{\nicefrac{1}{2}}$. The direct
product of the two spinors can be decomposed into the crystal point
group irreps as
\begin{equation}
D_{\nicefrac{1}{2}}\otimes D_{\nicefrac{1}{2}} =  A_1^+\oplus
A_2^+\oplus E^-.
\end{equation}
Hence the total exciton representation is given by
\begin{equation}
E^+_{\rm lattice} \otimes (A_1^+\oplus A_2^+\oplus E^-)_{\rm
  spin}=2E^+\oplus A_1^-\oplus A_2^-\oplus E^-.
\end{equation}
The $E^+$ irrep corresponds to the vector representation, and
therefore the two $E^+$ irreps correspond to excitons coupled to
in-plane polarized light. The $z$ coordinate transforms as the
$A_2^{-}$ irrep, and therefore the $A_2^{-}$ exciton is coupled to
out-of-plane polarized light, which involves a spin-flip process in
recombination.\cite{Urbaszek_2017} In the case of tungsten
dichalcogenides, the $A_2^{-}$ exciton is the ground-state exciton,
and results in photon emission at an energy that is lower than the
excited bright exciton by the spin-orbit splitting of the conduction
band $\Delta'$.  The $A_1^-$ and $E^-$ excitons are not coupled to
light.  A summary of the classification of exciton states is given in
Table \ref{table:classification} using a notation similar to that used
in Figs.\ \ref{fig:classification} and \ref{fig:classification-trion}.
Finally, we note that the spin-flip transition resulting in the
emission of out-of-plane polarized light corresponding to the $A_2^-$
exciton is also relevant for ground-state trions and biexcitons in
tungsten dichalcogenides, resulting in trion or biexciton emission at
a photon energy shifted downwards relative to the excited bright
states by $\Delta'$.

\begin{table}[!htbp]
\centering
\caption{Classification of exciton states into irreps of $D_{3h}''$
  and the polarization ($\parallel$ and $z$ for in-plane and
  out-of-plane, respectively) of the electric field to which the
  excitons are coupled. \label{table:classification}}
\begin{center}
\begin{tabular}{lcc}
\hline \hline

Irrep & Excitons & \multicolumn{1}{l}{Field}  \\ \hline

$E^+$ & $  {\rm X}^{K\uparrow}_{K\uparrow}$, ${\rm
  X}^{K'\downarrow}_{K'\downarrow}$  & \multirow{2}{*}{${\bf
    E}_{\parallel}$} \\

$E^+$ & ${\rm X}^{K\uparrow}_{K\downarrow}$, ${\rm
  X}^{K'\downarrow}_{K'\uparrow}$ & \\ \hline

$A_2^-$ & X (dark): $\{{\rm X}_{K\uparrow}^{K\downarrow}-{\rm
  X}_{K'\downarrow}^{K'\uparrow}\}$ & ${\bf E}_z$ \\ \hline

$E^-$ & ${\rm X}_{K\uparrow}^{K\downarrow}$, ${\rm
  X}_{K'\downarrow}^{K'\uparrow}$ & \multirow{2}{*}{--} \\

$A_1^-$ & $\{{\rm X}_{K\downarrow}^{K\uparrow}+{\rm
  X}_{K'\uparrow}^{K'\downarrow}\}$ & \\

\hline \hline
\end{tabular}
\end{center}
\end{table}

\section{Charge-carrier complexes in 2D
  semiconductors \label{sec:excitonic_properties}}

\subsection{Screened Coulomb interaction between charge carriers}

We model the charge carriers in a 2D semiconductor using a
Mott-Wannier model, in which small numbers of quasielectrons and
quasiholes are treated within the band effective mass approximation
and interact via an appropriately screened Coulomb interaction.  The
band effective masses for different 2D semiconductors are assumed to
be 2D-isotropic, and are discussed in Sec.\ \ref{sec:sensitivity}.
However, unlike quasi-2D electron(-hole) systems in GaAs/InAs
heterostructures, the form of the Coulomb interaction is profoundly
affected by the 2D nature of single-layer TMDCs, as we will now
discuss.

\begin{figure}[!htbp]
\begin{center}
\includegraphics[clip,width=0.45\textwidth]{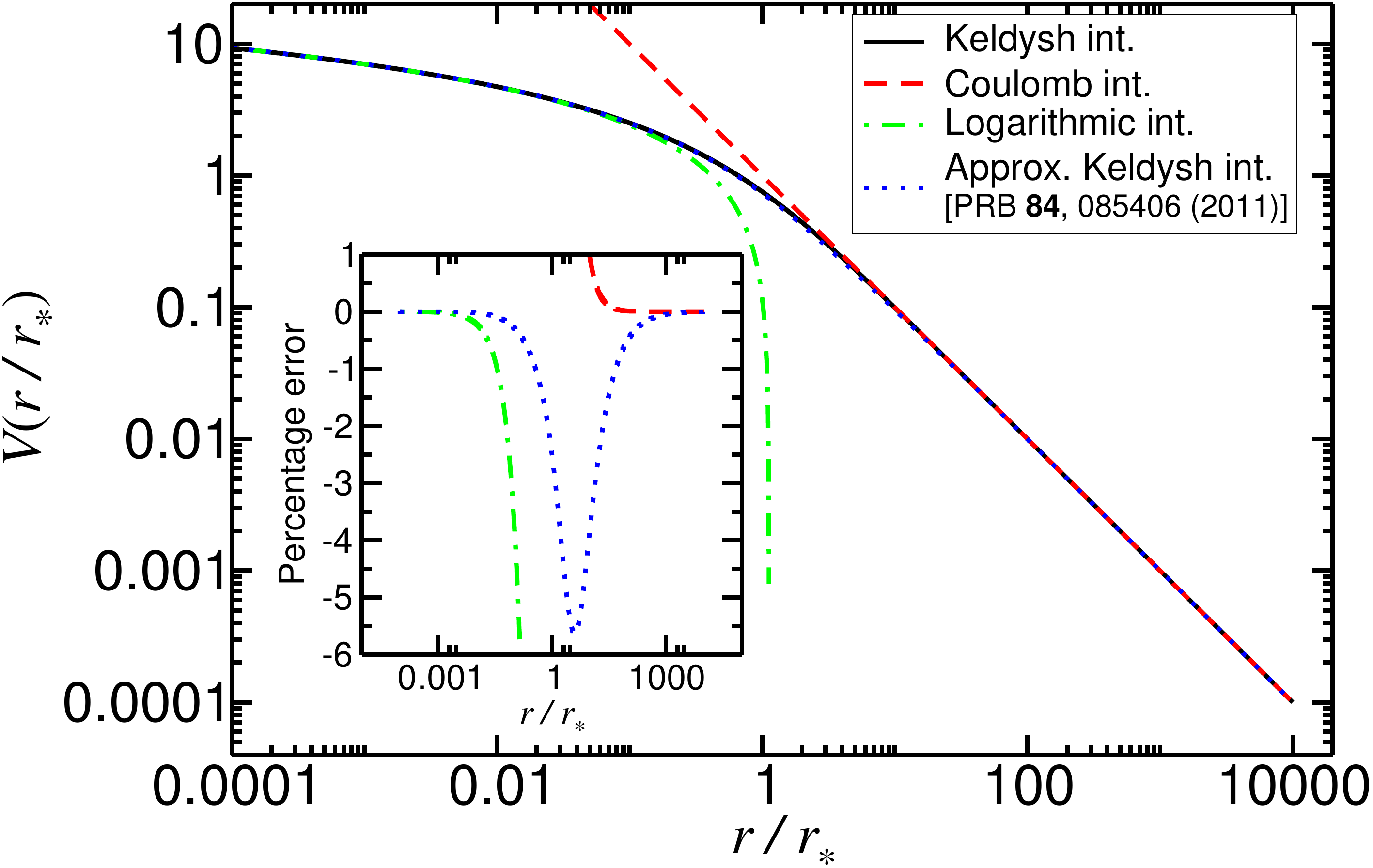}
\end{center}
\caption{(Color online) Dimensionless interaction potential between
  charge carriers in a 2D semiconductor, as defined in
  Eq.\ (\ref{eq:dimensionless_V}). The inset shows the percentage
  error in different approximations [Eqs.\ (\ref{eq:coulomb_int}),
    (\ref{eq:log_int}), and (\ref{eq:approx_keldysh_int})] to the
  Keldysh interaction of Eq.\ (\ref{eq:keldysh_int}).
 \label{fig:full_int}}
\end{figure}

Consider a charge density $\rho(x, y)\delta(z)$ in the $z = 0$ plane
of the 2D material, embedded in an isotropic medium of permittivity
$\epsilon$. The resulting electric displacement field is ${\bf D} = -
\epsilon \nabla \phi + {\bf P}_{\bot} (x, y) \delta (z) = - \epsilon
\nabla \phi - \kappa [\nabla \phi (x, y, 0)] \delta (z)$, where $\phi$
is the electrostatic potential, ${\bf P}_{\bot} (x, y)$ is the
in-plane polarization, and $\kappa$ is the in-plane susceptibility of
the material. By using Gauss's law, $\nabla \cdot {\bf D} = \rho
\delta (z)$, we obtain
\begin{equation}
  \epsilon \nabla^2 \phi = - \rho \delta (z) - \kappa [\nabla^2 \phi
    (x, y,0)] \delta (z).
\end{equation}
After taking the Fourier transform, denoting the wavevector in the
$(x, y)$ plane by ${\bf q}$ and the wavenumber in the $z$ direction by
$k$, we find
\begin{equation}
  \phi ({\bf q}, k) = \frac{\rho ({\bf q}) - \kappa q^2 \phi ({\bf q},
    z = 0)}{\epsilon (q^2 + k^2)}.
\end{equation}
However
\begin{eqnarray}
  \phi ({\bf q}, z = 0) & = & \frac{1}{2 \pi} \int \phi ({\bf q}, k)
  \, dk \nonumber\\ & = & \frac{1}{2\epsilon q} [\rho ({\bf q}) -
    \kappa q^2 \phi({\bf q}, z = 0)].
\end{eqnarray}
Rearranging, we find the in-plane electric potential to be
\begin{equation}
  \phi ({\bf q}, z = 0) = \frac{\rho ({\bf q})}{q (2\epsilon + q
    \kappa)}.
\end{equation}
Therefore the electrostatic potential energy between charges $q_i$ and
$q_j$ in a 2D semiconductor is
\begin{equation}
  v (q) = \frac{q_i q_j}{2\epsilon q (1 + r_{\ast} q)},
\end{equation}
where $r_\ast \equiv \kappa/(2\epsilon)$. After taking the Fourier
transform, the potential energy can be written as
\begin{equation} v(r) = \frac{q_iq_j}{4\pi \epsilon r_{\ast}} V
\left( \frac{r}{r_{\ast}}
\right), \label{eq:dimensionless_V} \end{equation} where $r$ is the
separation of the particles and
\begin{equation}
  V \left(r/r_\ast\right) = \frac{\pi}{2} \left[ H_0 \left(
    \frac{r}{r_{\ast}} \right) - Y_0 \left( \frac{r}{r_{\ast}} \right)
    \right],
\label{eq:keldysh_int}
\end{equation}
where $H_n(x)$ is a Struve function and $Y_n(x)$ is a Bessel function
of the second kind.  This result was first derived by
Keldysh,\cite{Keldysh_1979} and we refer to the interaction of
Eq.\ (\ref{eq:keldysh_int}) as the \textit{Keldysh interaction}.  At
long range ($r \gg r_{\ast}$) this potential becomes a Coulomb
interaction:
\begin{equation} V(r/r_\ast) \approx r_\ast / r, \label{eq:coulomb_int}
\end{equation}
while at short range ($r \ll r_{\ast}$) it is approximately
logarithmic:
\begin{equation} V(r/r_\ast) \approx \left[\log
  \left(2 r_\ast / r\right) - \gamma\right] = \log\left(
  \frac{2r_\ast}{\exp(\gamma)r}
  \right), \label{eq:log_int} \end{equation} where $\gamma$ is Euler's
constant.  We refer to the interaction potential of
Eq.\ (\ref{eq:log_int}) as the \textit{logarithmic interaction}. The
Keldysh interaction is plotted in Fig.\ \ref{fig:full_int}, along with
the Coulomb ($r_\ast=0$) and logarithmic ($r_\ast \to \infty$)
approximations.

The following approximation to Eq.\ (\ref{eq:keldysh_int}) was
introduced in Ref.\ \onlinecite{Cudazzo_2011}:
\begin{equation} V(r/r_\ast) \approx -
  \log\left(\frac{r/r_\ast}{1+r/r_\ast}\right)-\left[
    \gamma-\log(2)\right]
  e^{-r/r_\ast}. \label{eq:approx_keldysh_int} \end{equation} This
form of potential was used in the diffusion quantum Monte Carlo (DMC)
study of Ref.\ \onlinecite{Mayers_2015}. It is also plotted in
Fig.\ \ref{fig:full_int}, where it can be seen that the error in
Eq.\ (\ref{eq:approx_keldysh_int}) is as large as several percent in
the region $r \approx r_\ast$.  We compare DMC results obtained using
Eqs.\ (\ref{eq:keldysh_int}) and (\ref{eq:approx_keldysh_int}) in
Sec.\ \ref{sec:compare_ints}.

Finally, the Mott-Wannier-Keldysh Schr\"{o}dinger equation for a set
of charged quasiparticles in a 2D semiconductor is
\begin{equation}
  \left[ - \sum_i \frac{\hbar^2}{2 m_i} \nabla_i^2 + \sum_{i>j}
    \frac{q_i q_j}{4\pi \epsilon r_{\ast}} V \left( \frac{r_{i
        j}}{r_{\ast}} \right) \right] \psi = E
  \psi, \label{eq:mott_wannier_SE}
\end{equation}
where $m_i$ and $q_i$ are the band effective mass and charge of
particle $i$, $r_{ij}$ is the separation of particles $i$ and $j$, and
$E$ is the energy eigenvalue.

Now consider the situation in which the 2D semiconductor has a
dielectric medium of permittivity $\epsilon_{\rm a}$ above it and a
dielectric medium of permittivity $\epsilon_{\rm b}$ below it, as
would be the case for a 2D semiconductor deposited on a substrate.  In
general this is a more complicated problem than the situation
described above.  However, if we take $\epsilon \equiv (\epsilon_{\rm
  a}+\epsilon_{\rm b})/2$ in the expressions above, the correction to
the electrostatic energy of Eq.\ (\ref{eq:dimensionless_V}) is second
order in $\epsilon_{\rm a}-\epsilon_{\rm b}$. Hence the Keldysh
interaction remains valid when the permittivity $\epsilon$ is chosen
to be the average of the permittivities of the media on either side of
the 2D semiconductor, provided these permittivities are similar.

\subsection{Units and scaling \label{sec:units}}

\subsubsection{Excitonic units}

The energies of complexes interacting via the Keldysh or Coulomb
interactions are given in terms of the exciton Rydberg, $R_{\rm
  y}^\ast = \mu e^4/[2 (4 \pi \epsilon)^2 \hbar^2]$, and lengths are
given in terms of the exciton Bohr radius, $a_0^\ast = 4 \pi \epsilon
\hbar^2/(\mu e^2)$, where $\mu = m_{\rm e}m_{\rm h}/(m_{\rm e}+m_{\rm
  h})$ is the reduced mass of electron-hole pairs, with $m_{\rm e}$
and $m_{\rm h}$ being the electron and hole masses, respectively.

Let $\tilde{\bf r}_i={\bf r}_i/a_0^\ast$.  Then
Eq.\ (\ref{eq:mott_wannier_SE}) can be written as
\begin{equation} \left[ -\sum_i \frac{\mu}{m_i} \tilde{\nabla}_i^2+\sum_{i>j}
  \frac{2q_iq_ja_0^\ast}{e^2r_\ast}V\left( \frac{\tilde{r}_{ij}
    a_0^\ast}{r_\ast} \right) \right] \psi = {\cal E}
  \psi, \end{equation} where ${\cal E}=E/R_{\rm y}^\ast$.  Note that
$\mu/m_i$ only depends on the electron-hole mass ratio $\sigma \equiv
m_{\rm e}/m_{\rm h}$.  Hence for a fixed value of $r_\ast/a_0^\ast$,
the dimensionless energy eigenvalues ${\cal E}$ only depend on the
mass ratio, not on the absolute masses. Furthermore, for an exciton we
may write the Schr\"{o}dinger equation in terms of the difference
coordinate ${\bf r}_{\rm eh}$ as
\begin{equation} \left[ -\tilde{\nabla}_{\rm
  eh}^2-\frac{2a_0^\ast}{r_\ast}V\left( \frac{\tilde{r}_{\rm eh}
      a_0^\ast}{r_\ast} \right) \right] \psi = {\cal E}_{\rm X}
  \psi, \end{equation} so that for a given value of $r_\ast/a_0^\ast$,
the dimensionless exciton energy eigenvalues ${\cal E}_{\rm X}$ are
also independent of the mass ratio.  For the case of the Coulomb
interaction ($r_\ast=0$), the dimensionless ground-state energy of an
isolated exciton is ${\cal E}_{\rm X}=-4$, irrespective of the mass of
the electron or the hole.  The binding energies in excitonic Rydbergs
of donor-bound trions, biexcitons, and donor-bound biexcitons only
depend on $r_\ast/a_0^\ast$ and the electron-hole mass ratio $\sigma$.
Unfortunately, the energies of the different complexes go to zero in
these units in the limit that $r_\ast \to \infty$, and so a separate
set of units is required for the case of the logarithmic interaction,
as discussed in Sec.\ \ref{sec:units_log}.

\subsubsection{Logarithmic interaction \label{sec:units_log}}

For the limit $r_\ast \to \infty$, where the interaction is of
logarithmic form, we use the dimensionless units introduced in
Ref.\ \onlinecite{Ganchev_2015}.  The Schr\"{o}dinger equation for a
charge carrier complex with the logarithmic approximation to the
interaction [Eq.\ (\ref{eq:log_int})] is
\begin{equation} \left[ - \sum_i \frac{\hbar^2}{2m_i} \nabla_i^2
+ \sum_{i>j} \frac{q_iq_j}{4 \pi \epsilon r_\ast} \log\left(
\frac{2r_\ast}{\exp(\gamma)r} \right) \right] \psi = E
  \psi. \end{equation} Let
\begin{equation} r_0 = \sqrt{\frac{4\pi \epsilon r_\ast \hbar^2}{2e^2
  \mu}} \end{equation} and
\begin{equation} E_0=\frac{e^2}{4\pi \epsilon r_\ast}. \end{equation}
Defining dimensionless coordinates $\tilde{\bf r}_i={\bf r}_i/r_0$ and
a dimensionless energy ${\cal E}=E/E_0$, the Schr\"{o}dinger equation
can be written as
\begin{widetext}
\begin{equation} \left\{-\sum_i \frac{\mu}{m_i} \tilde{\nabla}_i^2 -\sum_{i>j}
  \frac{q_iq_j}{e^2}
  \left[\log(\tilde{r}_{ij})+\log\left(r_0/r_\ast\right)
    +\gamma-\log(2)\right] \right\} \psi = {\cal E}
  \psi. \end{equation} The only dependence of the dimensionless energy
${\cal E}$ of the complex on $r_\ast$ is through the pairwise additive
constant
\begin{equation} {\cal C} = -\sum_{i>j} \frac{q_iq_j}{e^2}
  \log\left(r_0/r_\ast\right). \end{equation}  Note that
\begin{equation}
  \sum_{i>j} \frac{q_i q_j}{e^2} = \left\{ \begin{array}{ll} -1 &
    \text{for an exciton or donor atom} \\ -1 & \text{for a trion or
      donor-bound exciton} \\ -2 & \text{for a biexciton or
      donor-bound trion} \\ -2 & \text{for a donor-bound biexciton}
    \\ \frac{(n_+ - n_-)^2 - n_+ - n_-}{2} & \text{for a complex of
      $n_+$ charges $+e$ and $n_-$ charges $-e$}
  \end{array} \right. .
\end{equation}
\end{widetext}
Hence the additive constant ${\cal C}$ cancels out of the binding
energies of the different charge-carrier complexes defined in
Sec.\ \ref{sec:binding_energies}.

For an isolated exciton, we may write the Schr\"{o}dinger equation in
terms of the difference coordinate $r_{\rm eh}$ and reduced mass,
giving
\begin{equation}
  \left[-\tilde{\nabla}_{\rm eh}^2+\log(\tilde{r}_{\rm
      eh})+\log\left(r_0/r_\ast\right) +\gamma-\log(2)\right] \psi =
       {\cal E}_{\rm X} \psi. \end{equation} The only dependence of
the dimensionless energy eigenvalue ${\cal E}_{\rm X}$ on the mass
ratio and $r_\ast$ comes from the constant term $\log(r_0/r_\ast)$ in
the Hamiltonian.  Hence we may write the ground-state dimensionless
energy as
\begin{equation} {\cal E}_{\rm X}={\cal
  E}_{\rm X0}-\log\left(r_\ast / r_0
  \right), \label{eq:X_energy} \end{equation} where ${\cal E}_{\rm X0}
= 0.41057747491(7)$ was evaluated by a finite-element method (see
Sec.\ \ref{sec:x_results}).

\subsection{Binding energies and spectra of charge-carrier
  complexes \label{sec:binding_energies}}

We define the binding energies $E_{\rm X^-}^{\rm b}$, $E_{\rm XX}^{\rm
  b}$, $E_{\rm D^{+}X}^{\rm b}$, $E_{\rm D^0X}^{\rm b}$, and $E_{\rm
  D^{+}XX}^{\rm b}$ of a trion, biexciton, donor-bound exciton,
donor-bound trion, and donor-bound biexciton, respectively, as
follows:
\begin{eqnarray}
E_{\rm X^-}^{\rm b} & = & E_{\rm X}-E_{\rm X^-} \label{eq:be_t_defn}
\\  E_{\rm XX}^{\rm b}  & = & 2E_{\rm X}-E_{\rm XX} \\ E_{\rm
  D^+X}^{\rm b}  & = & E_{\rm D^0}-E_{\rm D^+X} \\ E_{\rm D^0X}^{\rm
  b}  & = & E_{\rm D^0}+E_{\rm X}-E_{\rm D^0X} \\  E_{\rm D^+XX}^{\rm
  b} & = & E_{\rm D^0X}-E_{\rm D^+XX},
\label{eq:be_dbb_defn} \end{eqnarray}
where $E_{\rm X}$, $E_{\rm X^-}$, $E_{\rm XX}$, $E_{\rm D^0}$, $E_{\rm
  D^+X}$, $E_{\rm D^0X}$, and $E_{\rm D^+XX}$ are the ground-state
total energies of an exciton, trion, biexciton, donor atom,
donor-bound exciton, donor-bound trion, and donor-bound biexciton,
respectively.  These are the binding energies with respect to
dissociation into the most energetically competitive species.  With
the exceptions of the donor-bound exciton (D$^+$X) and donor-bound
biexciton (D$^+$XX), each of the complexes dissociates into an exciton
(X) plus one other complex.  Binding energies of charge-conjugated
complexes (such as positive trions, acceptor-bound trions, and
acceptor-bound biexcitons) are defined in an analogous fashion.  Note
that, under the definitions of
Eqs.\ (\ref{eq:be_t_defn})--(\ref{eq:be_dbb_defn}), a binding energy
is \textit{positive} for a bound complex.

\begin{figure*}[!htbp]
\begin{center}
\includegraphics[clip,width=0.95\textwidth]{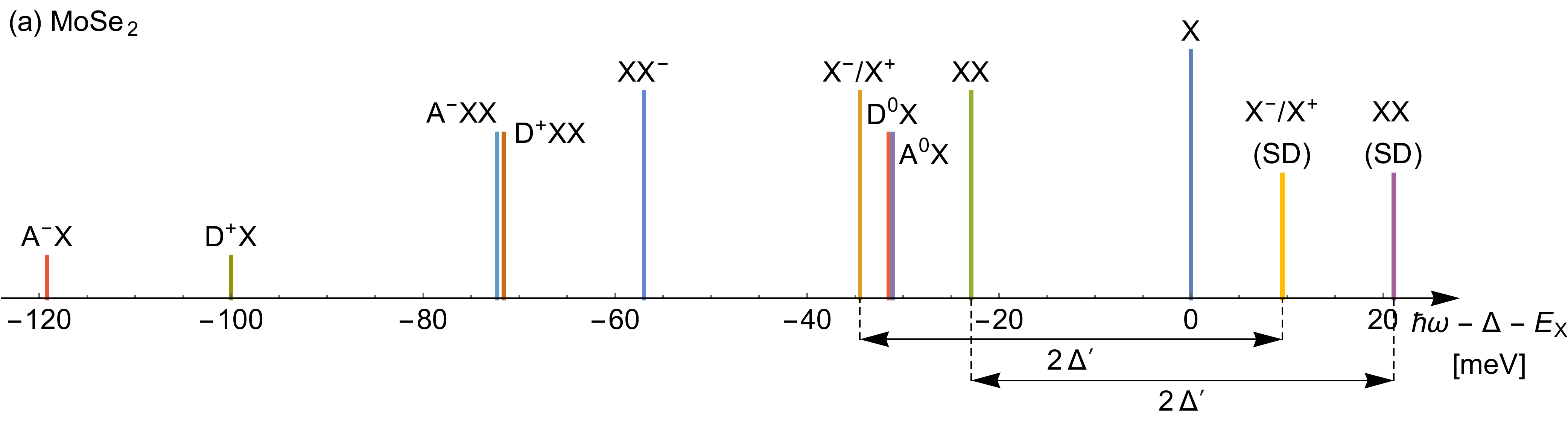}
\\ \includegraphics[clip,width=0.95\textwidth]{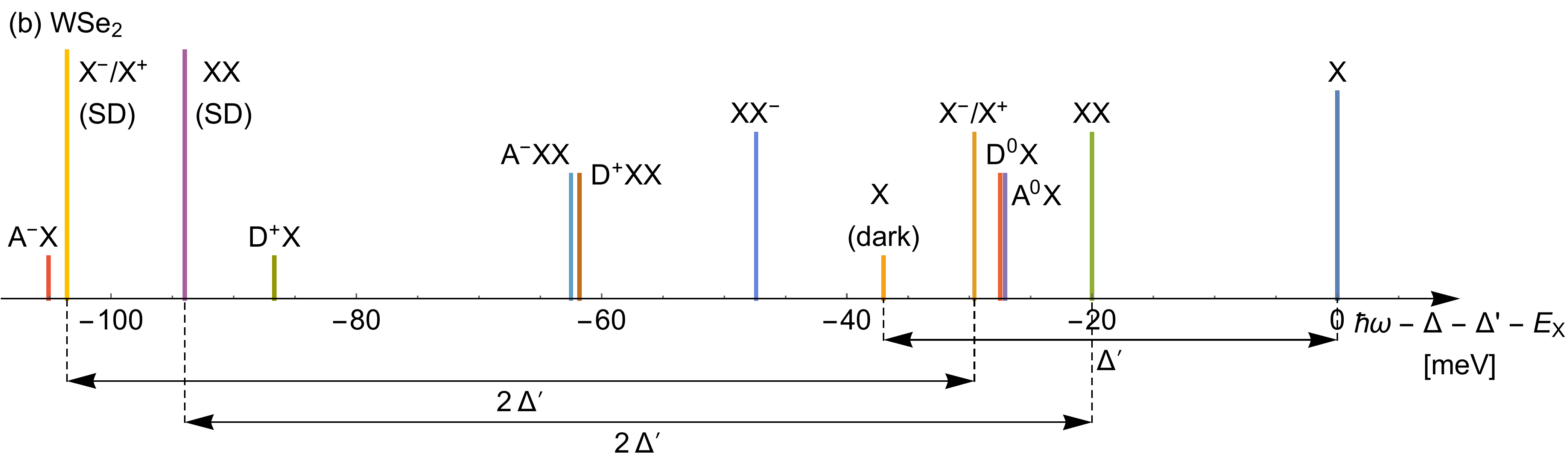}
\end{center}
\caption{(Color online) Expected photoemission spectra for (a)
  MoSe$_2$ and (b) WSe$_2$, showing lines for the different complexes
  studied in this work. $\Delta$ and $\Delta'$ are the quasiparticle
  band gap and the spin-splitting of the conduction band,
  respectively. The numerical values of $\Delta'$ are taken from
  density-functional-theory calculations with the
  Heyd-Scuseria-Ernzerhof (HSE06) hybrid
  functional.\cite{Kormanyos_2015} $E_{\rm X}$ is the total energy of
  an exciton. The lines show the frequency relative to the bright
  exciton peak arising due to the recombination of a single
  electron-hole pair in each complex: see
  Sec.\ \ref{sec:binding_energies}.  E.g., the D$^0$X line shows the
  frequency relative to the exciton peak of the process
  D$^0$X${}\to{}$D$^0+\gamma$.  The trion and biexciton peaks labelled
  ``SD'' arise from semidark complexes, and are offset by $2\Delta'$,
  as explained in Sec.\ \ref{sec:classification_T_XX}; the exciton
  peak labelled ``dark'' arises from the process described in
  Sec.\ \ref{sec:group_theory}; the other peaks arise from bright
  complexes.  Donor- and acceptor-bound exciton peaks are shown with
  very low intensity due to the marginal stability of these complexes.
\label{fig:spectrum}}
\end{figure*}

The energy difference between the exciton peak in a photoluminescence
experiment and the peak corresponding to a particular complex is equal
to the energy required to separate a single exciton from that complex.
Thus the energy difference between the exciton peak and the trion peak
is $E_{\rm X}-E_{\rm X^-}=E_{\rm X^-}^{\rm b}$; the energy difference
between the exciton peak and the biexciton peak is $2E_{\rm X}-E_{\rm
  XX}=E_{\rm XX}^{\rm b}$; and the energy difference between the
exciton peak and the donor-bound trion peak is $E_{\rm X}+E_{\rm
  D^0}-E_{\rm D^0X}=E_{\rm D^0X}^{\rm b}$.  On the other hand, the
energy difference between the exciton peak and the donor-bound exciton
peak is $E_{\rm X}-E_{\rm D^+X}=E_{\rm D^+X}^{\rm b} + E_{\rm
  X}-E_{\rm D^0}$, and the energy difference between the exciton peak
and the donor-bound biexciton peak is $E_{\rm X}+E_{\rm D^+X}-E_{\rm
  D^+XX}=E_{\rm D^+XX}^{\rm b}+E_{\rm D^0X}^{\rm b}-E_{\rm D^+X}^{\rm
  b}$.  Some of these peaks are shown in Fig.\ \ref{fig:spectrum}.  In
addition there are expected to be offsets to the peak positions due to
the spin-splitting of the conduction bands of TMDCs, as described in
Sec.\ \ref{sec:excitonic_properties}.

In Sec.\ \ref{sec:dbbiex_results} we report DMC binding energies for
quintons and other large charge-carrier complexes in tungsten and
molybdenum dichalcogenides.  In each of these cases the binding energy
is defined to be the energy required to remove an exciton from the
complex; this is the binding energy with respect to dissociation into
the most energetically competitive products.

\subsection{Contact and exchange interactions between charge
  carriers \label{sec:pdfs_theory}}

The Mott-Wannier model of a charge-carrier complex is valid provided
the complex extends over many unit cells of the underlying crystal.
However, when charge carriers are present at the same point in space
there is an energy contribution due to local exchange and correlation
effects.\cite{Szyniszewski_2017} Although the excitons in TMDCs are
Mott-Wannier-like, their wave functions only extend over a small
number of primitive unit cells, so that local exchange and correlation
effects are expected to be significant.  We may represent this effect
within a Mott-Wannier model by introducing additional pairwise contact
interaction potentials.  For example, for a biexciton the Hamiltonian
should include an additional term of the form
\begin{equation} A^{\rm ee}\delta({\bf r}_{\rm ee})+A^{\rm
  hh}\delta({\bf r}_{\rm hh}) + A^{\rm eh} \sum_{i=1}^2 \sum_{j=1}^2
  \delta({\bf r}_{{\rm e}_i{\rm h}_j}), \end{equation} where $A^{\rm
  ee}$, $A^{\rm hh}$, and $A^{\rm eh}$ are constants and ${\bf r}_{\rm
  ee}$, ${\bf r}_{\rm hh}$, and ${\bf r}_{{\rm e}_i{\rm h}_j}$ are the
electron-electron separation, the hole-hole separation, and the
separation of electron $i$ and hole $j$, respectively. Evaluating
$A^{\rm ee}$, $A^{\rm hh}$, and $A^{\rm eh}$ by \textit{ab initio}
calculations is challenging, and so we leave them as free parameters
to be determined in experiments or subsequent \textit{ab initio}
calculations.  If we evaluate the expectation value of this contact
interaction then we find that the first-order perturbative correction
to the total energy can be written as $A^{\rm eh}g_{\rm XX}^{\rm
  eh}({\bf 0})+A^{\rm ee} g_{\rm XX}^{\rm ee}({\bf 0})+A^{\rm
  hh}g_{\rm XX}^{\rm hh}({\bf 0})$, where the electron-electron,
hole-hole pair, and electron-hole PDFs are
\begin{eqnarray} g_{\rm XX}^{\rm
  ee}({\bf r}) & = & \left< \delta\left({\bf r}-{\bf r}_{\rm
    ee}\right) \right> \\ g_{\rm XX}^{\rm hh}({\bf r}) & = & \left<
  \delta\left({\bf r}-{\bf r}_{\rm hh}\right) \right> \\ g_{\rm
    XX}^{\rm eh}({\bf r}) & = & \left< \sum_{i=1}^2 \sum_{j=1}^2
  \delta\left({\bf r}-{\bf r}_{{\rm e}_i{\rm h}_j}\right)
  \right>, \end{eqnarray} respectively.  We report contact PDF data
within the Mott-Wannier model.

In addition to the role of the contact PDF in evaluating perturbative
corrections due to contact interactions, the PDF and contact PDF
contain a wealth of physical information.  The exciton recombination
rate of a charge-carrier complex is proportional to the electron-hole
contact PDF\@. Furthermore, the PDF gives a very direct indication of
the spatial size and shape of a charge-carrier complex.

The contact PDF also plays a role in the intervalley scattering of
carriers. As the intervalley scattering involves a large momentum
transfer of the order of the inverse lattice constant, the interaction
is short range and can be modelled by a contact interaction with both
carriers in the same position. In particular, the electron-electron
contact PDF for the semidark trion and biexciton in tungsten-based
TMDCs determines the coupling strength of the dark and bright states
as $\mu_{\rm bd}\propto g^{\rm ee}(0)$ and hence determines the
recombination rates of the semidark states.\cite{Danovich_2017}

\section{Computational methodology \label{sec:methodology}}

\subsection{Quantum Monte Carlo modelling of excitonic complexes
\label{sec:qmc_method}}

Our total-energy and PDF calculations were carried out using the
variational quantum Monte Carlo (VMC) and DMC
approaches.\cite{Ceperley_1980,Foulkes_2001} The ground-state wave
function for a set of interacting, distinguishable particles is
nodeless; hence the fixed-node DMC algorithm is exact for all the
systems studied in this work with the exception of biexcitons with
indistinguishable holes.  We used a numerical representation of the
potential of Eq.\ (\ref{eq:keldysh_int}) that is accurate to at least
eight significant figures.  Trial wave functions were optimized using
VMC with variance minimization\cite{Umrigar_1988,Drummond_2005} and
energy minimization.\cite{Umrigar_2007} The DMC calculations were
performed using time steps in the ratio $1:4$ with the corresponding
target configuration populations being in the ratio $4:1$.
Afterwards, the energies were extrapolated linearly to zero time step
and hence, simultaneously, to infinite population.  To perform all our
calculations, the \textsc{casino} code was used.\cite{Needs_2010}

QMC methods have previously been used to study 2D trions with nonlocal
screening\cite{Ganchev_2015,Mayers_2015} and the Coulomb
interaction\cite{Spink_2015} and 2D biexcitons with the Coulomb
interaction (including indirect biexcitons in coupled-quantum-well
heterostructures)\cite{Bressanini_1998,Tan_2005,Lee_2009,Maezono_2013}
and in TMDCs with nonlocal screening.\cite{Mayers_2015,Kylanpaa_2015}
In a recent work some of the present authors have investigated the
binding energies of trions and biexcitons using DMC for a range of
susceptibility parameters $r_\ast$ and effective masses, and have
represented the DMC data using simple interpolation
formulas.\cite{Szyniszewski_2017} It was shown that for the applicable
range of $r_\ast$ values, 2D semiconductors are expected to show
larger trion binding energies than biexciton binding energies, in
contrast to the situation in quasi-2D systems such as GaAs/InAs
quantum wells.  Here we extend this work to include extreme cases and
donor/acceptor-bound carrier complexes.

\subsection{Wave functions for complexes of distinguishable charge
  carriers \label{sec:ex2D_jas}}

Our trial wave functions for complexes of distinguishable charge
carriers were of the Jastrow form $\Psi = \exp[J({\bf R})]$, where
${\bf R}$ is the vector of all the particle coordinates. The Jastrow
exponent $J({\bf R})$ included a pairwise sum of terms of the
form\footnote{For biexcitons with distinguishable particles, we used
  the Jastrow form introduced in the Supplemental Material to
  Ref.\ \onlinecite{Ganchev_2015}.}
\begin{equation}
u_{\rm ex2D}(r) = \frac{[c_1+\Gamma^\prime \log(r)+c_2r]r^2}{1+c_3r^2}
\label{eq:pairwise_term}
\end{equation}
for the Keldysh and logarithmic interactions, where $r$ is
interparticle distance, $c_1$, $c_2 \leq 0$, and $c_3 \geq 0$ are
optimizable parameters, and
\begin{equation}
  \Gamma^\prime=-\frac{q_iq_jm_im_j}{2a_0^\ast \mu e^2
    r_\ast(m_i+m_j)} \label{eq:Gammap}
\end{equation}
for distinguishable pairs of particles of charge $q_i$ and $q_j$ and
mass $m_i$ and $m_j$.  Different constants $c_i$ are used for each
type of particle pair.  This form satisfies the analog of the Kato
cusp conditions,\cite{Kato_1957,Pack_1966} i.e., it ensures that the
local energy $\Psi^{-1}\hat{H}\Psi$ is nondivergent at coalescence
points, where $\hat{H}$ is the Hamiltonian operator.

Where the interaction between the charge carriers was of Coulomb form,
we used pairwise terms of the form
\begin{equation}
u_{\rm ex2D}(r) = \frac{\Gamma r+c_1r^2}{1+c_2r}
\end{equation}
in the Jastrow exponent, where $c_1 \leq 0$ and $c_2 \geq 0$ are
optimizable parameters, and
\begin{equation}
  \Gamma=\frac{2q_iq_jm_im_j}{a_0^\ast \mu
    e^2(m_i+m_j)} \label{eq:Gamma} \end{equation} for distinguishable
pairs of particles of mass $m_i$ and $m_j$ and charge $q_i$ and $q_j$.
This form satisfies the Kato cusp
conditions.\cite{Kato_1957,Pack_1966}

Donor ions and other infinitely heavy particles were fixed point
charges in our calculations.  In this case $u_{\rm ex2D}$ provided a
one-body Jastrow term between the free particles and the fixed
particles that satisfies the Kato cusp conditions.  In addition,
cuspless one-body, two-body and three-body polynomial terms truncated
at finite range were used in our Jastrow
factor.\cite{Drummond_2004,Lopez_2012}

\subsection{Wave functions for biexcitons with indistinguishable holes}

For biexcitons with indistinguishable holes we used the trial wave
function
\begin{equation} \Psi=\exp[J({\bf R})] x_{\rm hh}'({\bf R}), \end{equation}
where $J$ is of the form described in Sec.\ \ref{sec:ex2D_jas}.  For
indistinguishable particles of mass $m$ and charge $q$ interacting via
the logarithmic or Keldysh interactions, Eq.\ (\ref{eq:Gammap}) must
be replaced by $\Gamma^\prime=-q^2m/(8a_0^\ast \mu e^2 r_\ast)$, while
for indistinguishable pairs of particles interacting via the Coulomb
interaction, Eq.\ (\ref{eq:Gamma}) must be replaced by
$\Gamma=q^2m/(2a_0^\ast \mu e^2)$.  $x_{\rm hh}'$ is the $x$-component
of
\begin{eqnarray} {\bf r}_{\rm hh}' & = & {\bf r}_{\rm hh} + \eta_{\rm hh}(r_{\rm
  hh}){\bf r}_{\rm hh} \nonumber \\ & & {} + \eta_{\rm eh}(r_{\rm
    e_1h_1}){\bf r}_{\rm e_1h_1} + \eta_{\rm eh}(r_{\rm e_1h_2}){\bf
    r}_{\rm e_1h_2} \nonumber \\ & & {} - \eta_{\rm eh}(r_{\rm
    e_2h_1}){\bf r}_{\rm e_2h_1} - \eta_{\rm eh}(r_{\rm e_2h_2}){\bf
    r}_{\rm e_2h_2}, \label{eq:bf_like} \end{eqnarray} where
$\eta_{\rm hh}$ and $\eta_{\rm eh}$ are smoothly truncated
polynomials, with optimizable expansion coefficients, and ${\bf
  r}_{\rm hh}$ and ${\bf r}_{{\rm e}_i{\rm h}_j}$ are the hole-hole
and electron-hole relative positions, respectively.  Equation
(\ref{eq:bf_like}) is effectively a
backflow\cite{Feynman_1954,Feynman_1956} transformation:
$\Psi=\exp(J)x_{\rm hh}$ introduces the correct nodal topology for the
state that we want to consider and Eq.\ (\ref{eq:bf_like}) maps the
particle coordinates $\{{\bf r}\}$ to quasiparticle coordinates
$\{{\bf r}^\prime\}$ without changing the nodal topology.  In
Eq.\ (\ref{eq:bf_like}),
\begin{equation} \eta_{\rm hh}(r)=\sum_{n=2}^{N_\eta^{\rm hh}} a_n r^n
  (r-L)^C\Theta (L-r) \label{eq:eta_hh} \end{equation} and
\begin{equation} \eta_{\rm eh}(r)=\sum_{m=0}^{N_\eta^{\rm eh}} b_m r^m(r-L)^C\Theta
  (L-r) \label{eq:eta_eh} \end{equation} are smoothly truncated
polynomials with optimizable parameters $\{a_n\}$ and $\{b_n\}$. $L$
is a cutoff length, $N_\eta^{\rm hh}$ and $N_\eta^{\rm eh}$ determine
the amount of variational freedom, $C=3$ to ensure smooth behavior at
the cutoffs, and $\Theta$ denotes the Heaviside function. We require
$b_1=Cb_0/L$ to ensure that $\eta$ does not affect the Kato cusp
conditions, which are enforced by the Jastrow factor. We optimized the
free parameters in our antisymmetric wave function using energy
minimization.\cite{Umrigar_2007}

For different values of $N_\eta^{\rm hh}$ and $N_\eta^{\rm eh}$ in
Eqs.\ (\ref{eq:eta_hh}) and (\ref{eq:eta_eh}), we compare the VMC
ground-state energy, variance, and DMC energy of biexcitons with
indistinguishable electrons interacting via the logarithmic
interaction in Table \ref{table:test_log_eta}. Analogous results for
biexcitons interacting via the Keldysh interaction at finite $r_\ast$
are shown in Table \ref{table:test_2d_eta}. Our results show that
increasing $N_\eta^{\rm hh}$ and $N_\eta^{\rm eh}$ slightly decreases
the variances; nevertheless, the VMC and DMC energies are independent
of the number of free parameters when $N_\eta^{\rm hh},N_\eta^{\rm eh}
\geq 2$ to within our statistical error bars. We have used
$N_\eta^{\rm hh}=N_\eta^{\rm eh}=3$ in our production calculations.

\begin{table}[!htbp]
\caption{Effect of changing the number of free parameters in $\eta$
  [i.e., the values of $N_\eta^{\rm hh}$ and $N_\eta^{\rm eh}$ in
    Eqs.\ (\ref{eq:eta_hh}) and (\ref{eq:eta_eh})] on the VMC
  ground-state energy ($E_{\rm VMC}$), VMC energy variance, and DMC
  energy ($E_{\rm DMC}$) for biexcitons with indistinguishable holes
  interacting via the logarithmic interaction. The mass ratio is
  $\sigma=0.1$ and the reduced mass is $\mu=0.5m_0$, where $m_0$ is
  the bare electron mass. In each case $r_\ast=r_0$.
  \label{table:test_log_eta}}
\begin{center}
  \begin{tabular}{lr@{}lr@{}lr@{}l}
  \hline \hline

$N_\eta^{\rm hh}=N_\eta^{\rm eh}$ &\multicolumn{2}{c}{$E_{\rm VMC}$
    ($E_0$)} & \multicolumn{2}{c}{Variance ($E_0^2$)} &
  \multicolumn{2}{c}{$E_{\rm DMC}$ ($E_0$)} \\ \hline

  2 & $0.$&$7604(3)$ & $~~~~0.$&$00920$ & $0.$&$7585(2)$ \\

  3 & $0.$&$7602(3)$ & $~~~~0.$&$00908$ & $0.$&$7584(2)$ \\

  4 & $0.$&$7605(3)$  & $~~~~0.$&$00914$ & $0.$&$7579(2)$ \\

  6 & $0.$&$7606(3)$ & $~~~~0.$&$00927$ & $0.$&$7580(2)$ \\

\hline \hline
\end{tabular}
\end{center}
\end{table}

\begin{table}[!htbp]
\caption{Effect of changing the number of free parameters in $\eta$
  [i.e., the values of $N_\eta^{\rm hh}$ and $N_\eta^{\rm eh}$ in
    Eqs.\ (\ref{eq:eta_hh}) and (\ref{eq:eta_eh})] on the VMC
  ground-state energy ($E_{\rm VMC}$), VMC energy variance, and DMC
  energy ($E_{\rm DMC}$) of biexcitons with indistinguishable holes
  interacting via the Keldysh interaction, with an electron-hole mass
  ratio of $\sigma=0.1$.  $r_\ast=0$ corresponds to the Coulomb
  interaction.
  \label{table:test_2d_eta}}
\begin{center}
  \begin{tabular}{lcr@{}lr@{}lr@{}l}
  \hline \hline $r_\ast/a^\ast_0$ & $N_\eta^{\rm hh}=N_\eta^{\rm eh}$
  &\multicolumn{2}{c}{$E_{\rm VMC}$ ($R_{\rm y}^\ast$)} &
  \multicolumn{2}{c}{Variance (${R_{\rm y}^\ast}^2$)} &
  \multicolumn{2}{c}{$E_{\rm DMC}$ ($R_{\rm y}^\ast$)} \\ \hline

  0 & 2 & $-8.$&$608(1)$ & $~~~~0.$&$1709$ & $-8.$&$6100(4)$ \\

  0 & 3 & $-8.$&$608(1)$ & $~~~~0.$&$1658$ & $-8.$&$6112(4)$ \\

  0 & 4 & $-8.$&$606(1)$ & $~~~~0.$&$1718$ & $-8.$&$6108(4)$ \\

  0 & 6 & $-8.$&$608(1)$ & $~~~~0.$&$1719$ & $-8.$&$6108(4)$ \\

  \hline

  8 & 2 & $-0.$&$6304(1)$ & $~~~0.$&$0008$ & $-0.$&$6308(1)$ \\

  8 & 3 & $-0.$&$63020(8)$ & $~~~0.$&$0007$ & $-0.$&$6306(2)$ \\

  8 & 4 & $-0.$&$6301(1)$ & $~~~0.$&$0007$ & $-0.$&$6308(1)$ \\

  8 & 6 & $-0.$&$63024(8)$ & $~~~0.$&$0007$ & $-0.$&$6309(1)$ \\

\hline \hline
\end{tabular}
\end{center}
\end{table}

Biexcitons with distinguishable electrons and indistinguishable holes
can trivially be mapped onto biexcitons with indistinguishable
electrons and distinguishable holes by charge conjugation.

\subsection{Time-step and population-control errors}

\begin{figure}[!htbp]
\begin{center}
\includegraphics[clip,width=0.45\textwidth]{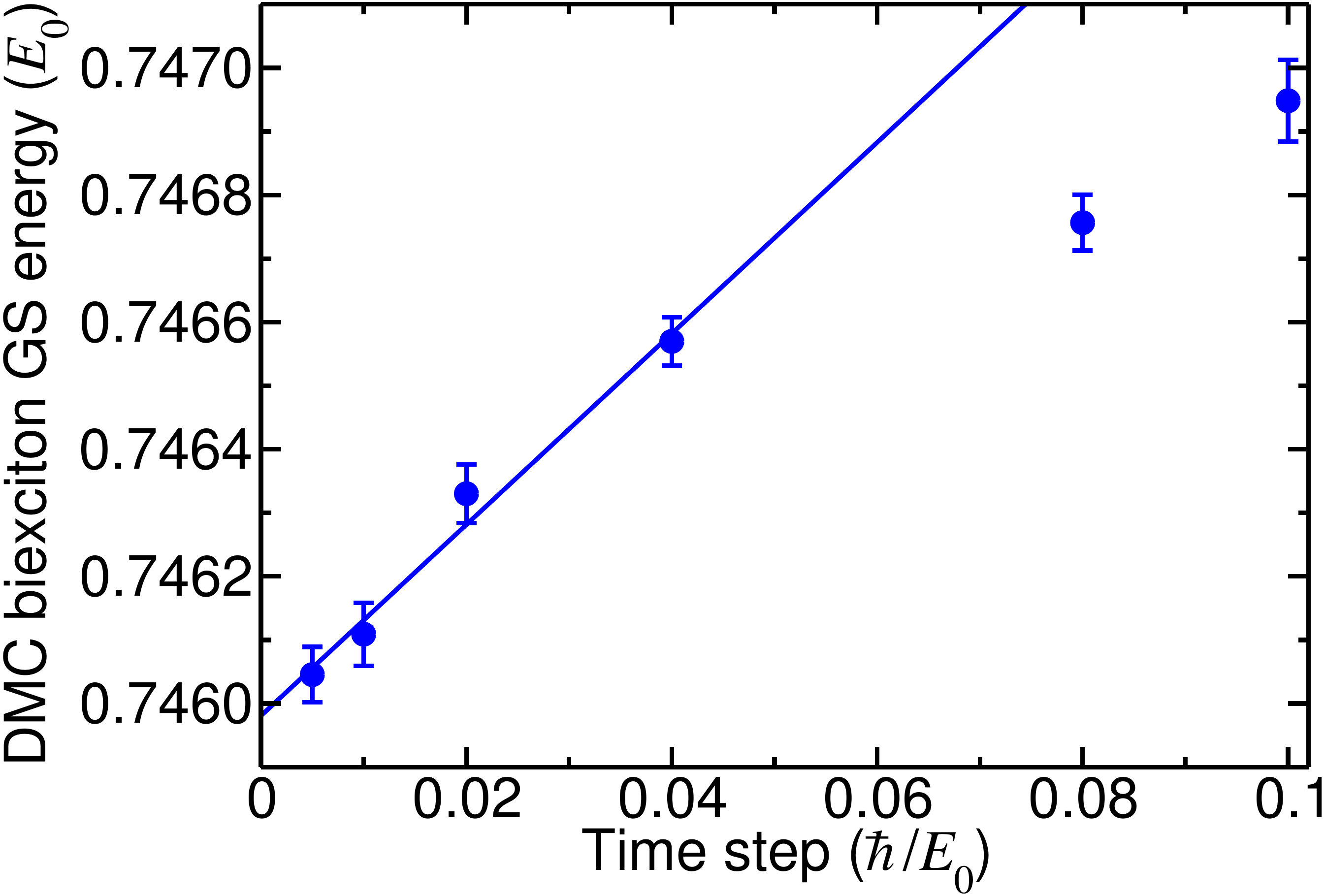}
\end{center}
\caption{(Color online) DMC ground-state (GS) energy of a biexciton
  with distinguishable particles at mass ratio $\sigma=1$ against time
  step, with the logarithmic interaction between charges. The
  configuration population was varied in inverse proportion to the
  time step.  The reduced mass is $\mu=0.5m_0$ and $r_\ast=r_0$.
\label{fig:time_test}}
\end{figure}

We chose our DMC time steps such that the root-mean-square distance
diffused by each particle in a single time step was much less than
$r_0$ for the logarithmic interaction, much less than $a_0^\ast$ for
the Coulomb interaction, and much less than $\min\{r_0,a^*_0\}$ for
the Keldysh interaction at finite $r_\ast$. In
Fig.\ \ref{fig:time_test} we plot the DMC total energy of a biexciton
with distinguishable particles against time step. The figure confirms
that the linear extrapolation scheme described in
Sec.\ \ref{sec:qmc_method} largely eliminates the effects of time-step
bias, provided the time steps used are sufficiently small. For the
logarithmic interaction with $\sigma=1$ and $r_\ast=r_0$, the time
step should evidently be rather less than $0.04\hbar/E_0$.

\begin{figure}[!htbp]
\begin{center}
\includegraphics[clip,width=0.45\textwidth]{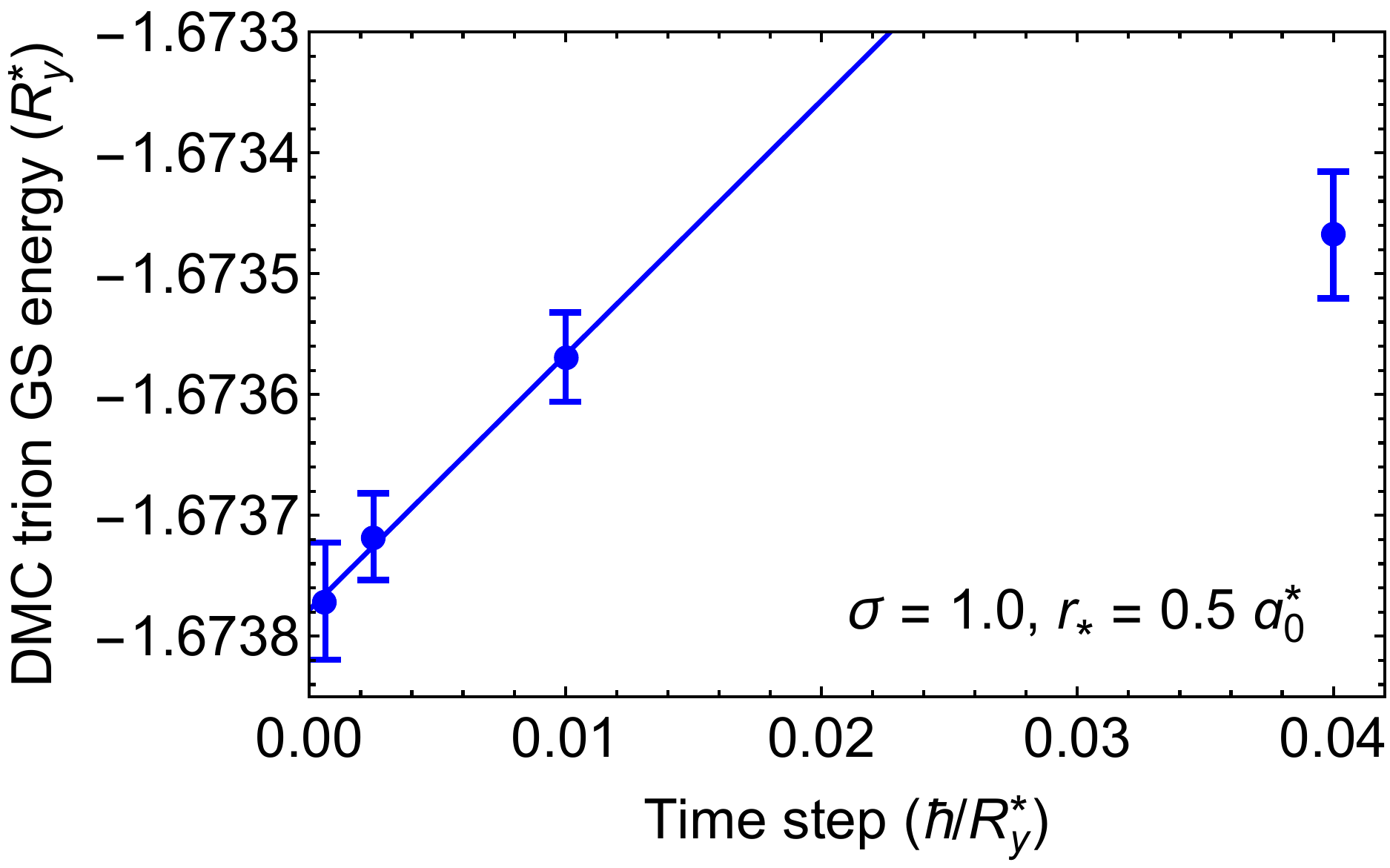}
\end{center}
\caption{(Color online) DMC ground-state (GS) energy of a negative
  trion at mass ratio $\sigma=1$ and $r_\ast=0.5a_0^\ast$ against time
  step, with the Keldysh interaction between charges. The
  configuration population was varied in inverse proportion to the
  time step.
\label{fig:trion_time_test}}
\end{figure}

\begin{figure}[!htbp]
\begin{center}
\includegraphics[clip,width=0.45\textwidth]{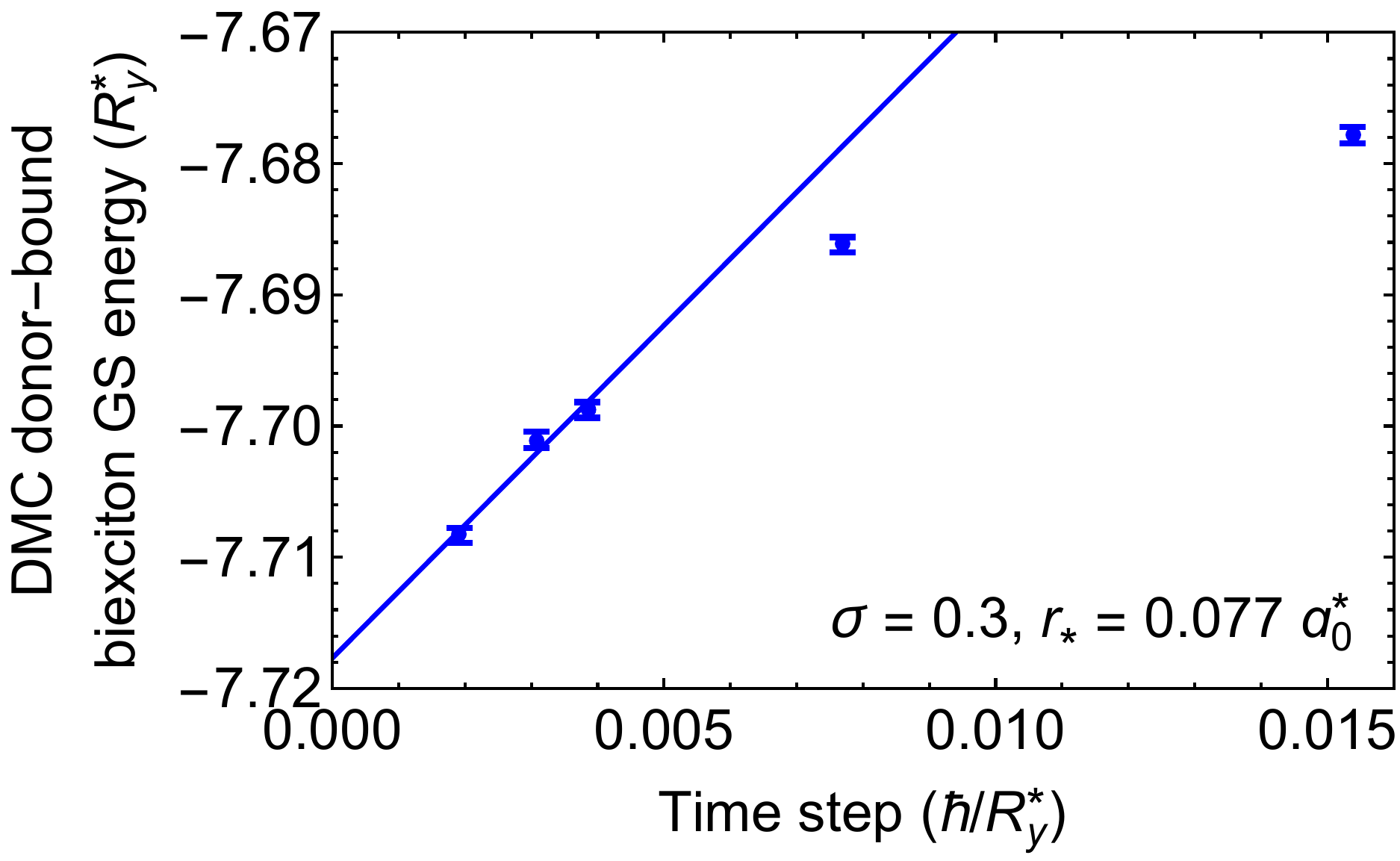}
\end{center}
\caption{(Color online) DMC ground-state (GS) energy of a donor-bound
  biexciton at $\sigma=0.3$ and $r_\ast=0.077a_0^\ast$ against time
  step, with the Keldysh interaction between charges. The
  configuration population was varied in inverse proportion to the
  time step.
\label{fig:dxx_time_test}}
\end{figure}

Figures \ref{fig:trion_time_test} and \ref{fig:dxx_time_test} show
similar time-step tests performed for a negative trion and a
donor-bound biexciton with the Keldysh interaction.  For $r_{\ast}
\geq 0.25 a_0^{\ast}$, one should use time steps of less than $0.01
\hbar / R_{\rm y}^{\ast}$ to be in the linear time-step bias regime,
while for $r_{\ast} < 0.25 a_0^{\ast}$, time steps of less than
$0.0025 \hbar / R_{\rm y}^{\ast}$ are required.

\subsection{PDF calculations \label{sec:pdf_extrap}}

The PDFs defined in Sec.\ \ref{sec:pdfs_theory} were evaluated by
binning the interparticle distances sampled in VMC and DMC
calculations.  The errors in the VMC and DMC PDFs are linear in the
error in the trial wave function; however, the error in the
extrapolated estimate (twice the DMC estimate minus the VMC estimate)
is quadratic in the error in the trial wave
function.\cite{Ceperley_1979} Our reported PDFs were obtained by
extrapolated estimation.

Contact PDF data have been calculated by extrapolating electron-hole
and electron-electron PDFs to zero separation for each $r_\ast$ value
and mass ratio considered.  To perform the extrapolation we fitted
$\exp[\tilde{g}(r)]$ to our PDF data at short
range,\cite{Drummond_2011} where
\begin{equation}
  \tilde{g}(r) = a_0^\prime + 2\Gamma^\prime r^2\log(r) + a_2^\prime
  r^2 + a_3^\prime r^3 + \cdots + a_6^\prime
  r^6\label{eq:contactPDFformula}
\end{equation}
for the Keldysh and logarithmic interactions and
\begin{equation}
  \tilde{g}(r) = a_0 + 2\Gamma r + a_2 r^2 + \cdots + a_6 r^6
\end{equation}
for the Coulomb interaction ($r_\ast=0$), where $\Gamma^\prime$ and
$\Gamma$ are defined in Eqs.\ (\ref{eq:Gammap}) and (\ref{eq:Gamma})
and $a_0^\prime$, $a_2^\prime$, \dots, $a_6^\prime$ and $a_0$, $a_2$,
\dots, $a_6$ are fitting parameters.  These forms satisfy (the analog
of) the Kimball cusp conditions.\cite{Kimball_1973} The model
functions were fitted to our PDF data at small $r$, with the data
being weighted by $2\pi r$.

\subsection{Sensitivity of binding energy to the form of screened
  interaction \label{sec:compare_ints}}

We have investigated whether the approximation to the Keldysh
interaction given in Eq.\ (\ref{eq:approx_keldysh_int}), which has
been used in previous QMC studies of excitonic
complexes,\cite{Mayers_2015} leads to significant errors.  For an
exciton with $r_\ast=a_0^\ast/2$, the DMC total energies are $E_{\rm
  X}=-1.5358899(2)R_{\rm y}^\ast$ and $-1.4668074(3)R_{\rm y}^\ast$
with the Keldysh interaction [Eq.\ (\ref{eq:keldysh_int})] and the
approximate Keldysh interaction [Eq.\ (\ref{eq:approx_keldysh_int})],
respectively.  This is a difference of about 4.5\%, which is small but
non-negligible.  The DMC binding energies of trions with
$r_\ast=a_0^\ast/2$ and mass ratio $\sigma=1$ using the exact and
approximate Keldysh interactions are $0.1377(4)R_{\rm y}^\ast$ and
$0.1335(3)R_{\rm y}^\ast$, respectively, so the error in the binding
energy due to the approximate Keldysh interaction is about 3\%.  Since
these errors are easily avoidable, we have used the exact Keldysh
interaction in our production calculations.

\section{Numerical results \label{sec:results}}

\subsection{Excitons \label{sec:x_results}}

\begin{figure}[!htbp]
\begin{center}
\includegraphics[clip,width=0.45\textwidth]{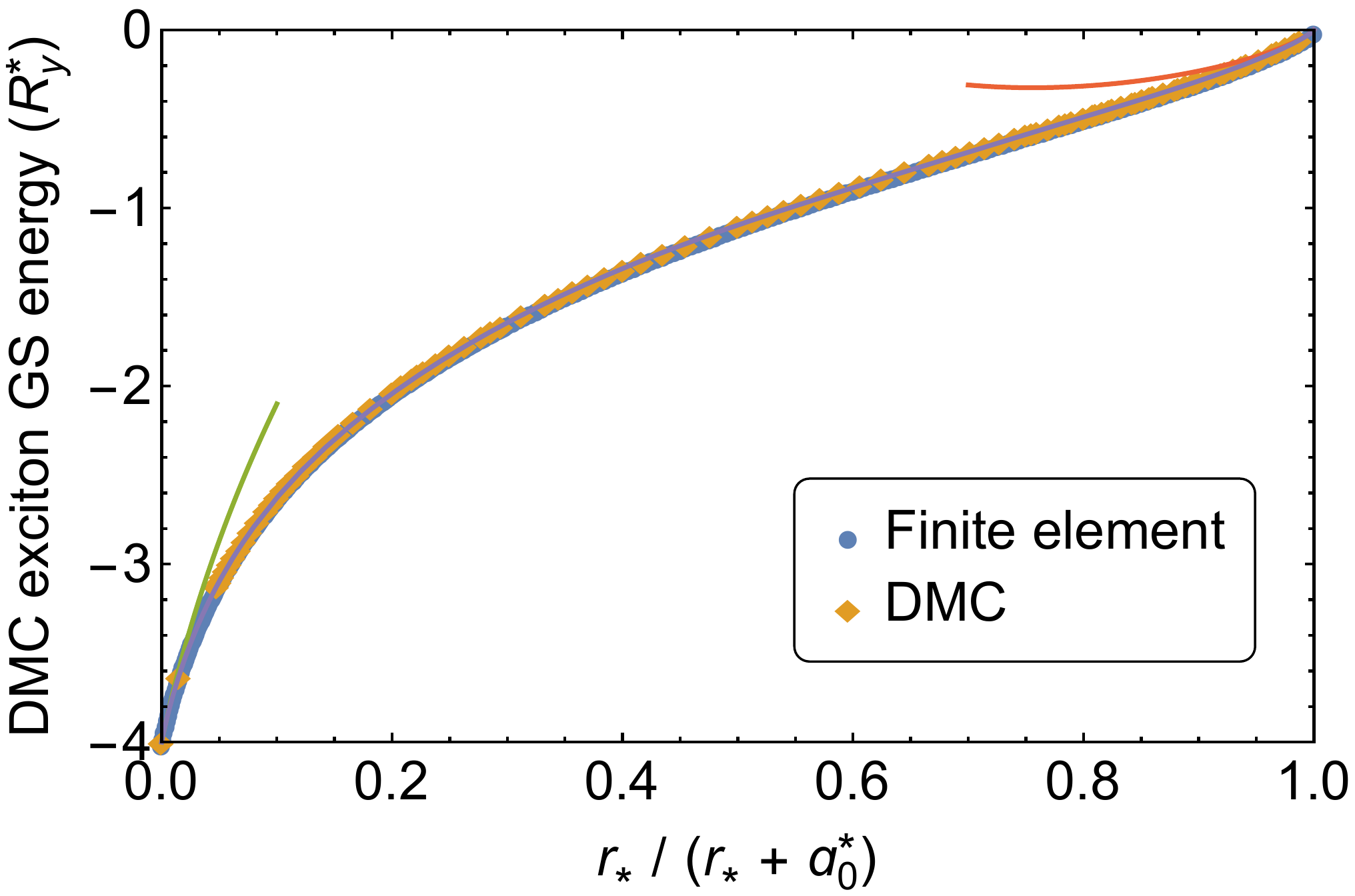}
\end{center}
  \caption{(Color online) Exciton ground-state (GS) energy evaluated
    using DMC and a finite-element method. The plot also shows the
    approximations to the ground-state energy obtained by first-order
    perturbation theory about the Coulomb limit (green) and by using
    the logarithmic approximation to the Keldysh potential
    (red).\label{fig:be_x}}
\end{figure}

The exciton ground-state energy is presented in
Fig.\ \ref{fig:be_x}. Our DMC data are in agreement with the results
of finite-element calculations as implemented in the Mathematica
software.\cite{Mathematica} In excitonic units, the energy of an
exciton is independent of the effective masses: see
Sec.\ \ref{sec:units}. In the Coulomb limit, one recovers the
well-known excitonic energy of $-4 R_{\rm y}^{\ast}$. We can determine
the behavior of the energy near the Coulomb limit by evaluating the
first-order perturbative correction
\begin{eqnarray}
  \frac{\langle \Delta v \rangle}{R_{\rm y}^{\ast}} & = & \frac{32
    r_{\ast} (a_0^\ast + 4r_{\ast})}{(a_0^\ast)^2 + 16r^2_{\ast}}
  \nonumber \\ & & {} - \frac{128 a_0^\ast r_{\ast}^2\left[
      \text{csch}^{-1}(4r_{\ast}/a_0^\ast) +
      \sinh^{-1}(4r_{\ast}/a_0^\ast) \right]}{\left[(a_0^\ast)^2 +
      16r_{\ast}^2\right]^{3/2}} \nonumber \\ & \approx &
  32r_{\ast}/a_0^\ast + O((r_{\ast}/a_0^\ast)^2),
\end{eqnarray}
where $\Delta v = v_{\text{Keldysh}} - v_{\text{Coulomb}}$ is the
difference between the Keldysh potential of
Eq.\ (\ref{eq:keldysh_int}) and the Coulomb potential of
Eq.\ (\ref{eq:coulomb_int}), and the expectation value is taken with
respect to the exact ground-state wave function for the Coulomb
interaction $\Psi=\exp(-2r/a_0^\ast)$. The correction is shown in
Fig.\ \ref{fig:be_x} as a green line.

\begin{figure}[!htbp]
\begin{center}
\includegraphics[clip,width=0.45\textwidth]{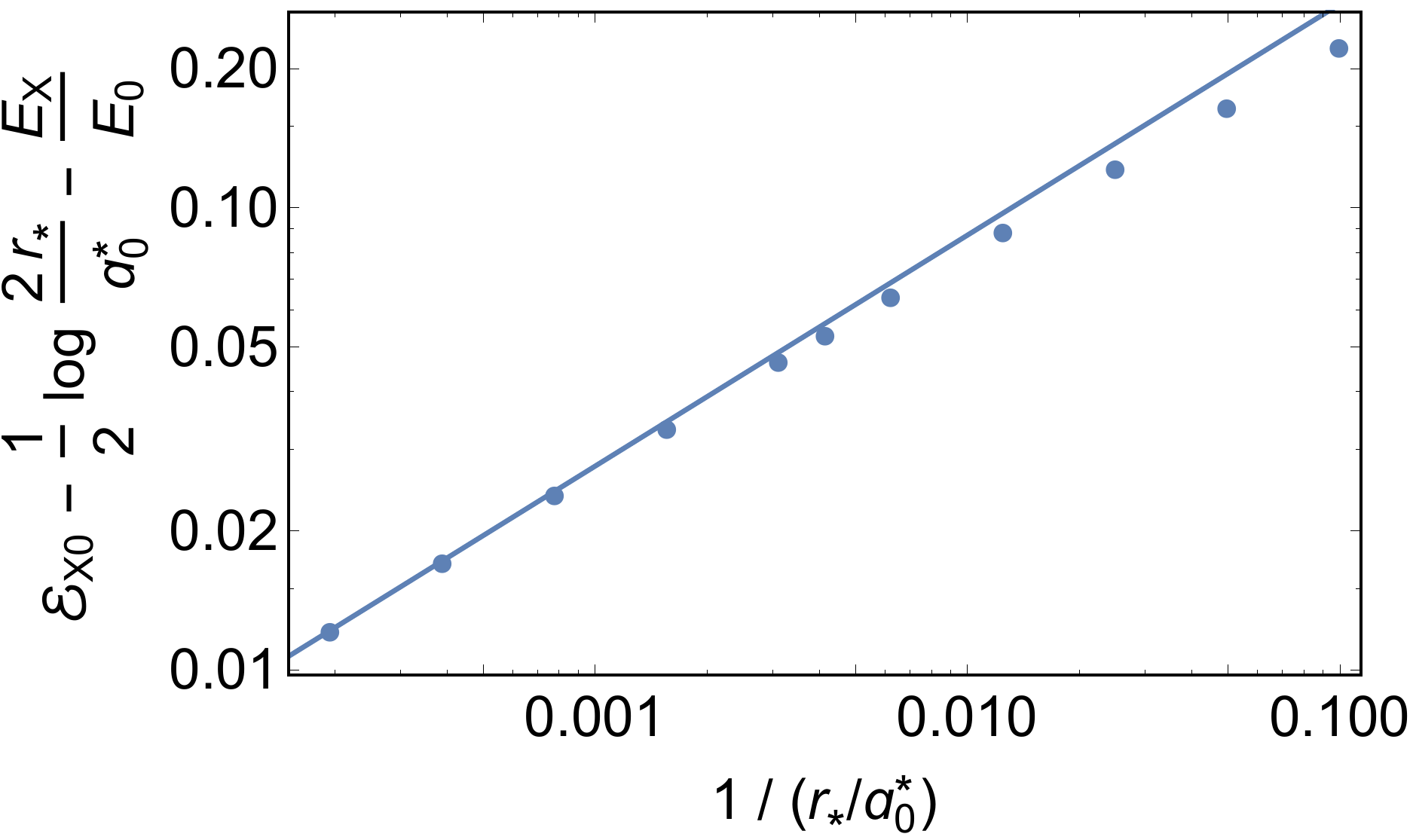}
\end{center}
\caption{(Color online) Difference of dimensionless exciton energies
  with the Keldysh interaction and the logarithmic approximation to
  the Keldysh interaction, calculated using first-order perturbation
  theory within VMC\@.  The solid line is a fit of $a \sqrt{a_0^\ast /
    r_{\ast}}$ to the VMC data, with $a=0.871(2)$. \label{fig:x_log}}
\end{figure}

We have numerically evaluated the dimensionless constant
$\mathcal{E}_{\text{X0}}$ in Eq.\ (\ref{eq:X_energy}) to be
$\mathcal{E}_{\text{X0}} = 0.41057739(7)$ using DMC and
$\mathcal{E}_{\text{X0}} = 0.41057747491(7)$ using the finite-element
method. These results confirm the expected accuracy of the DMC method.
The logarithmic-limit behavior from Eq.\ (\ref{eq:X_energy}) is also
shown in Fig.\ \ref{fig:be_x} (red line) and matches the DMC data near
$r_{\ast} \to \infty$. The difference $\Delta E_{\rm X}/E_0$ between
the exciton energies in units of $E_0$ with the Keldysh and
logarithmic interactions at large $r_\ast$ was calculated
numerically. Using the optimized ground-state wave function for the
logarithmic interaction, we used VMC to evaluate the first-order
perturbative approximation $\Delta E_{\rm X}/E_0 \approx \langle
v_{\text{Keldysh}} - v_{\text{logarithmic}} \rangle$. The results are
presented in Fig.\ \ref{fig:x_log} and show that the leading-order
error in the exciton energy due to the logarithmic interaction goes as
$\sqrt{a_0^\ast / r_{\ast}}$.

We fitted the function
\begin{equation}
  \frac{E_X}{R_{\rm y}^\ast} = \frac{(1 - y) \left[ \begin{array}{l} -
        4 + 33 y + a_1 y^{3 / 2} \\ \hspace{3em} {} + \sum_{k = 2}^5
        a_k y^k + \log (1 - y) \end{array} \right]}{1 + (1 - y) y^3
    (b_1 + b_2 y)},
\label{eq:EX_fit} \end{equation}
with $a_5 = - 29 + 2\mathcal{E}_{\text{X} 0} - a_1 - a_2 - a_3 - a_4 -
\log 2$ to our DMC exciton energy data, where
$y=r_\ast/(a_0^\ast+r_\ast)$ and the remaining $\{a_i\}$ and $\{b_i\}$
are six free fitting parameters.  The fractional error in the fit of
Eq.\ (\ref{eq:EX_fit}) to our DMC data is everywhere less than 0.5\%.

\begin{figure}[!htbp]
\begin{center}
\includegraphics[clip,width=0.45\textwidth]{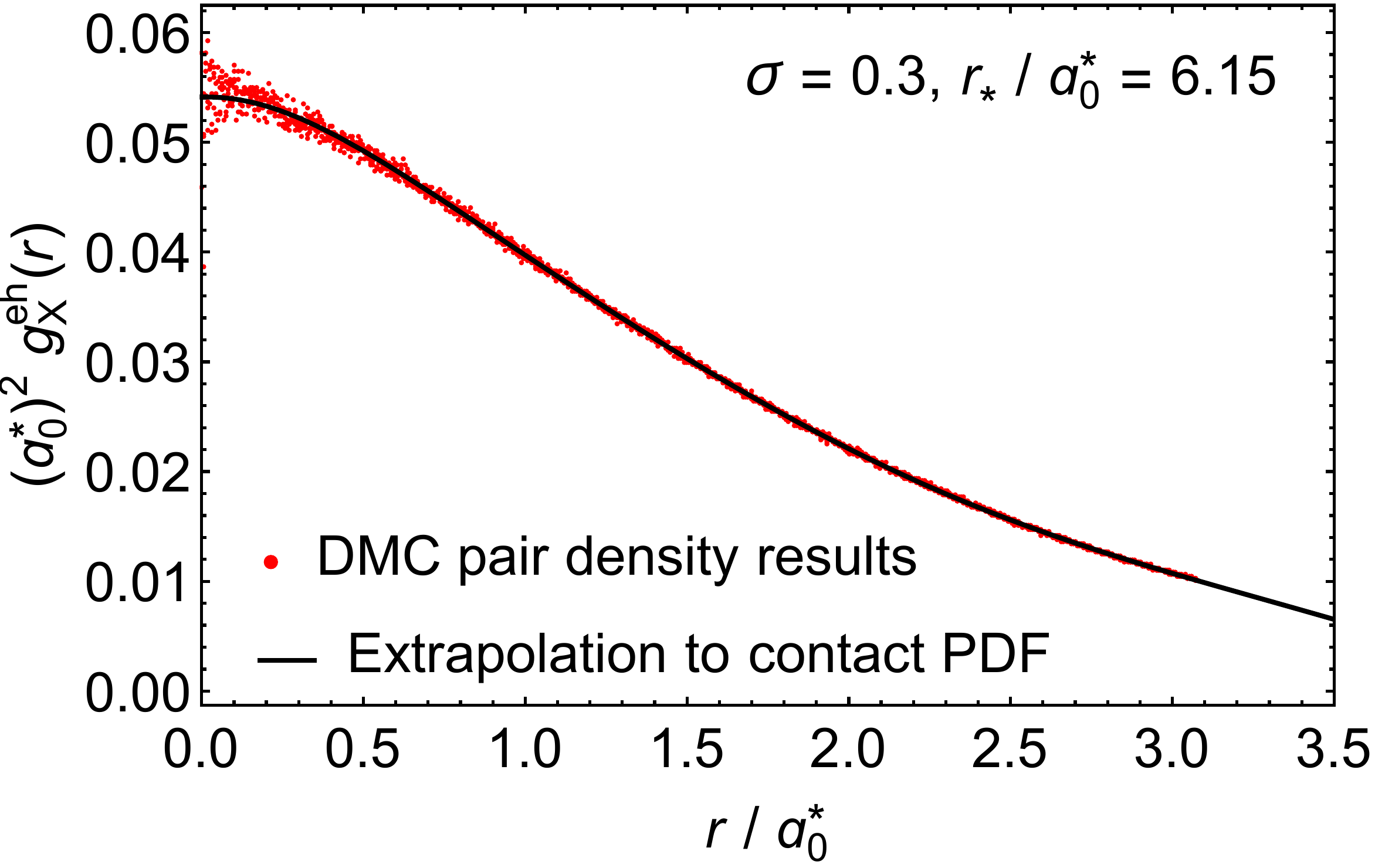}
\end{center}
\caption{(Color online) PDF of an exciton with $\sigma=0.3$ and
  $r_*=6.15a_0^\ast$. The contact PDF is extracted by fitting the
  numerical results to Eq.\ (\ref{eq:contactPDFformula}).
  \label{fig:pdf_x_extrapolation}}
\end{figure}

\begin{figure}[!htbp]
\begin{center}
\includegraphics[clip,width=0.45\textwidth]{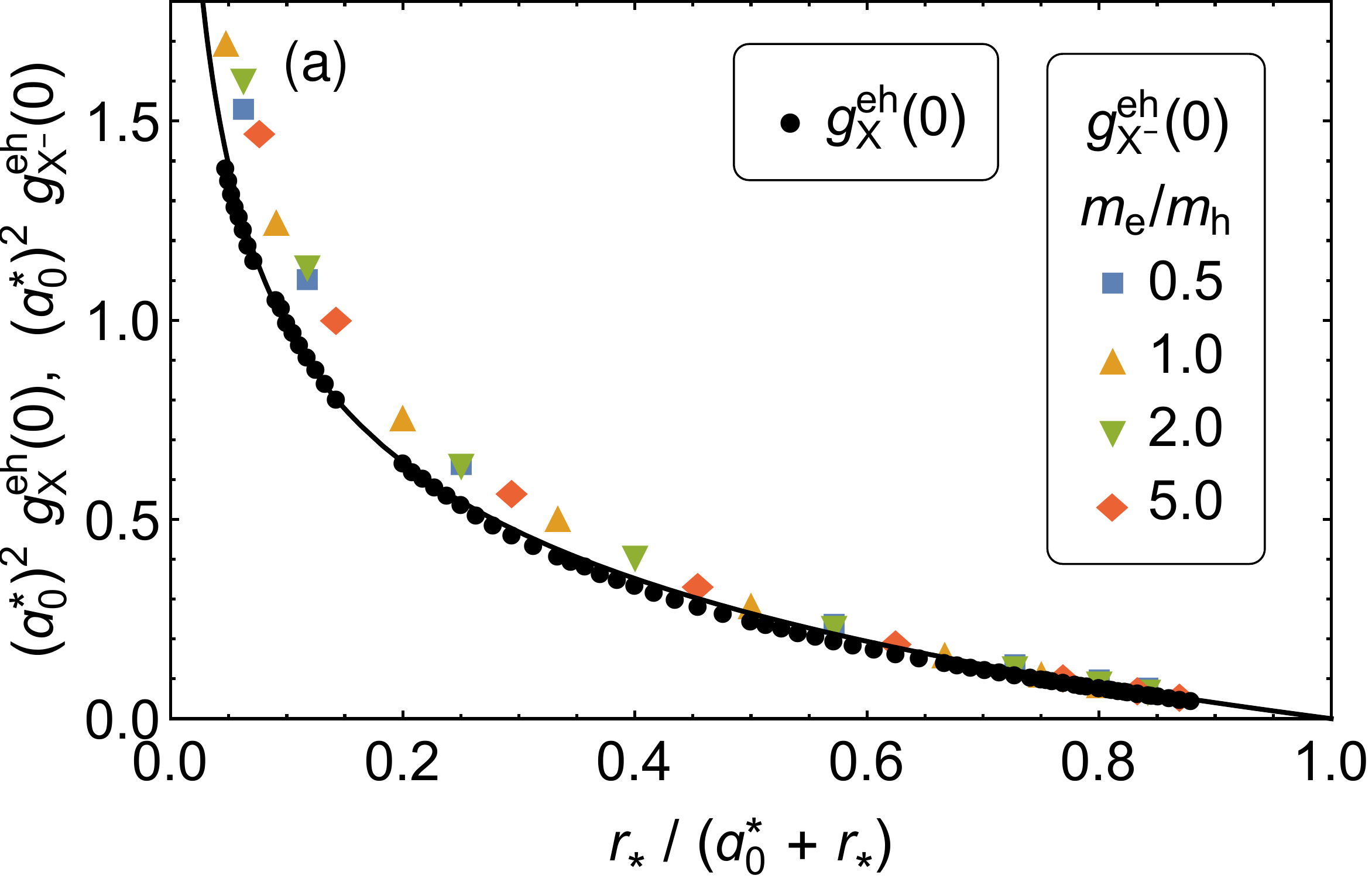}
\\ \includegraphics[clip,width=0.45\textwidth]{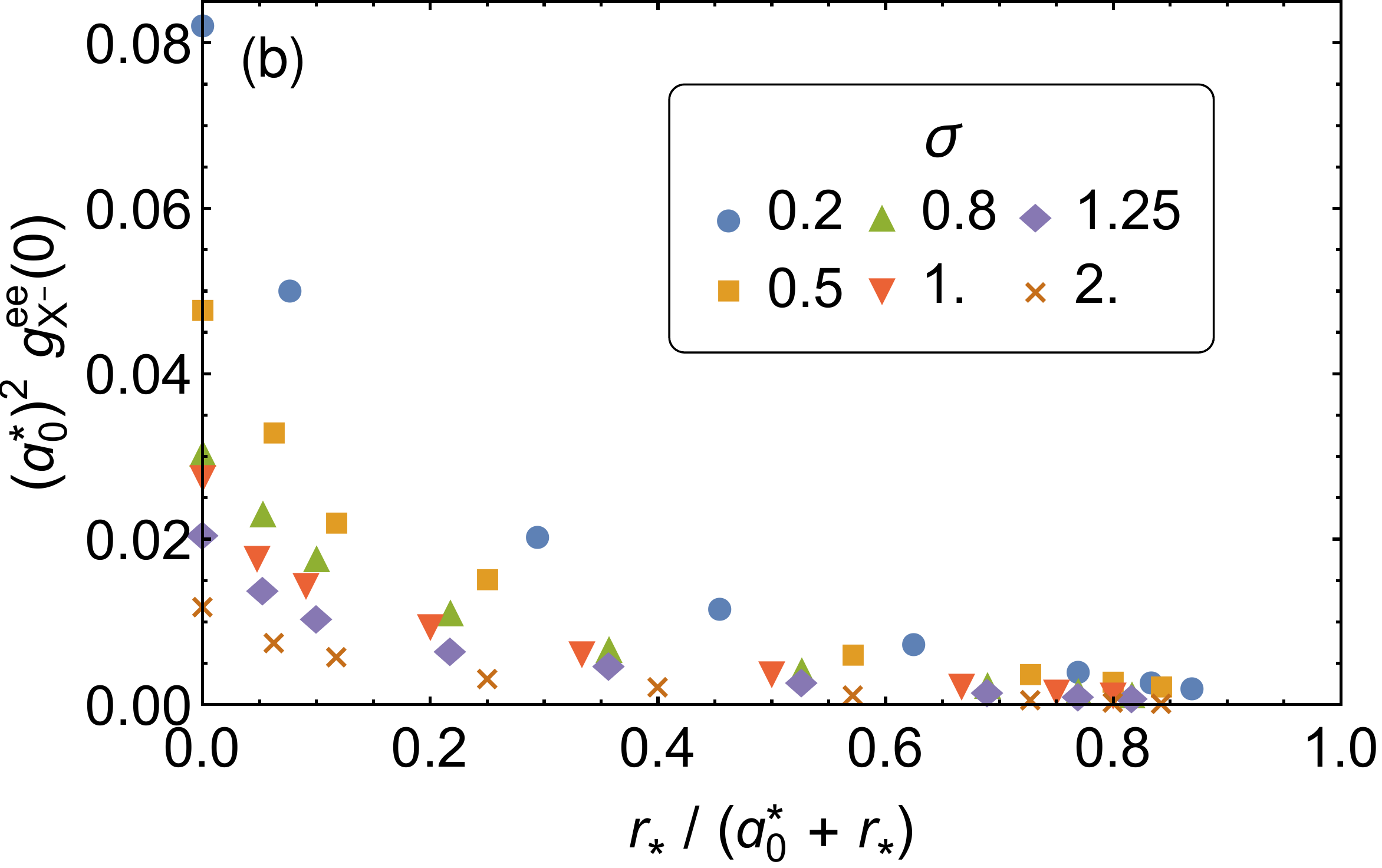}
\end{center}
\caption{(Color online) (a) Electron-hole contact PDFs of an exciton
  (in black) and a negative trion (in color).  (b) Electron-electron
  contact PDFs of a negative trion.  These data were presented in
  Ref.\ \onlinecite{Szyniszewski_2017}, and are shown here for
  completeness.
  \label{fig:pdf_trion}}
\end{figure}

Contact PDFs were extracted as described in
Sec.\ \ref{sec:pdf_extrap}. An example of a fit to Monte Carlo-sampled
PDF data is shown in Fig.\ \ref{fig:pdf_x_extrapolation}, and our
contact PDF results are shown in Fig.\ \ref{fig:pdf_trion}(a).  In all
our plots of contact PDFs the statistical error bars from the Monte
Carlo calculation are smaller than the symbols.  Unlike the DMC mixed
estimate of the energy, the extrapolated estimate of the PDF depends
on the stochastically optimized trial wave function and hence in some
cases slight noise in the $g(0)$ data is visible.

In the Supplemental Material we provide a program for evaluating our
fit to the total energy of an exciton [Eq.\ (\ref{eq:EX_fit})], as
well as fits to the binding energies of biexcitons, trions,
donor-bound excitons, donor-bound trions, and donor-bound
biexcitons.\cite{supplemental} In addition, the program reports fits
to contact PDFs for the different clusters.

\subsection{Biexcitons}

\subsubsection{Binding energies}\label{sec:results_biexcitons_bindingenergies}

\begin{figure}[!htbp]
\begin{center}
\includegraphics[clip,width=0.45\textwidth]{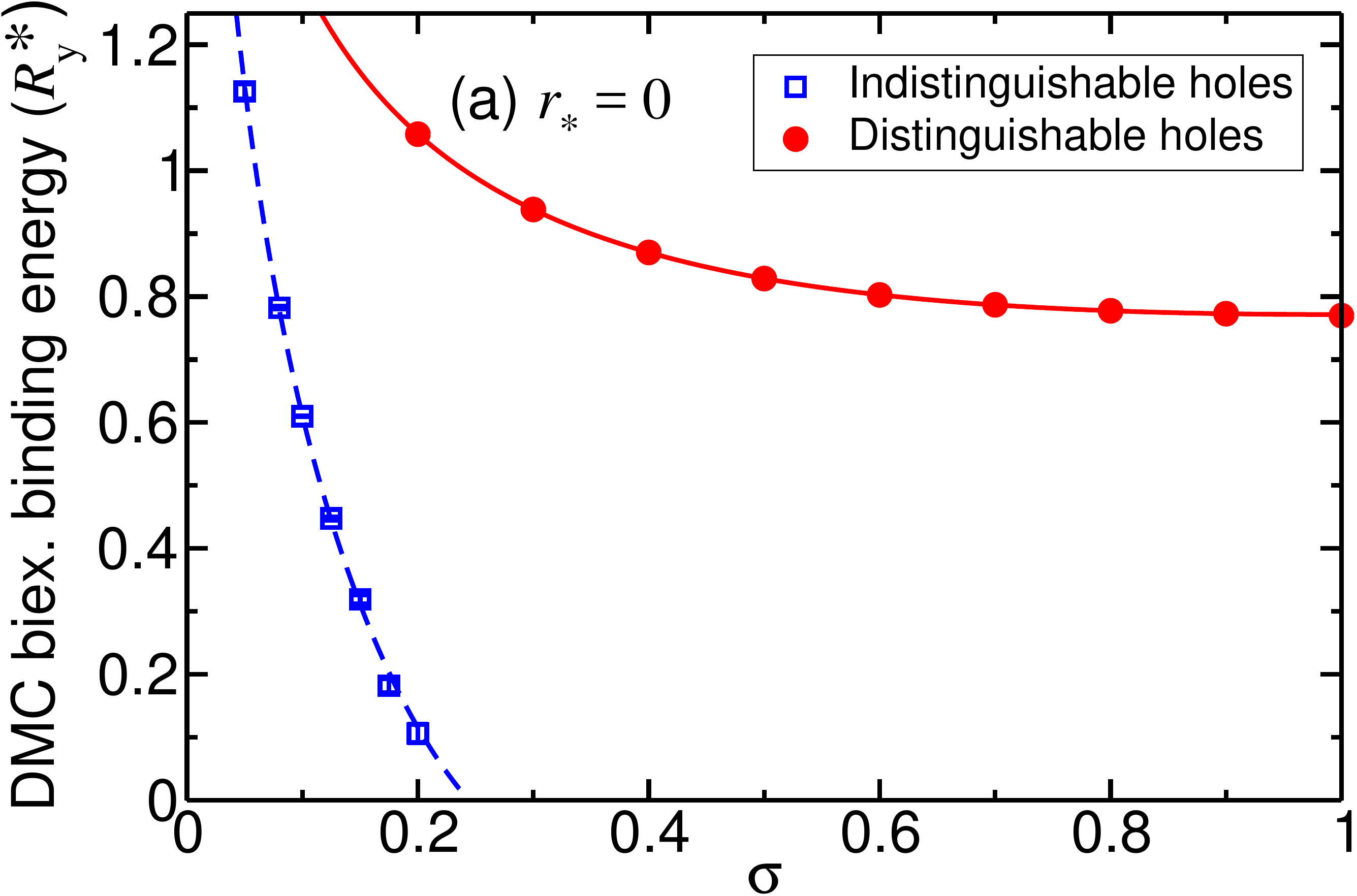}
\\ \includegraphics[clip,width=0.45\textwidth]{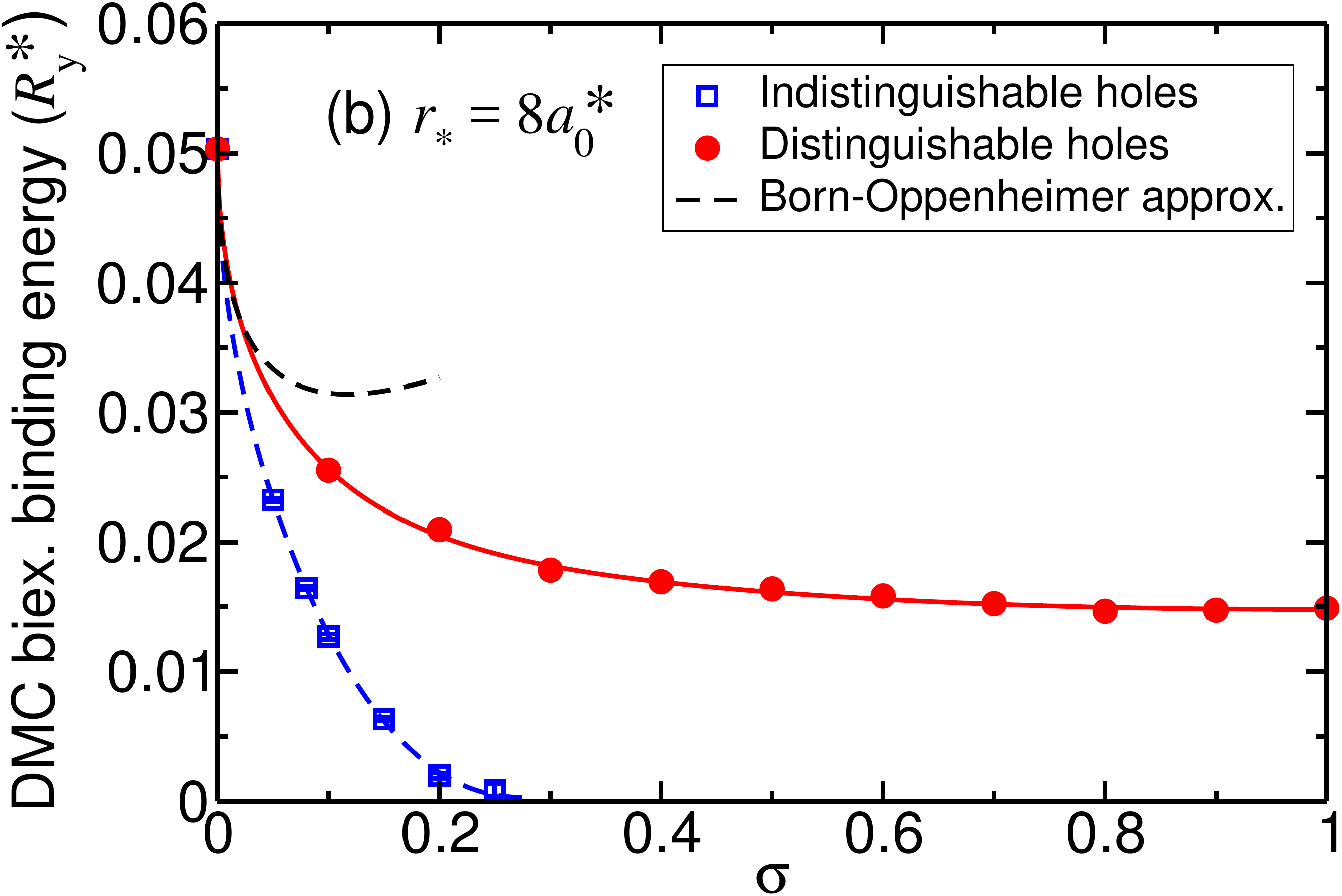}
\\ \includegraphics[clip,width=0.45\textwidth]{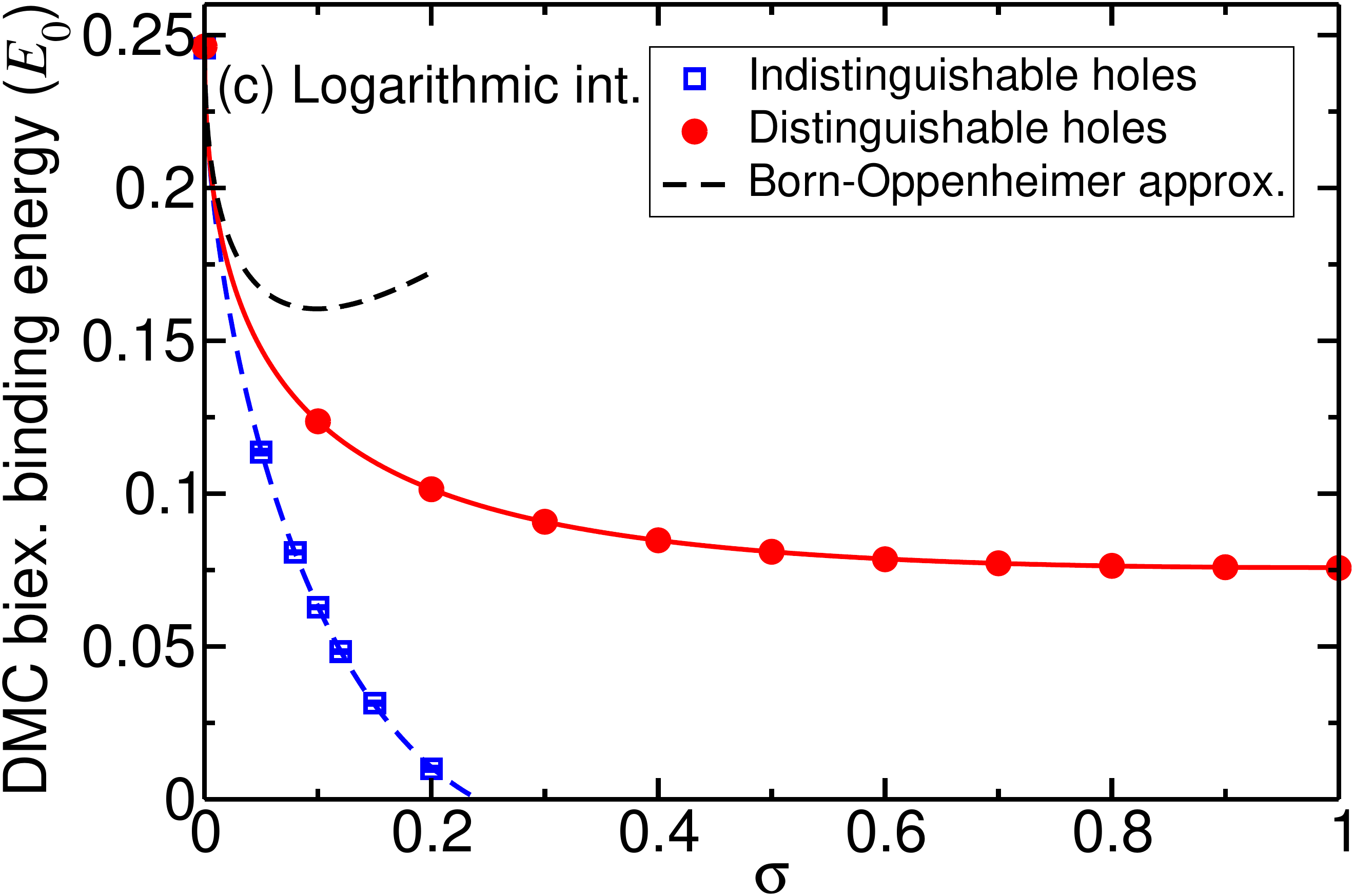}
\end{center}
\caption{(Color online) DMC binding energies of biexcitons with
  distinguishable electrons and distinguishable holes and biexcitons
  with distinguishable electrons and indistinguishable holes against
  mass ratio $\sigma$ with (a) the Coulomb interaction ($r_\ast=0$),
  (b) the Keldysh interaction with $r_\ast=8a^*_0$, and (c) the
  logarithmic interaction [Eq.\ (\ref{eq:log_int})] between charge
  carriers.
  \label{fig:be_2d_compare}}
\end{figure}

\begin{figure}[!htbp]
\begin{center}
\includegraphics[clip,width=0.45\textwidth]{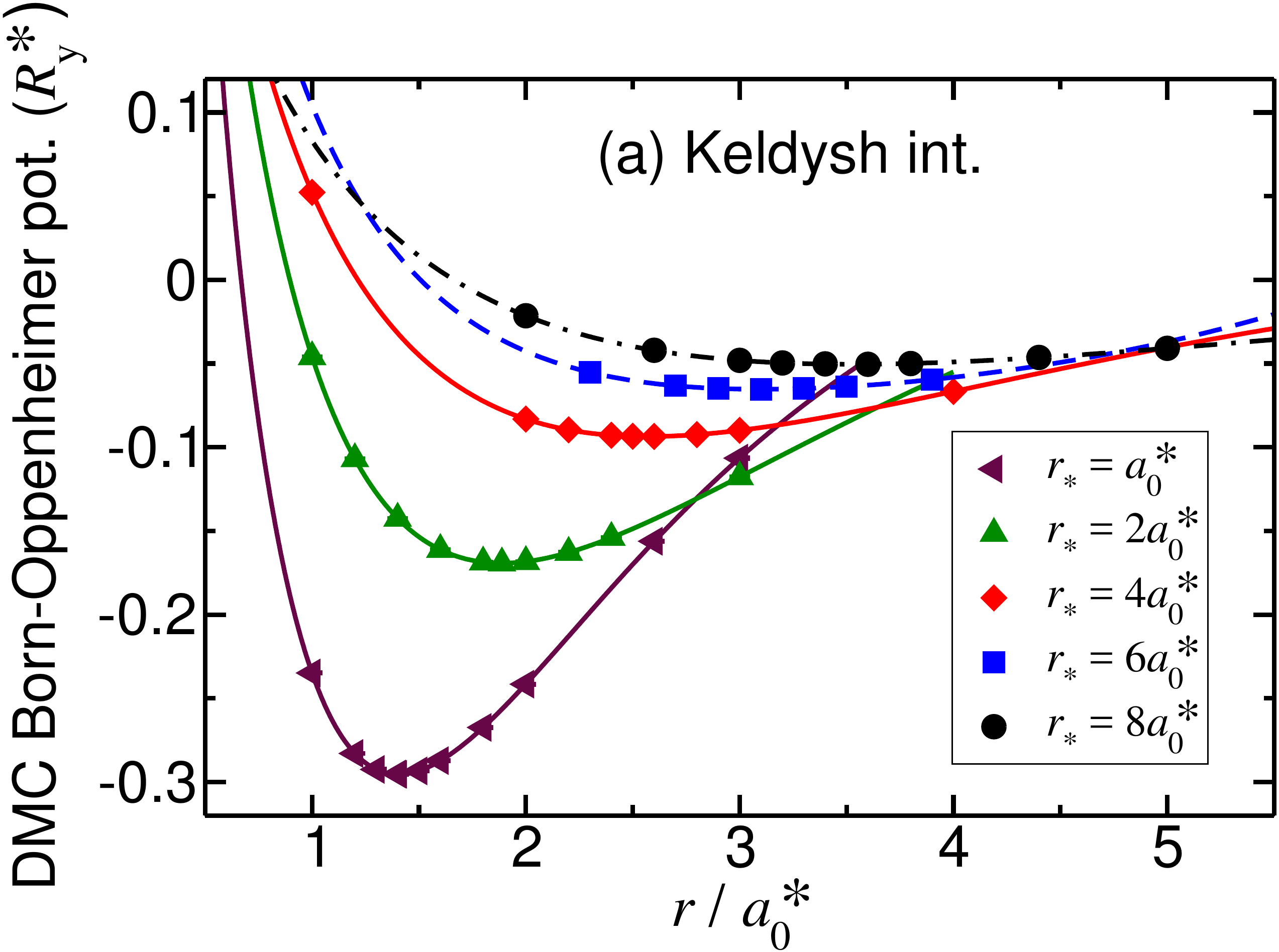}
\\ \includegraphics[clip,width=0.45\textwidth]{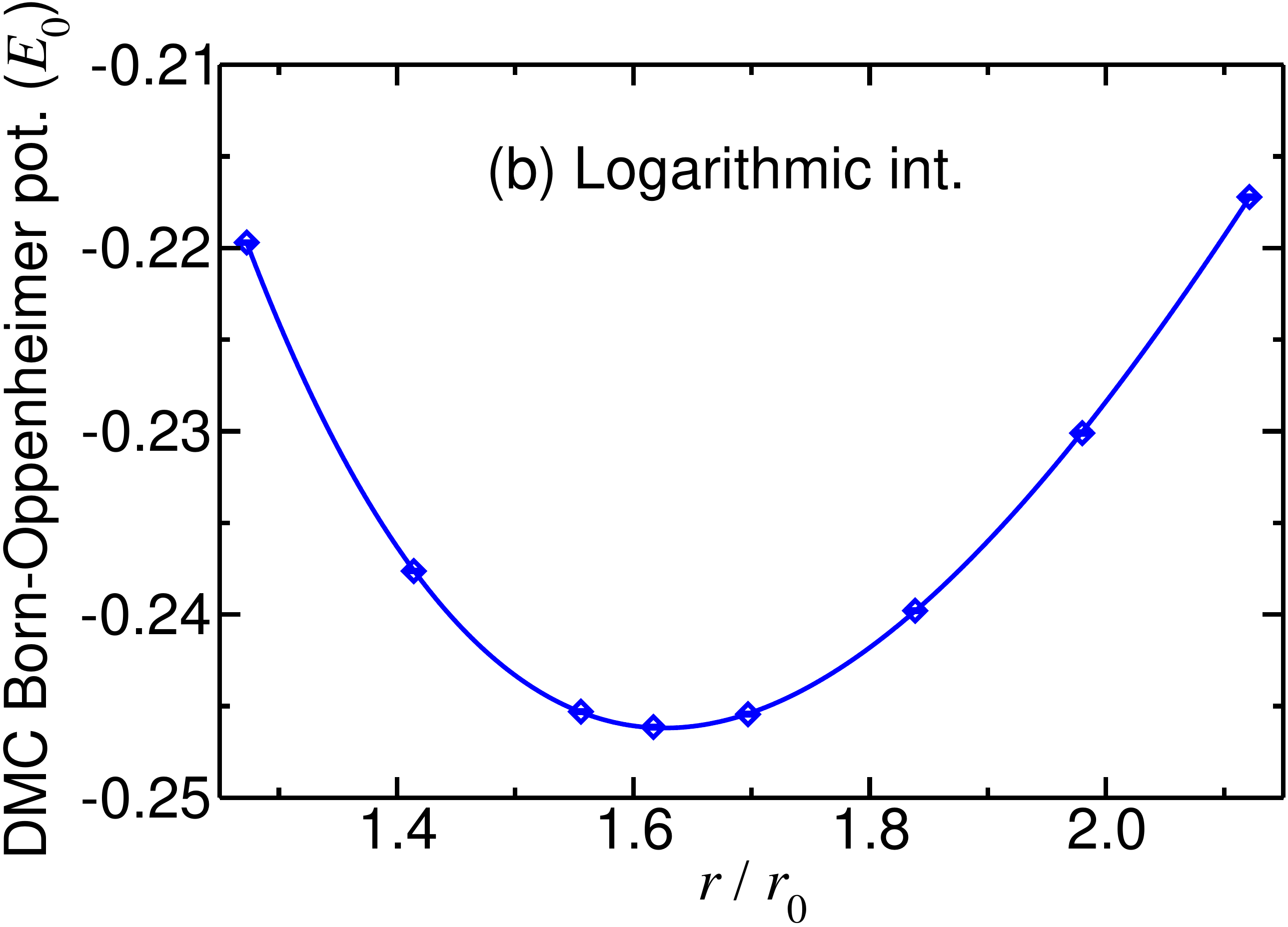}
\end{center}
\caption{(Color online) DMC Born-Oppenheimer potential energy of a
  heavy-hole biexciton with distinguishable electrons against the hole
  separation for (a) the Keldysh interaction with $r_\ast=a_0^\ast$,
  $2a_0^\ast$, $4a_0^\ast$, $6a_0^\ast$, and $8 a^*_0$; and (b) the
  logarithmic interaction between charge carriers.  The zero of the
  Born-Oppenheimer potential energy in the plot is twice the isolated
  exciton energy.
\label{fig:be_2d_dist}}
\end{figure}

We compare the stability of biexcitons with distinguishable and
indistinguishable holes in the limit of the Coulomb interaction
($r_\ast=0$) in Fig.\ \ref{fig:be_2d_compare}(a) and at $r_\ast=8
a^*_0$ in Fig.\ \ref{fig:be_2d_compare}(b). We find that biexcitons
with indistinguishable holes are unbound for $\sigma \gtrsim 0.3$,
while biexcitons consisting of distinguishable particles are bound at
all mass ratios.  The binding energies at $\sigma=0$ are obtained
using the Born-Oppenheimer potentials as a function of heavy-hole
separation $r$ plotted in Fig.\ \ref{fig:be_2d_dist}.  We fitted
$U(r)=\alpha+\beta\sqrt{r}+\gamma r+\delta r^2$, where $\alpha$,
$\beta$, $\gamma$, and $\delta$ are fitting parameters, to our DMC
data to find the minimum and the curvature about the minimum of the
Born-Oppenheimer potential.  For the logarithmic interaction we fitted
$U(r)=\zeta+\eta\exp(-r/d)+\kappa\log(r)$ to our data, where $\zeta$,
$\eta$, $d$, and $\kappa$ are fitting parameters.  The
Born-Oppenheimer approximation in Fig.\ \ref{fig:be_2d_compare}(b) for
heavy holes is in agreement with our DMC calculations at small
$\sigma$.  Analogous results obtained with the logarithmic interaction
are shown in Fig.\ \ref{fig:be_2d_compare}(c).  For $\sigma \gtrsim
0.2$, only biexcitons with distinguishable holes are stable.  Hence it
is only at extreme mass ratios, where exchange effects between the
heavy particles are negligible, that biexcitons with indistinguishable
particles are stable.

\begin{figure}[!htbp]
\begin{center}
\includegraphics[clip,width=0.45\textwidth]{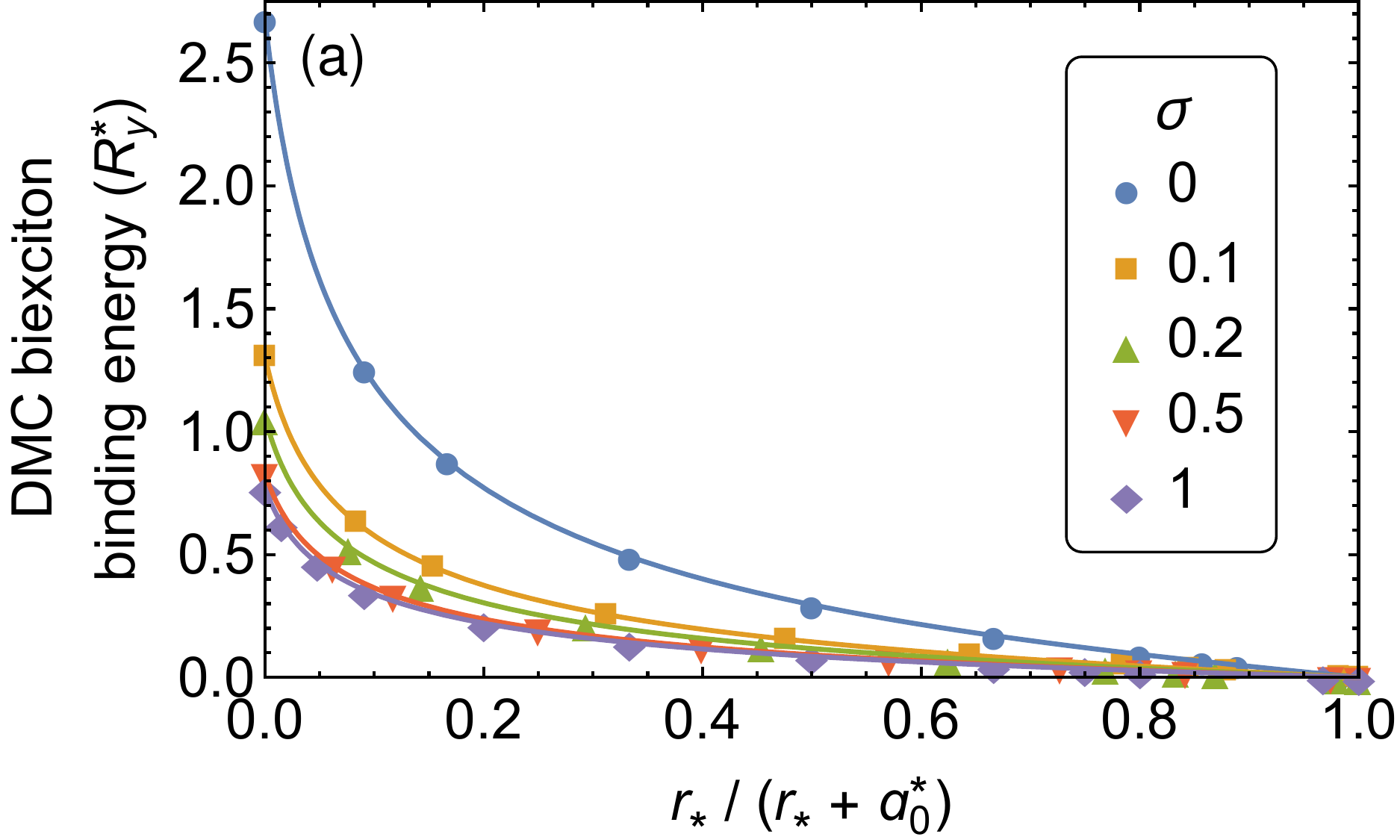}
\\ \includegraphics[clip,width=0.45\textwidth]{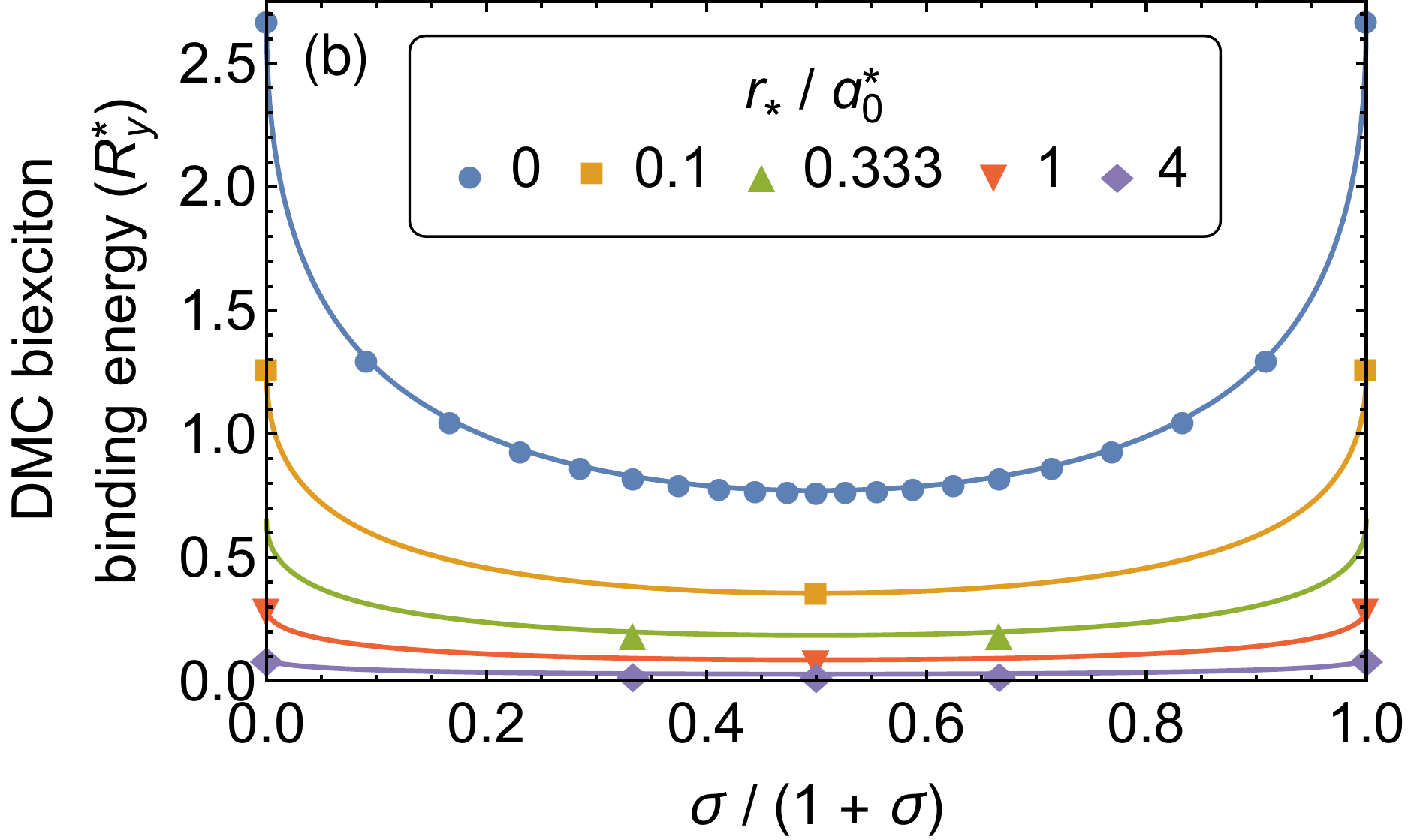}
\\ \includegraphics[clip,width=0.45\textwidth]{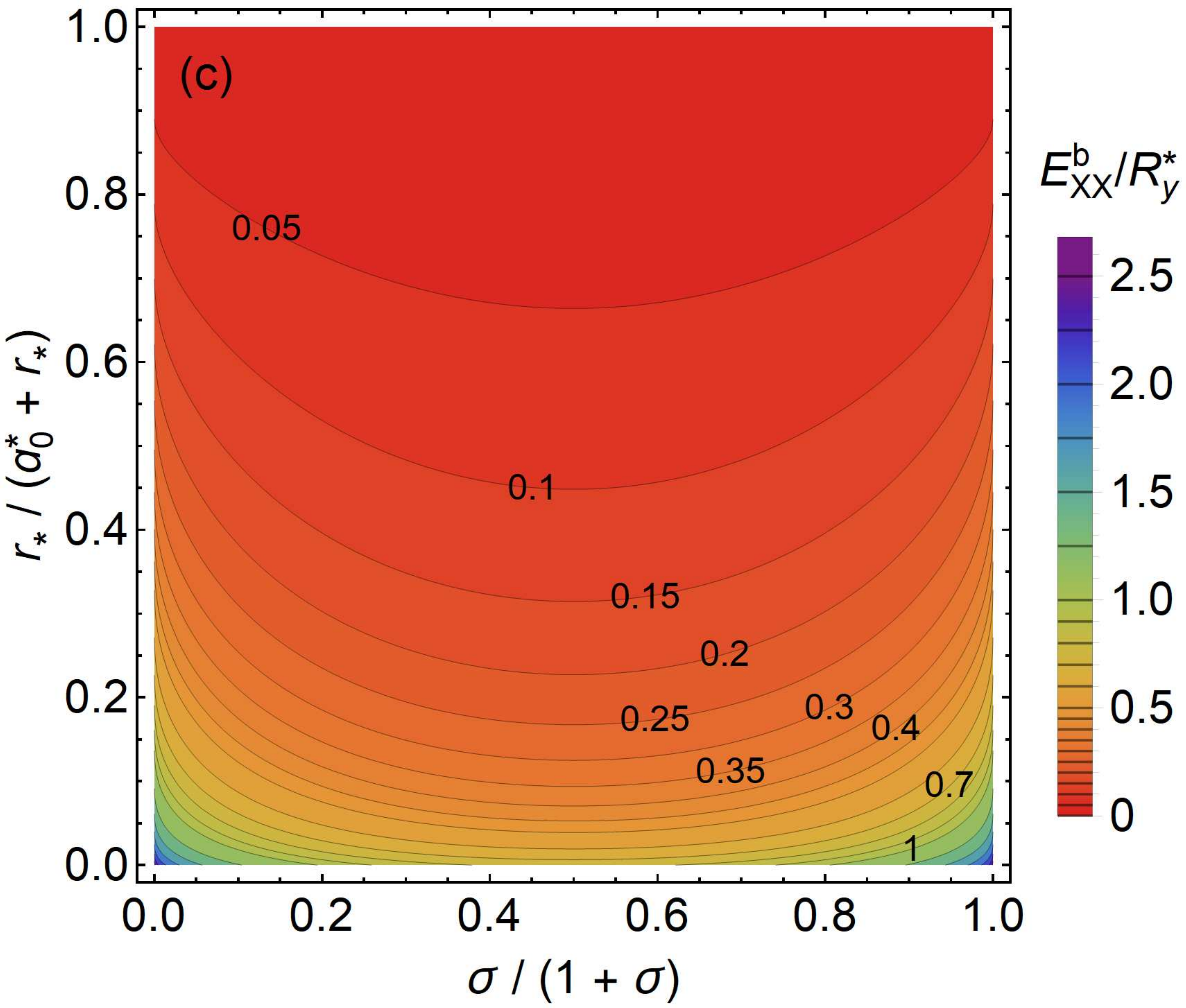}
\end{center}
\caption{(Color online) (a) DMC binding energies of biexcitons with
  distinguishable particles against rescaled susceptibility
  $r_\ast/(a^\ast_0+r_\ast)$. (b) DMC binding energies of biexcitons
  with distinguishable particles against rescaled mass ratio
  $\sigma/(1+\sigma)$. (c) DMC binding energies of biexcitons with
  distinguishable particles against rescaled susceptibility and
  rescaled mass ratio. The DMC results for distinguishable particles
  were reported in Ref.\ \onlinecite{Szyniszewski_2017}.
\label{fig:be_xx}}
\end{figure}

\begin{figure}[!htbp]
\begin{center}
\includegraphics[clip,width=0.45\textwidth]{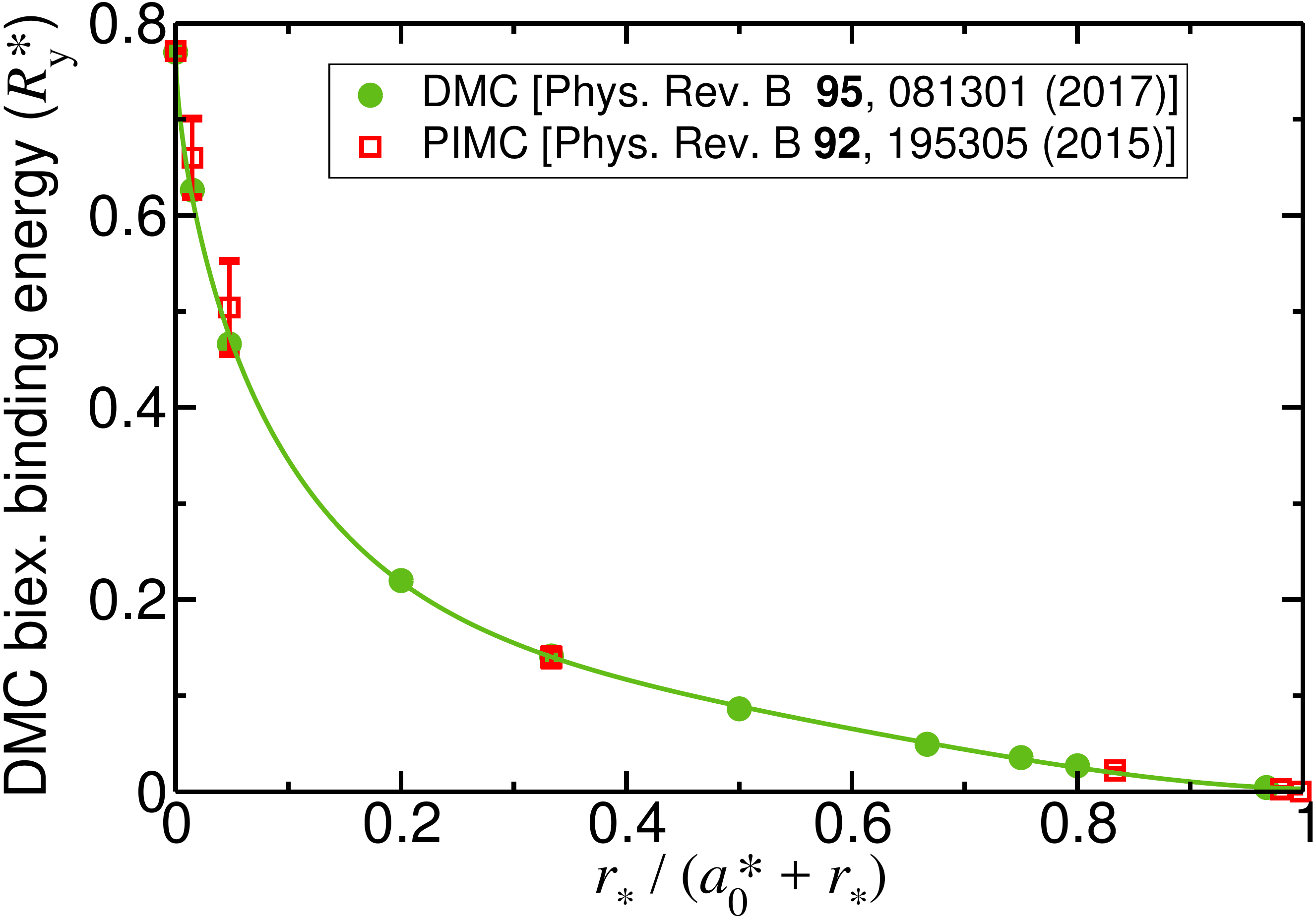}
\end{center}
\caption{(Color online) Binding energies of biexcitons with
  distinguishable particles of equal mass ($\sigma=1$) against
  rescaled susceptibility $r_\ast/(a^\ast_0+r_\ast)$, as calculated
  using DMC\cite{Szyniszewski_2017} and PIMC\@.\cite{Velizhanin_2015}
\label{fig:be_xx_cf}}
\end{figure}

Figure \ref{fig:be_xx} shows DMC binding energies for biexcitons with
distinguishable particles interacting via the Keldysh interaction as a
function of $x=\sigma/(1+\sigma)$ and rescaled in-plane susceptibility
$y=r_\ast/(a^*_0+r_\ast)$.  Our results are in agreement with
path-integral Monte Carlo (PIMC) data at finite $r_\ast$, as shown in
Fig.\ \ref{fig:be_xx_cf}.\cite{Velizhanin_2015} However, the PIMC data
obtained by Velizhanin and Saxena have much larger statistical errors
and they quoted a previous DMC result\cite{Stebe_1989} at $r_\ast=0$
due to the infeasibility of PIMC in this case.  The function
\begin{equation} \frac{E_{\rm XX}^{\rm b}}{R_{\rm y}^\ast}=\frac{(1 - y)
\sum_{i, j} a_{ij} \left[ x^{i / 2} + (1 - x)^{i / 2} \right] y^j}{1 +
    \sum_{i, j} b_{ij} \left[ x^{i / 2} + (1 - x)^{i / 2} \right] y^j}
 \label{eq:EbXX_fit}
 \end{equation}
containing $17$ fitting parameters $\{a_{ij}\}$ and $\{b_{ij}\}$, was
fitted to our DMC binding-energy data, giving a fractional error of
less than 1.5\% everywhere.  This choice of fitting function exhibits
the correct behavior as $\sigma \to 0$, as derived in
App.\ \ref{app:bo_biex}, and is also invariant under charge
conjugation ($m_{\rm e} \leftrightarrow m_{\rm h}$).  Equation
(\ref{eq:EbXX_fit}) accurately reproduces the DMC biexciton binding
energies over the whole space of possible susceptibility and
mass-ratio parameters, unlike the simple fitting functions reported in
Ref.\ \onlinecite{Szyniszewski_2017}. The latter are by construction
only valid in the currently experimentally relevant region and,
because of the relative simplicity of the fitting function, give
significantly larger fractional errors (up to 5\%) than
Eq.\ (\ref{eq:EbXX_fit}).  The fitted binding energy can be evaluated
using the program supplied in the Supplemental
Material.\cite{supplemental} Binding-energy results in the limit of
large $r_\ast$, where the interaction is of logarithmic form, are
given in Sec.\ \ref{sec:logint_results}.

In Table \ref{table:be_ex_theo}, we compare the DMC binding energies
of biexcitons in monolayer TMDCs with experiment and with previous
theoretical works.  Our DMC binding energies are in good agreement
with previous DMC binding energies where available,\cite{Mayers_2015}
and also with PIMC calculations.\cite{Kylanpaa_2015} The small
differences between DMC results in the literature must be due to the
use of different effective masses, etc. Unfortunately, the theoretical
biexciton binding energies are up to three times smaller than those
reported in experimental
works.\cite{Sie_2015,Mai_2015,Plechinger_2015,You_2015} There is also
a striking, qualitative disagreement with the experimental works
regarding the trion and biexciton binding energies: the Mott-Wannier
model with the Keldysh interaction predicts that the trion has a
larger binding energy than the
biexciton,\cite{Mayers_2015,Szyniszewski_2017} while the experimental
studies report that the biexciton peak occurs at lower energies than
the trion peak in photoluminescence spectra (i.e., that the biexciton
has a larger binding energy). The theoretical results are reported for
a free-standing monolayer; any screening by the substrate and
environment would further exacerbate the disagreement with experiment.

\begin{table*}
\caption{Total energies of excitons (X) and binding energies of
  biexcitons (XX) and trions (X$^-$ and X$^+$) with distinguishable
  particles for different monolayer TMDCs suspended in vacuum
  ($\epsilon=\epsilon_0$). We compare our results with values reported
  in the literature obtained by DMC, PIMC, hyperspherical harmonics
  (HH), stochastic variational (SV), and variational (V) methods. For
  each complex we use the values of $m_{\rm e}$, $m_{\rm h}$, and
  $r_\ast$ shown in bold in Table \ref{table:be_mass_rst} to evaluate
  the fits of Eqs.\ (\ref{eq:EbXX_fit}) and (\ref{eq:EbXm_fit}). Note
  that the ditellurides adopt a 2H stacking arrangement in bulk and
  few-layer samples, which may complicate comparison with experiment.
 \label{table:be_ex_theo}}
\begin{center}
\begin{tabular}{lcccccccc}
\hline \hline

& \multicolumn{2}{c}{Ex.\ tot.\ en.\ $E_{\rm X}$ (meV)} &
\multicolumn{6}{c}{Biexciton binding energy $E_{\rm XX}^{\rm b}$ (meV)} \\

\raisebox{1.5ex}[0pt]{TMDC} & ~~~Eq.\ (\ref{eq:EX_fit}) &
V\cite{Berkelbach_2013} & Eq.\ (\ref{eq:EbXX_fit}) &
DMC\cite{Mayers_2015} & PIMC\cite{Kylanpaa_2015} & Exp. &
SV\cite{Zhang_2015,Kidd_2016} & HH\cite{Kezerashvili_2016}\\\hline

MoS$_2$ & $-546.5$ & $-540$ & $23.5$ & $22.7(3)$ & $22.7(5)$ &
$70$,\cite{Mai_2014} $40$--$60$\cite{Sie_2015} & $22.5$ & $22.1$ \\

MoSe$_2$ & $-543.3$ & $-470$ & $22.9$ & $17.7(3)$ & $19.3(5)$ &
$\simeq 20$\cite{Hao_2017} & $18.4$ & $17.9$ \\

MoTe$_2$ & $-385.7$ & & $13.4$ & & $14.4(4)$ & &  \\

WS$_2$ & $-511.9$ & $-500$ & $23.4$ & $23.3(3)$ & $23.9(5)$ &
$65$,\cite{Plechinger_2015} $65$--$70$,\cite{Paradisanos_2017}
$45$\cite{Shang_2015} & $23.6$ & $23.1$ \\

WSe$_2$ & $-461.1$ & $-450$ & $20.0$ & $20.2(3)$ & $20.7(5)$ &
$52$\cite{You_2015} & $20.2$ & $19.8$ \\

WTe$_2$ & $-454.3$ & & $18.9$ & & & & &\\ \hline \hline
\end{tabular}

\vspace{1em}

\begin{tabular}{lcccccccccc}
\hline \hline

& \multicolumn{7}{c}{Negative trion binding energy $E_{\rm X^-}^{\rm b}$
  (meV)} & \multicolumn{3}{c}{Positive trion bind.\ en.\ $E_{\rm
    X^+}^{\rm b}$ (meV)} \\

\raisebox{1.5ex}[0pt]{TMDC} & Eq.\ (\ref{eq:EbXm_fit}) & DMC\cite{Mayers_2015}
& PIMC\cite{Kylanpaa_2015} & SV\cite{Zhang_2015,Kidd_2016} &
HH\cite{Kezerashvili_2016} & V\cite{Berkelbach_2013} & Exp. &
Eq.\ (\ref{eq:EbXm_fit}) & PIMC\cite{Kylanpaa_2015} & Exp. \\ \hline

MoS$_2$ & $35.0$ & $33.8(3)$ & $32.0(3)$ & $33.7$ & $32.8$ & $26$ &
$40$,\cite{Lin_2014,Lui_2014} $18.0(15)$,\cite{Mak_2013} $43$\cite{Zhang_2015_1}
& $34.9$ & $31.6(3)$ & \\

MoSe$_2$ & $34.5$ & $28.4(3)$ & $27.7(3)$ & $28.2$ & $27.6$ & $21$ &
$30$\cite{Ross_2013,Singh_2014} & $34.4$ & $27.8(3)$ &
$30$\cite{Ross_2013} \\

MoTe$_2$ & $22.5$ & $21.0(2)$ & & & & & $25$\cite{Lezama_2015} & $22.6$ &
$20.9(3)$ & \\

WS$_2$ & $33.5$ & $34.0(3)$ & $33.1(3)$ & $33.8$ & $33.1$ & $26$ &
$34$,\cite{Zhu_2015} $36$,\cite{Chernikov_2014}
$10$--$15$,\cite{Shang_2015} $30$,\cite{Plechinger_2015}
$45$\cite{Zhu_2014} & $33.6$ & $33.5(4)$ & \\

WSe$_2$ & $29.6$ & $29.5(3)$ & $28.5(3)$ & $29.5$ & $28.3$ & $22$ &
$30$\cite{Srivastava_2015,Jones_2013} & $29.6$ & $28.5(4)$ &
$30$,\cite{Srivastava_2015} $24$\cite{Jones_2013} \\

WTe$_2$ & $28.7$ & & & & & &  & $28.6$ & & \\

\hline \hline
\end{tabular}
\end{center}
\end{table*}

The ground-state wave function of a system of distinguishable
particles is nodeless, and so DMC provides exact solutions to
Mott-Wannier models of excitonic complexes.  Hence the disagreement
with experiment regarding the binding energies of biexcitons in 2D
semiconductors could only arise for one or more of the following four
reasons: (i) the 2D Mott-Wannier model with the Keldysh interaction
between charge carriers is incorrect or incomplete; (ii) the
parameters (band effective masses and $r_\ast$ values) used in the
model are incorrect; (iii) the exciton that remains after exciton
recombination in a biexciton is not in its ground
state;\cite{Zhang_2015} or (iv) the experimental spectra have been
misinterpreted or the peaks have been misclassified.

As explained in Sec.\ \ref{sec:pdfs_theory}, there should be an
additional contact interaction between charge carriers; however, the
Mott-Wannier model with the Keldysh interaction apparently provides a
good description\cite{Mayers_2015,Szyniszewski_2017} of the energies
of excitons and trions, and there is no obvious reason to believe that
contact interactions should be more important in a biexciton than in a
trion or exciton.  Moreover, it is unlikely that the contact
interactions could be responsible for the threefold difference between
the theoretical and experimental biexciton binding energies.

The second possibility is that the Mott-Wannier model is in principle
correct, but the band effective masses and in-plane susceptibilities
used in the model are incorrect.  These are taken from \textit{ab
  initio} calculations, which might not provide a sufficiently
accurate description of the electronic band structure. However, as
shown in Sec.\ \ref{sec:sensitivity}, the different mass ratios and
in-plane susceptibilities reported in the literature do not
significantly affect the binding energy; in fact the mass ratios and
in-plane susceptibilities would need to be in error by more than an
order of magnitude to explain the difference with experiment. Finally,
if inappropriate model parameters are responsible for the disagreement
with experiment regarding the biexciton binding energy, it is not
clear why the Mott-Wannier model with the same parameters apparently
provides a good description of excitons and trions.

We believe that the exciton that remains after exciton recombination
in a biexciton is unlikely to be in an excited state, because the
parent biexciton is in its nodeless ground state, which strongly
overlaps with the product of the ground states of the two daughter
excitons.

The misclassification of the experimental results may offer at least a
partial explanation of the disagreement.  By considering the behavior
of the photoluminescence emission intensity, it has been argued that
the observed peaks do indeed correspond to trions and
biexcitons.\cite{You_2015,Plechinger_2015} However, another
possibility is that they could correspond to charge-carrier complexes
involving donor or acceptor ions.  In particular, the energies
required to remove excitons from donor-bound biexcitons (see
Sec.\ \ref{sec:dbbiex_results}) are similar to the experimentally
observed ``biexciton'' binding energies. If donor-bound biexcitons are
responsible for the experimentally observed ``biexciton peak'' then we
might expect the intensity of the peak to depend strongly on the
doping of the sample. It is possible that other large charge-carrier
complexes could also contribute to the spectra.

None of these options offers an entirely satisfactory explanation of
the discrepancy.  Further experimental and theoretical modeling work
is required in order to understand the excitonic properties of 2D
semiconductors.

\subsubsection{Sensitivity of binding energies to effective masses and in-plane
  dielectric susceptibility \label{sec:sensitivity}}

In Table \ref{table:be_mass_rst} we compare the DMC binding energies
of biexcitons with distinguishable particles for a variety of
effective masses and in-plane screening lengths obtained by different
first-principles methods. Since a range of masses are reported in the
literature, we have taken the average of the reported masses that were
supposedly obtained using the same method.  The different model
parameters in the literature lead to a spread of about 1 meV in the
theoretical binding energies.

\begingroup \squeezetable
\begin{table}[!htbp]
\caption{Comparison of electron and hole effective masses and $r_\ast$
  values obtained by different \textit{ab initio} methods in the
  literature [many-body $GW$ calculations, and density functional
    theory either in the local density approximation (LDA) or using
    the Perdew-Burke-Ernzerhof (PBE) or Heyd-Scuseria-Ernzerhof
    (HSE06) exchange-correlation functionals]. Where multiple results
  are available using a given method, we have taken the average of the
  published results. The reported $r_\ast$ values and effective masses
  are not necessarily obtained using the same method. We assume that
  the materials are suspended in vacuum, i.e., that
  $\epsilon=\epsilon_0$.  The effective masses are reported in units
  of the bare electron mass $m_0$.  The binding energies $E^{\rm
    b}_{\rm XX}$ are calculated using Eq.\ (\ref{eq:EbXX_fit}).  The
  effective masses and $r_\ast$ values shown in bold are used to
  evaluate the binding energies reported in Tables
  \ref{table:be_ex_theo} and \ref{table:be_db_ab_complexes}.
 \label{table:be_mass_rst}}
\begin{center}
 \begin{tabular}{llr@{}lr@{}lr@{}lr@{}l} \hline \hline

 & \multicolumn{5}{c}{Effective masses} & & & & \\

 {\raisebox{1.5ex}[0pt]{TMDC}} & Method & \multicolumn{2}{c}{$m_{\rm
     e}/m_0$} & \multicolumn{2}{c}{$m_{\rm h}/m_0$} &
 \multicolumn{2}{c}{\raisebox{1.5ex}[0pt]{$r_\ast$ ({\AA})}} &
 \multicolumn{2}{c}{\raisebox{1.5ex}[0pt]{$E_{\rm XX}^{\rm b}$ (meV)}}
 \\ \hline

 \multirow{5}{*}{MoS$_2$} & $GW$\cite{Cheiwchanchamnangij_2012} &
 ${\bf 0}$&${\bf .35}$ & ${\bf 0}$&${\bf .428}$ & ${\bf 38}$&${\bf
   .62}$\cite{Cheiwchanchamnangij_2012} & $23$&$.5$ \\

& $G_0W_0$\cite{Ramasubramaniam_2012} & ~~$0$&$.60$ & ~$0$&$.54$ &
 $38$&$.62$\cite{Cheiwchanchamnangij_2012} & ~~~~~$24$&$.2$ \\

  &
 LDA\cite{Kadantsev_2012,Jin_2014,Chang_2014,Scalise_2014,Yu_2015,Kormanyos_2015}
 & $0$&$.495$ & $0$&$.576$ & $36$&$.28$\cite{Kumar_2012} & $25$&$.5$
 \\

  &
 PBE\cite{Wickramaratne_2014,Dong_2014,Yun_2012,Kormanyos_2015,Kylanpaa_2015}
 & $0$&$.470$ & $0$&$.575$ & $44$&$.69$\cite{Kylanpaa_2015} &
 $21$&$.1$ \\

 & HSE06\cite{Peelaers_2012} & $0$&$.37$ & $0$&$.44$ &
 $38$&$.62$\cite{Cheiwchanchamnangij_2012} & $23$&$.6$ \\

 \hline

  \multirow{4}{*}{MoSe$_2$} & $GW$\cite{Shi_2013} & ${\bf 0}$&${\bf
    .38}$ & ${\bf 0}$&${\bf .44}$ & $51$&$.71$\cite{Berkelbach_2013} &
  $18$&$.0$ \\

 & $G_0W_0$\cite{Ramasubramaniam_2012} & $0$&$.70$ & $0$&$.55$ &
  $51$&$.71$\cite{Berkelbach_2013} & $18$&$.7$ \\

 & LDA\cite{Jin_2014,Chang_2014,Kormanyos_2015} & $0$&$.59$ &
  $0$&$.686$ & ${\bf 39}$&${\bf .79}$\cite{Kumar_2012} & $23$&$.7$ \\

  & PBE\cite{Wickramaratne_2014,Liu_2011,Kormanyos_2015,Kylanpaa_2015}
  & $0$&$.546$ & $0$&$.643$ & $53$&$.16$\cite{Kylanpaa_2015} &
  $18$&$.1$ \\

 \hline

 \multirow{3}{*}{MoTe$_2$} & $G_0W_0$\cite{Ramasubramaniam_2012} &
 ${\bf 0}$&${\bf .69}$ & ${\bf 0}$&${\bf .66}$ & ${\bf 73}$&${\bf
   .61}$\cite{Kylanpaa_2015} & $13$&$.4$ \\

  & LDA\cite{Chang_2014}  & $0$&$.64$ &$0$&$.78$ &
 $73$&$.61$\cite{Kylanpaa_2015} & $15$&$.5$ \\

  & PBE\cite{Liu_2011} & $0$&$.575$ & $0$&$.702$ &
 $73$&$.61$\cite{Kylanpaa_2015} & $13$&$.4$ \\

 \hline

 \multirow{4}{*}{WS$_2$} & $GW$\cite{Shi_2013} & ${\bf 0}$&${\bf .27}$
 & ${\bf 0}$&${\bf .32}$ & ${\bf 37}$&${\bf
   .89}$\cite{Berkelbach_2013} & $23$&$.4$ \\

  & $G_0W_0$\cite{Ramasubramaniam_2012} & $0$&$.44$ & $0$&$.45$ &
 $37$&$.89$\cite{Berkelbach_2013} & $24$&$.1$ \\

  & LDA\cite{Jin_2014,Chang_2014,Kormanyos_2015} & $0$&$.312$ &
 $0$&$.422$ & $32$&$.42$\cite{Kumar_2012} & $27$&$.7$ \\

  & PBE\cite{Wickramaratne_2014,Liu_2011,Kormanyos_2015,Kylanpaa_2015}
 & $0$&$.328$ & $0$&$.402$ & $40$&$.17$\cite{Kylanpaa_2015} &
 $22$&$.6$ \\

 \hline

 \multirow{4}{*}{WSe$_2$} & $GW$\cite{Shi_2013} & ${\bf 0}$&${\bf
   .29}$ & ${\bf 0}$&${\bf .34}$ & ${\bf 45}$&${\bf
   .11}$\cite{Berkelbach_2013} & $20$&$.0$ \\

 & $G_0W_0$\cite{Ramasubramaniam_2012}  & $0$&$.53$ & $0$&$.52$ &
 $45$&$.11$\cite{Berkelbach_2013} & $20$&$.8$ \\

& LDA\cite{Jin_2014,Chang_2014,Kormanyos_2015} & $0$&$.36$ &
 $0$&$.476$ & $34$&$.72$\cite{Kumar_2012} & $26$&$.2$ \\

& PBE\cite{Wickramaratne_2014,Kormanyos_2015,Kylanpaa_2015} &
 $0$&$.342$ & $0$&$.428$ & $47$&$.57$\cite{Kylanpaa_2015} & $19$&$.4$
 \\

 \hline

 \multirow{2}{*}{WTe$_2$} & LDA\cite{Kormanyos_2015} & ${\bf 0}$&${\bf
   .325}$ & ${\bf 0}$&${\bf .460}$ & ${\bf 49}$&${\bf
   .56}$\cite{Kumar_2012,Lu_2016} & $18$&$.9$ \\

 & PBE\cite{Lv_2015} & $0$&$.307$ & $0$&$.51$ &
 $49$&$.56$\cite{Kumar_2012,Lu_2016} & $19$&$.3$ \\

 \hline \hline
\end{tabular}
\end{center}
\end{table}
\endgroup

The sensitivities of the exciton total energy and the trion and
biexciton binding energies to the model parameters are reported in
Table \ref{table:sensitivity}.  The energies depend relatively weakly
on the in-plane permittivity $r_\ast$; the errors arising from the
uncertainty in the effective mass almost certainly dominate errors
arising from the uncertainty in $r_\ast$.  The sensitivity of the
exciton energy to the effective masses is an order of magnitude larger
than the sensitivity of the trion binding energy, which is in turn an
order of magnitude larger than the sensitivity of the biexciton
binding energy.  To account for the 30--40 meV disagreement with
experiment over the biexciton binding energy the effective masses
would have to be more than an order of magnitude larger than the
\textit{ab initio} values reported in Table \ref{table:be_mass_rst}
and/or the $r_\ast$ value would have to be an order of magnitude
smaller.  While there is still appreciable uncertainty in the
\textit{ab initio} effective mass and $r_\ast$ values, it seems very
unlikely that both density functional theory and many-body $GW$
calculations would be in error by more than an order of magnitude.

\begin{table*}
\caption{Sensitivity of binding energies to the three parameters that
  characterize the Mott-Wannier-Keldysh model of excitonic complexes
  in 2D semiconductors suspended in vacuum.  The derivatives are
  evaluated using the effective mass and in-plane permittivity
  parameters reported in bold for different TMDCs in Table
  \ref{table:be_mass_rst}. $m_0$ is the bare electron
  mass.  \label{table:sensitivity}}
\begin{tabular}{lr@{}lr@{}lr@{}lr@{}lr@{}lr@{}lr@{}lr@{}lr@{}l}
\hline \hline

\multirow{3}{*}{TMDC} & \multicolumn{2}{c}{$\frac{\partial E_{\rm
      X}}{\partial m_{\rm e}}$} & \multicolumn{2}{c}{$\frac{\partial
    E_{\rm X}}{\partial m_{\rm h}}$} &
\multicolumn{2}{c}{$\frac{\partial E_{\rm X}}{\partial r_\ast}$} &
\multicolumn{2}{c}{$\frac{\partial E^{\rm b}_{\rm XX}}{\partial m_{\rm
      e}}$} & \multicolumn{2}{c}{$\frac{\partial E^{\rm b}_{\rm
      XX}}{\partial m_{\rm h}}$} & \multicolumn{2}{c}{$\frac{\partial
    E^{\rm b}_{\rm XX}}{\partial r_\ast}$} &
\multicolumn{2}{c}{$\frac{\partial E^{\rm b}_{\rm X^-}}{\partial
    m_{\rm e}}$} & \multicolumn{2}{c}{$\frac{\partial E^{\rm b}_{\rm
      X^-}}{\partial m_{\rm h}}$} & \multicolumn{2}{c}{$\frac{\partial
    E^{\rm b}_{\rm X^-}}{\partial r_\ast}$} \\

& \multicolumn{2}{c}{(meV/$m_0$)} & \multicolumn{2}{c}{(meV/$m_0$)} &
\multicolumn{2}{c}{(meV/{\AA})} & \multicolumn{2}{c}{(meV/$m_0$)} &
\multicolumn{2}{c}{(meV/$m_0$)} & \multicolumn{2}{c}{(meV/{\AA})} &
\multicolumn{2}{c}{(meV/$m_0$)} & \multicolumn{2}{c}{(meV/$m_0$)} &
\multicolumn{2}{c}{(meV/{\AA})} \\

\hline

MoS$_2$ & ~~~$-240$& & ~~~$-160$& & ~~~$10$& & ~~~~$1$&$.2$ &
~~~~$5$&$.4$ & ~~$-0$&$.56$ & ~~~~$10$& & ~~~~$9$&$.2$ & ~~$-0$&$.72$
\\

MoSe$_2$ & $-210$& & $-160$& & $9$&$.9$ & $1$&$.4$ & $4$&$.3$ &
$-0$&$.53$ & $8$&$.8$ & $8$&$.7$ & $-0$&$.70$ \\

WS$_2$ & $-310$& & $-220$& & $9$&$.5$ & $2$&$.1$ & $6$&$.7$ &
$-0$&$.56$ & $13$& & $12$& & $-0$&$.70$ \\

WSe$_2$ & $-240$& & $-180$& & $7$&$.3$ & $1$&$.6$ & $5$&$.1$ &
$-0$&$.41$ & $10$& & $9$&$.6$ & $-0$&$.52$ \\

\hline \hline
\end{tabular}
\end{table*}

\subsubsection{PDFs}\label{sec:NumRes_Biex_PDF}

\begin{figure}[!htbp]
\begin{center}
\includegraphics[clip,width=0.45\textwidth]{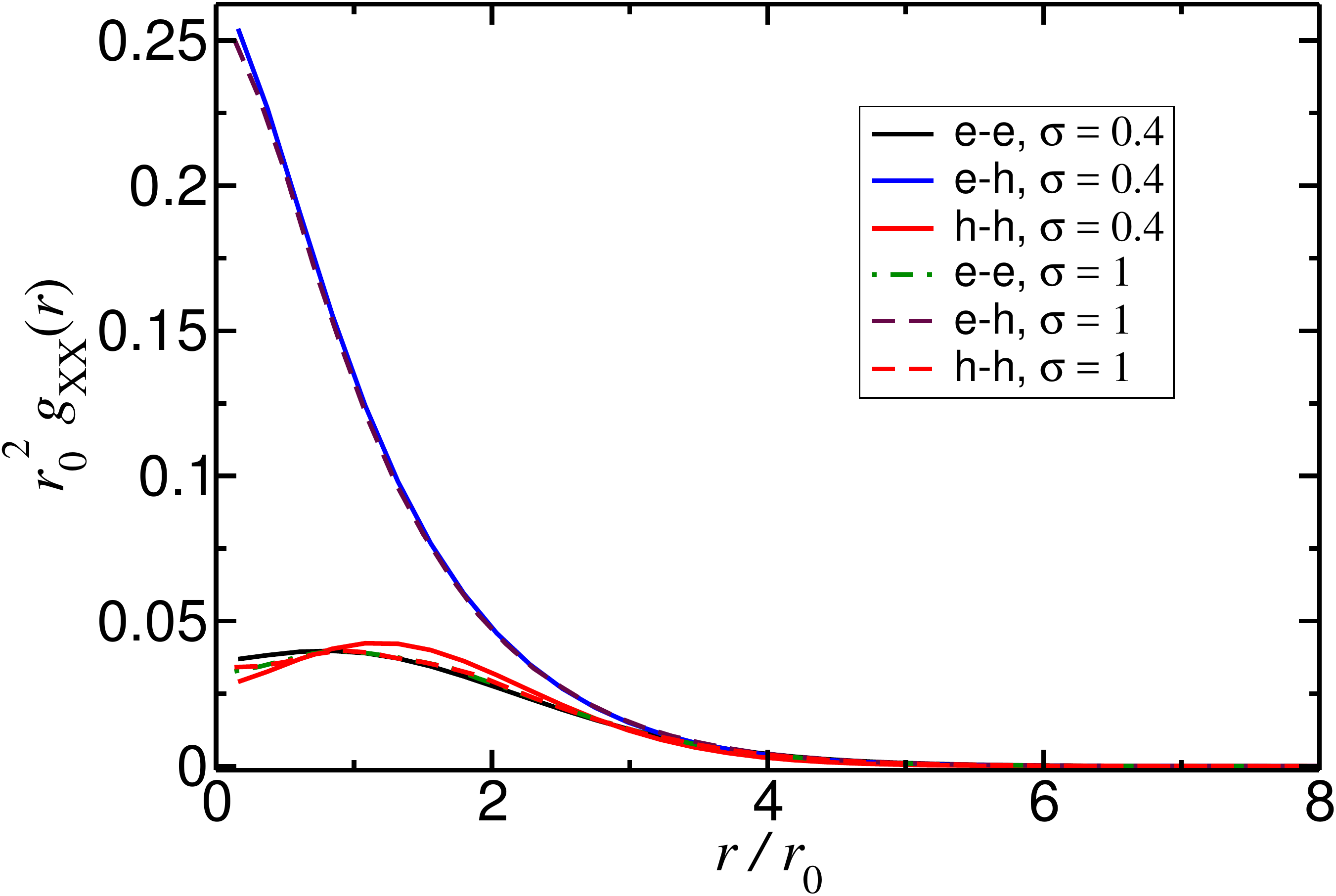}
\end{center}
\caption{(Color online) PDF $g_{\rm XX}(r)$ of a biexciton with
  distinguishable particles interacting via the logarithmic
  interaction plotted against interparticle separation at two
  different electron-hole mass ratios $\sigma$.
\label{fig:log_pcf}}
\end{figure}

In Fig.\ \ref{fig:log_pcf}, we show the PDFs of biexcitons with
distinguishable particles interacting via the logarithmic interaction
for two different mass ratios, $\sigma=0.4$ and $\sigma=1$. The
long-range biexciton wave function is relatively independent of the
mass ratio.  However, at short range the electron-hole PDF shows a
peak near the separation that corresponds to the minimum of the
Born-Oppenheimer potential-energy surface, which gets more pronounced
at extreme mass ratios.  As expected, the physical size of the
biexciton is a low multiple of $r_0$.

\begin{figure}[!htbp]
\begin{center}
\includegraphics[clip,width=0.45\textwidth]{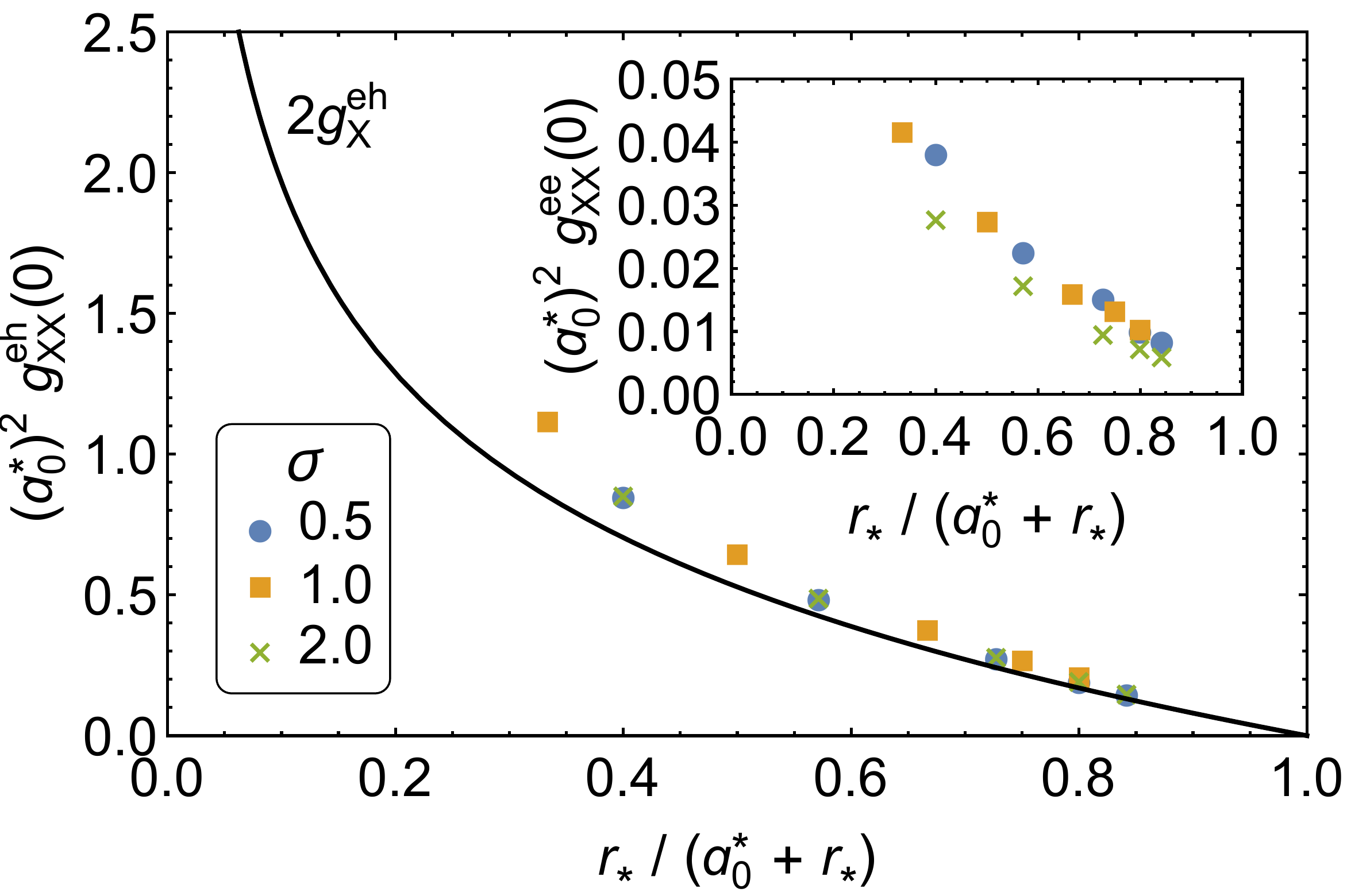}
\end{center}
\caption{(Color online) Electron-hole contact PDF of a biexciton with
  distinguishable particles against rescaled susceptibility. The black
  line indicates twice the exciton electron-hole contact PDF\@. The
  inset shows the electron-electron contact PDF\@.  These data were
  presented in Ref.\ \onlinecite{Szyniszewski_2017}, and are shown
  here for completeness.  \label{fig:pdf_xx}}
\end{figure}

Figure \ref{fig:pdf_xx} presents the electron-hole and
electron-electron contact $(r=0)$ PDFs for a biexciton. Notice that
$g_{\rm XX}^{\rm eh}\approx 2g^{\rm eh}_{\rm X}$.  Fits to the contact
PDFs can be evaluated using the program supplied as Supplemental
Material.\cite{supplemental}

\subsection{Trions}

\begin{figure}[!htbp]
\begin{center}
\includegraphics[clip,width=0.45\textwidth]{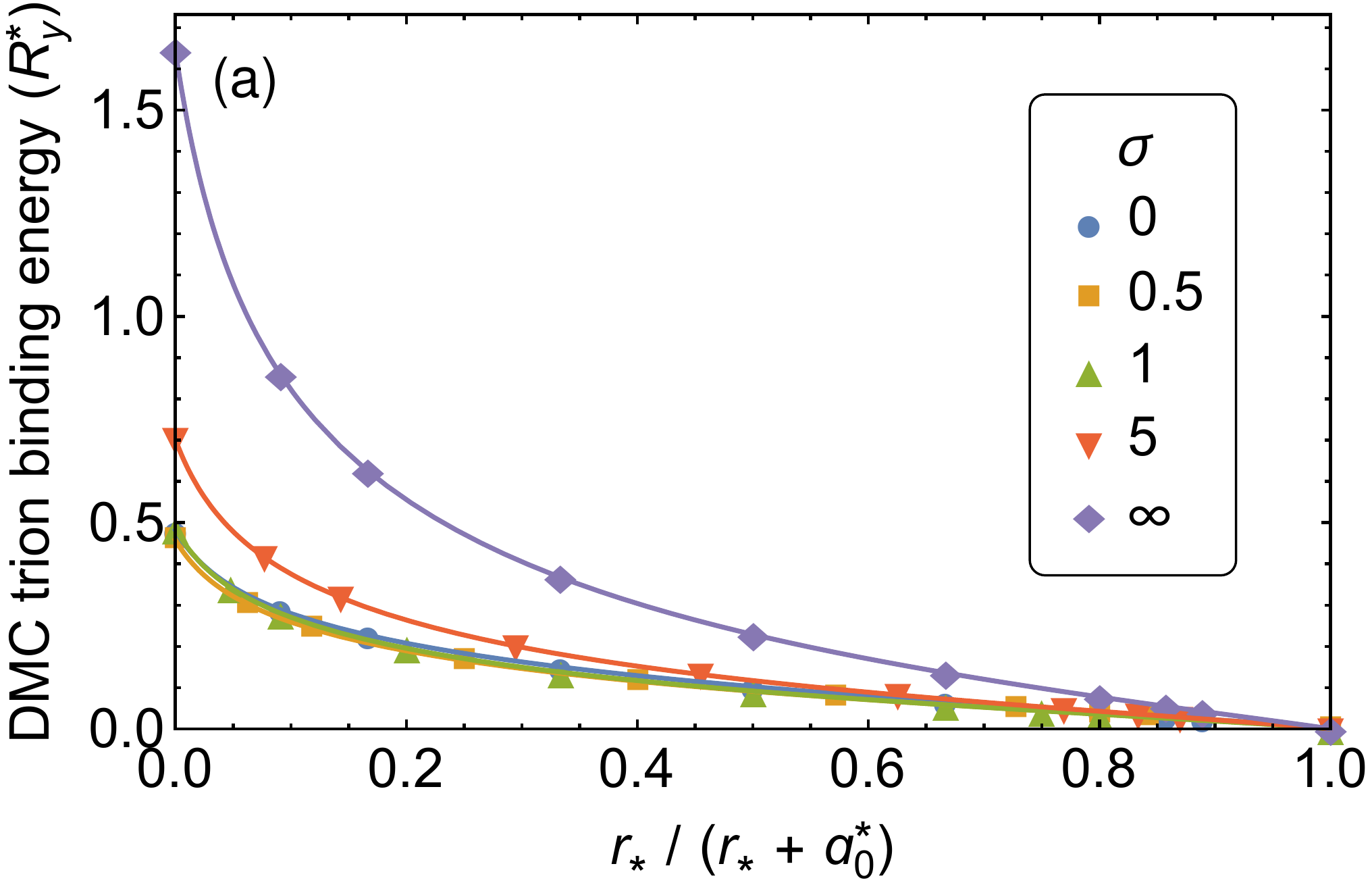}
\\ \includegraphics[clip,width=0.45\textwidth]{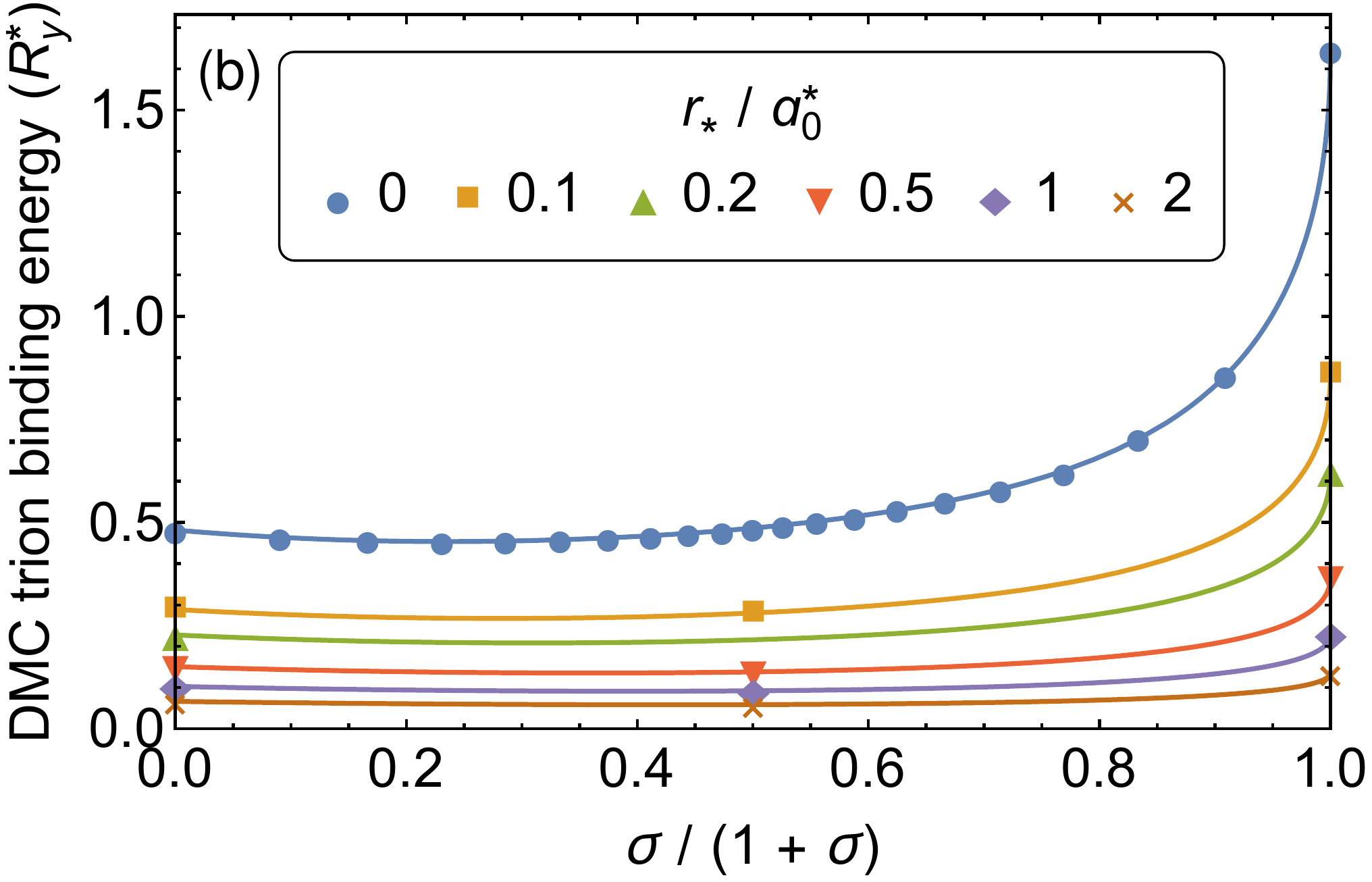}
\\ \includegraphics[clip,width=0.45\textwidth]{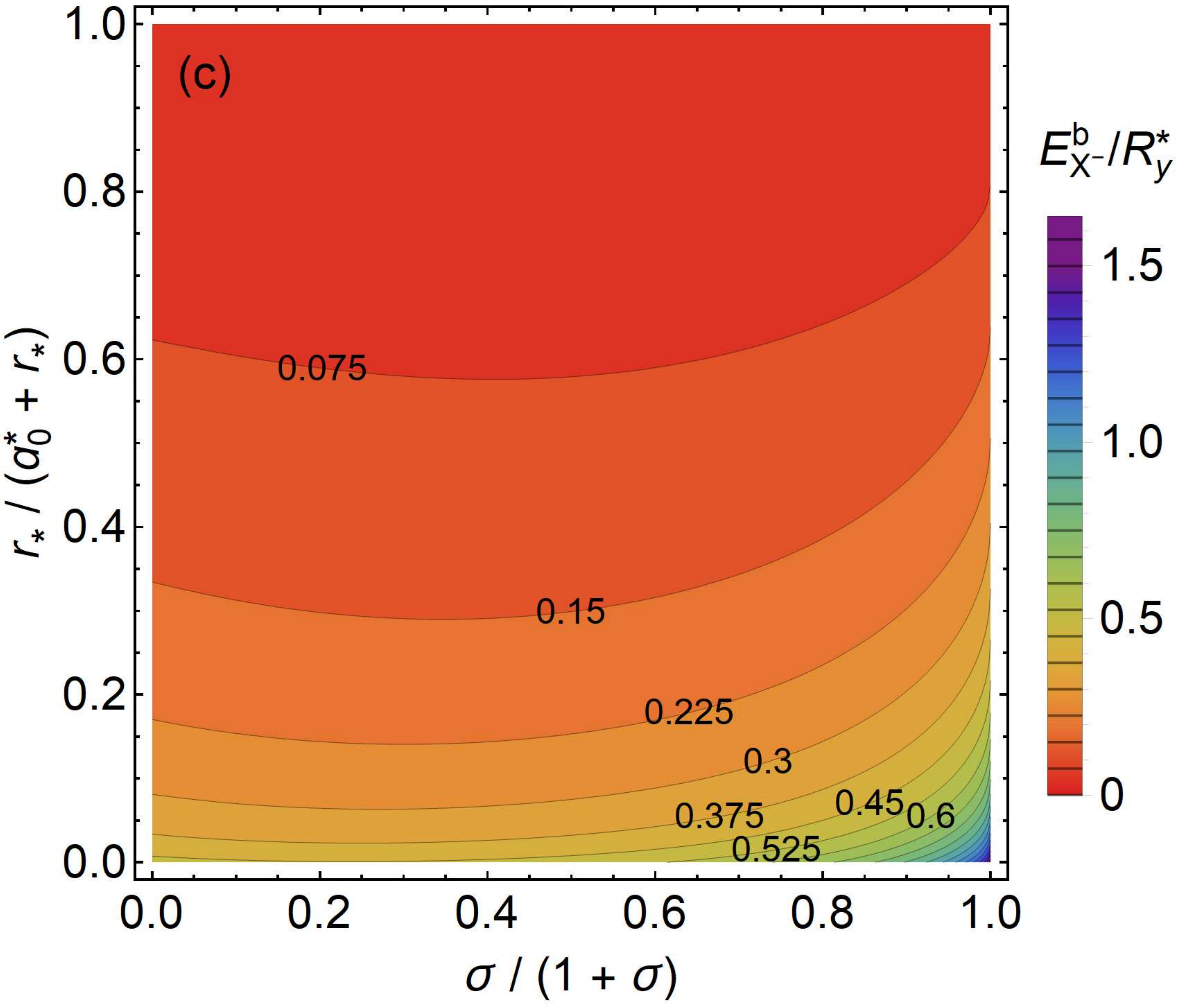}
\end{center}
\caption{(Color online) (a) DMC binding energies of trions with
  distinguishable particles against rescaled susceptibility
  $r_*/(a_0^*+r_*)$. (b) DMC binding energies of trions with
  distinguishable particles against rescaled mass ratio
  $\sigma/(1+\sigma)$. (c) DMC binding energies of trions with
  distinguishable particles against rescaled susceptibility and
  rescaled mass ratio. These data were presented in
  Ref.\ \onlinecite{Szyniszewski_2017}, and are shown here for
  completeness.
  \label{fig:be_trion}}
\end{figure}

The binding energies of negative trions are presented in
Fig.\ \ref{fig:be_trion}. We have fitted the function
\begin{equation}
  \frac{E^{\text{b}}_{\text{X}^-}}{R_{\rm y}^{\ast}} = \frac{(1 - y)
    \sum_{i, j} a_{ij} (1 - x)^{i / 2} y^j}{1 + \sum_{i, j} b_{ij} (1
    - x)^{i / 2} y^j}, \label{eq:EbXm_fit}
\end{equation}
where $x = \sigma / (1 + \sigma)$, $y = r_{\ast} / \left( r_{\ast} +
a_0^{\ast} \right)$, and the $\{a_{ij}\}$ and $\{b_{ij}\}$ are fitting
parameters, to the DMC trion binding energies. Equation
(\ref{eq:EbXm_fit}) satisfies the limiting behavior described in
App.\ \ref{app:trion_extreme}, has 31 free fitting parameters, and the
fractional error in the fit to our DMC data is everywhere less than
1\%. Positive trion binding energies can be obtained by charge
conjugating the corresponding negative trion.  The program included in
the Supplemental Material can be used to evaluate
Eq.\ (\ref{eq:EbXm_fit}).  The resulting trion binding energies for
various TMDCs are shown in Table \ref{table:be_ex_theo}.  It can be
seen that, in contrast to the biexciton binding energies, the trion
binding energies are in excellent agreement with the available
experimental results.  As shown in Table \ref{table:sensitivity},
trion binding energies are significantly more sensitive to the
effective mass values than biexciton binding energies; nevertheless,
the \textit{ab initio} effective masses would need to be in error by
an implausibly large amount to change the trion binding energies by
more than a few meV\@.  Binding-energy results in the limit of large
$r_\ast$, where the interaction is of logarithmic form, are given in
Sec.\ \ref{sec:logint_results}.

Figures \ref{fig:pdf_trion}(a) and \ref{fig:pdf_trion}(b) present the
electron-hole and electron-electron contact PDFs of trions. The
fitting functions can be found in the program supplied as Supplemental
Material.\cite{supplemental}

\subsection{Donor/acceptor-bound excitons}

\begin{figure}[!htbp]
\begin{center}
\includegraphics[clip,width=0.45\textwidth]{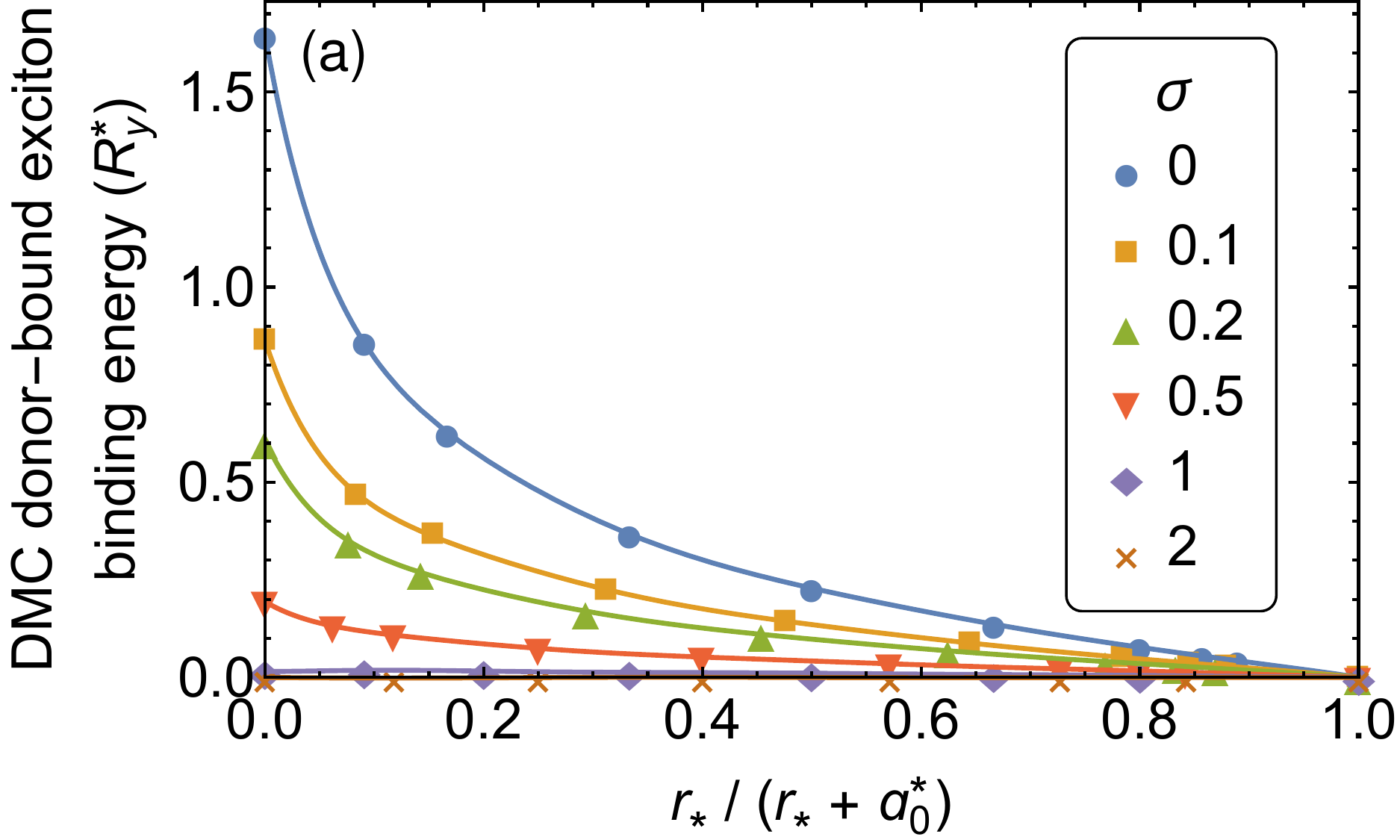}
\\ \includegraphics[clip,width=0.45\textwidth]{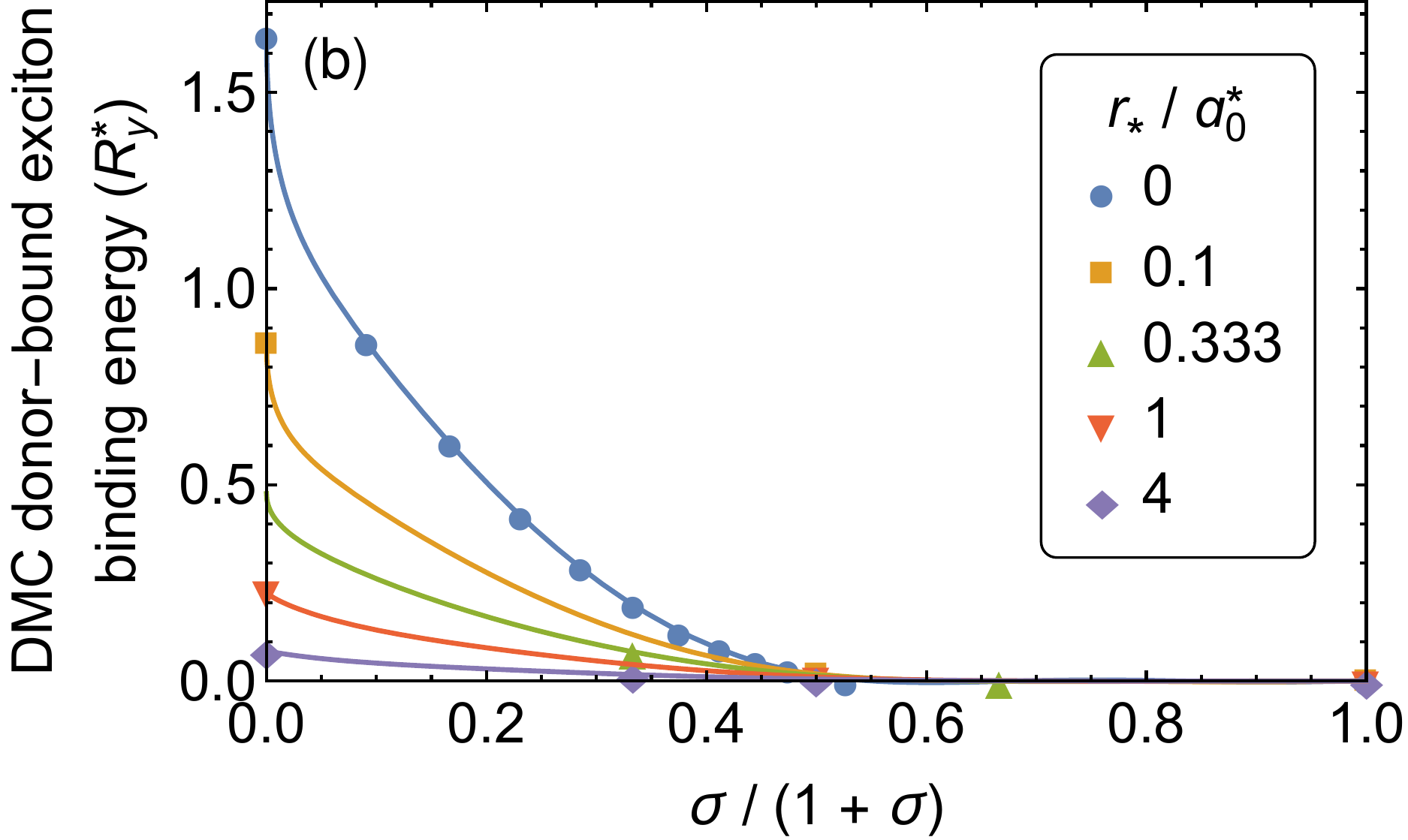}
\\ \includegraphics[clip,width=0.45\textwidth]{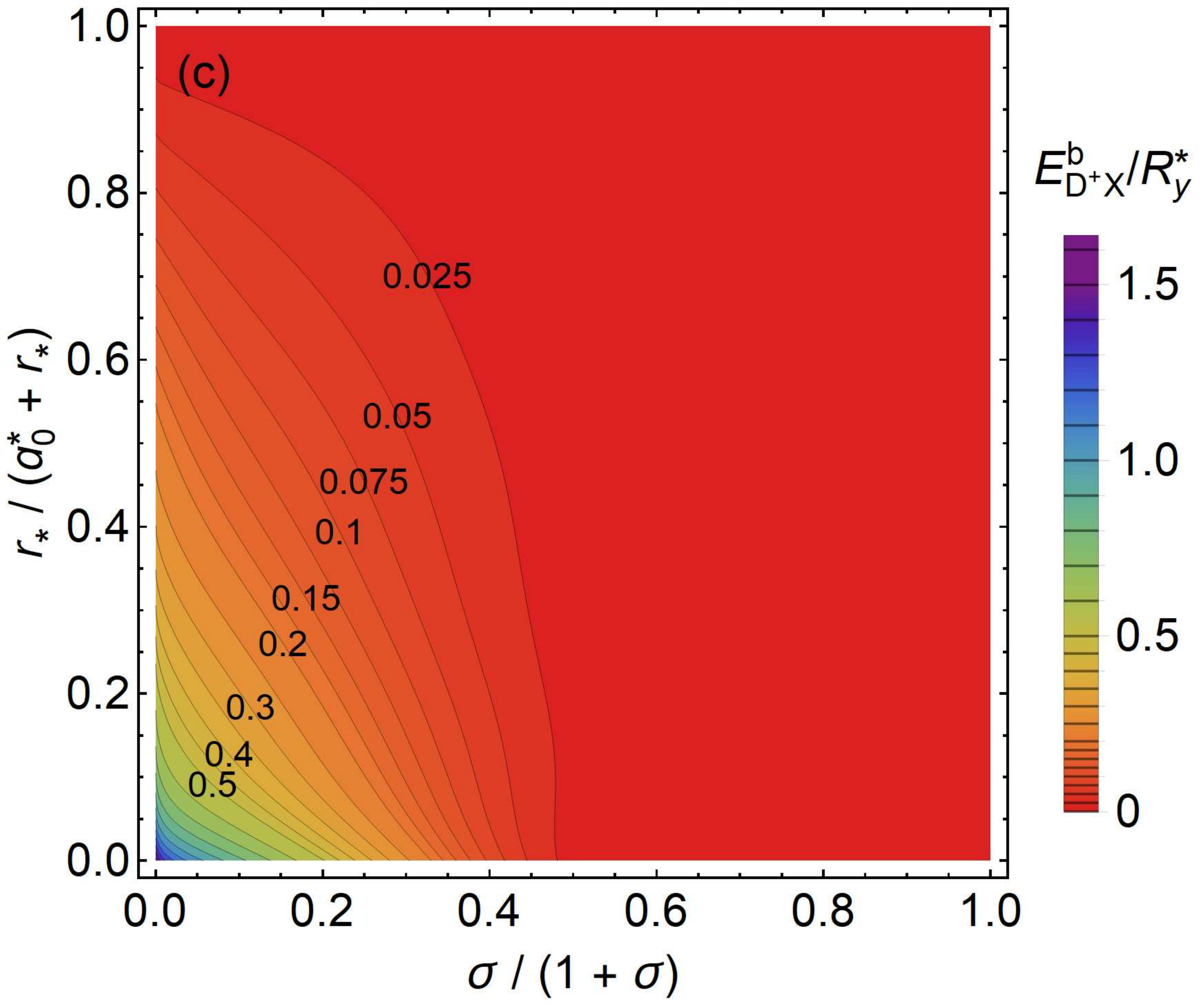}
\end{center}
\caption{(Color online) (a) DMC binding energies of donor-bound
  excitons with against rescaled susceptibility $r_*/(a_0^*+r_*)$.
  (b) DMC binding energies of donor-bound excitons against rescaled
  mass ratio $\sigma/(1+\sigma)$. (c) DMC binding energies of
  donor-bound excitons against rescaled susceptibility and rescaled
  mass ratio.
  \label{fig:be_dx}}
\end{figure}

We present the binding energies of donor-bound excitons in
Fig.\ \ref{fig:be_dx}. For $\sigma \gtrsim 1$, the binding energy is
close to zero. In this region, the calculations were especially
difficult, since the complex tends to unbind very easily. Therefore,
during the wave function optimization, the cutoff lengths for the
Jastrow factor were fixed at small values, to force the complex to be
bound. In the limit $\sigma\to\infty$, the complex is expected to be
unbound (see App.\ \ref{app:dx_extreme}), which is consistent with our
results.  Indeed, over a broad range of large electron-hole mass
ratios and large $r_\ast$ values, the DMC binding energy of the
donor-bound exciton is either zero or extremely small, such that the
binding energy cannot easily be resolved in DMC calculations.  The
following 50-parameter fitting formula has a fractional error that is
mostly less than 2\% in fits to our DMC data:
\begin{equation}
E^{\text{b}}_{\text{D}^+\text{X}} = (1-x)^2(1-y) \left[ \sum_{i, j}
  a_{ij} x^i y^j - \left( \sum_{k=0}^4 b_k y^k \right)^2 \sqrt{x}
  \right].
\label{eq:EbDplusX_fit}
\end{equation}
In this expression $x=\sigma/(1+\sigma)$ and
$y=r_\ast/(a_0^\ast+r_\ast)$, while the $\{a_{ij}\}$ are fitting
parameters.  Our fitting function can be evaluated using the program
in the Supplemental Material.\cite{supplemental} We summarize our
theoretical predictions for the binding energies of
donor/acceptor-bound excitons in various TMDCs in Table
\ref{table:be_db_ab_complexes}.  Binding-energies in the limit of
large $r_\ast$, where the interaction is of logarithmic form, are
given in Sec.\ \ref{sec:logint_results}.

\begin{table*}[!htbp]
\caption{As Table \ref{table:be_ex_theo} (using the $r_\ast$ values
  and effective masses shown in bold in Table
  \ref{table:be_mass_rst}), but for donor atoms (D$^0$), acceptor
  atoms (A$^0$), donor-bound excitons (D$^+$X), acceptor-bound
  excitons (A$^-$X), donor-bound trions (D$^0$X), acceptor-bound
  trions (A$^0$X), donor-bound biexcitons (D$^+$XX), and
  acceptor-bound biexcitons (A$^-$XX)\@.  The binding-energy results
  are our theoretical predictions using Eqs.\ (\ref{eq:EbDplusX_fit}),
  (\ref{eq:EbDplusXm_fit}), and (\ref{eq:EbDplusXX_fit}), while the
  energies of donor and acceptor atoms are calculated using
  Eq.\ (\ref{eq:EX_fit}) with infinite hole and electron masses,
  respectively. Note that the binding energy is defined with respect
  to dissociation into the most energetically favorable products,
  which do not always include an exciton: see the definitions in
  Sec.\ \ref{sec:binding_energies}. \label{table:be_db_ab_complexes}}
\begin{center}
\begin{tabular}{lcccccccc}
\hline \hline

& \multicolumn{2}{c}{Energy (meV)} & \multicolumn{6}{c}{Binding energy
  (meV)} \\

\raisebox{1.5ex}[0pt]{TMDC} & $E_{\rm D^0}$ & $E_{\rm A^0}$ & $E^{\rm
  b}_{\rm D^+X}$ & $E^{\rm b}_{\rm A^-X}$ & $E^{\rm b}_{\rm D^0X}$ &
$E^{\rm b}_{\rm A^0X}$ & $E^{\rm b}_{\rm D^+XX}$ & $E^{\rm b}_{\rm
  A^-XX}$ \\ \hline

MoS$_2$ & $-638.5$ & $-670.1$ & $7.2$ & $2.7$ & $32.4$ & $31.7$ &
$51.8$ & $48.0$ \\

MoSe$_2$ & $-636.8$ & $-659.3$ & $6.5$ & $3.2$ & $31.5$ & $31.1$ &
$50.8$ & $48.1$ \\

MoTe$_2$ & $-447.8$ & $-443.9$ & $3.9$ & $4.6$ & $17.9$ & $18.0$ &
$32.4$ & $32.9$ \\

WS$_2$ & $-606.8$ & $-633.6$ & $6.2$ & $2.7$ & $32.0$ & $31.5$ &
$48.6$ & $45.3$ \\

WSe$_2$ & $-542.3$ & $-563.6$ & $5.5$ & $2.6$ & $27.5$ & $27.1$ &
$43.4$ & $40.8$ \\

WTe$_2$ & $-519.0$ & $-562.1$ & $7.9$ & $1.6$ & $26.1$ & $25.3$ &
$44.0$ & $38.8$ \\

\hline \hline
\end{tabular}
\end{center}
\end{table*}

\begin{figure}[!htbp]
\begin{center}
\includegraphics[clip,width=0.45\textwidth]{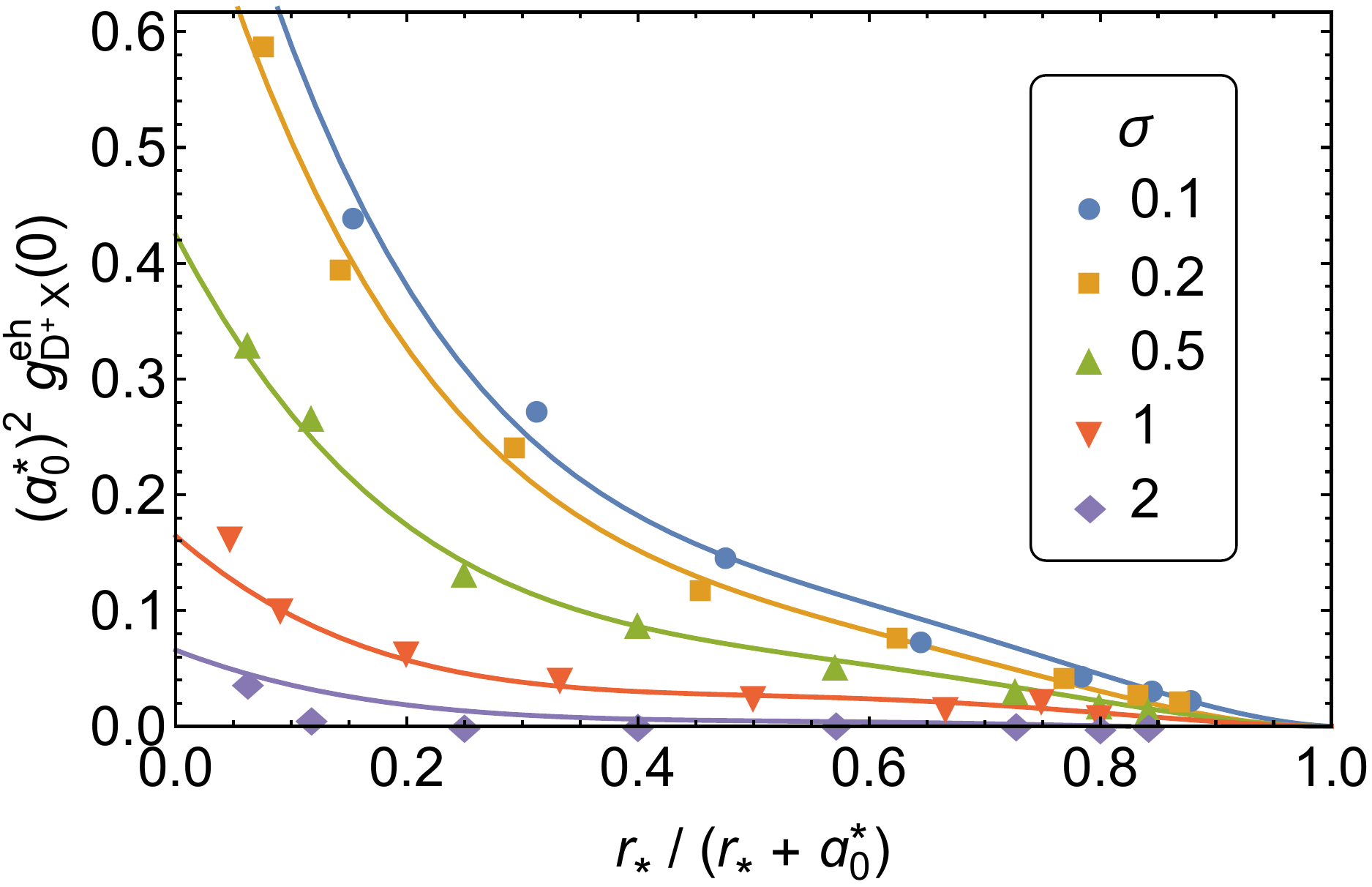}
\end{center}
\caption{(Color online) Electron-hole contact PDF of a donor-bound
  exciton. The solid lines were obtained using the fitting function
  reported in the Supplemental
  Material.\cite{supplemental} \label{fig:pdf_dx}}
\end{figure}

We have also calculated the electron-hole contact PDFs of donor-bound
excitons, which are presented in Fig.\ \ref{fig:pdf_dx}. Our results
confirm that the contact PDFs decrease to zero as $\sigma\to\infty$,
as expected, because the light hole becomes unbound in this limit.
Contact PDFs can be evaluated using the program supplied as
Supplemental Material.\cite{supplemental}

\subsection{Donor/acceptor-bound trions}

\begin{figure}[!htbp]
\begin{center}
\includegraphics[clip,width=0.45\textwidth]{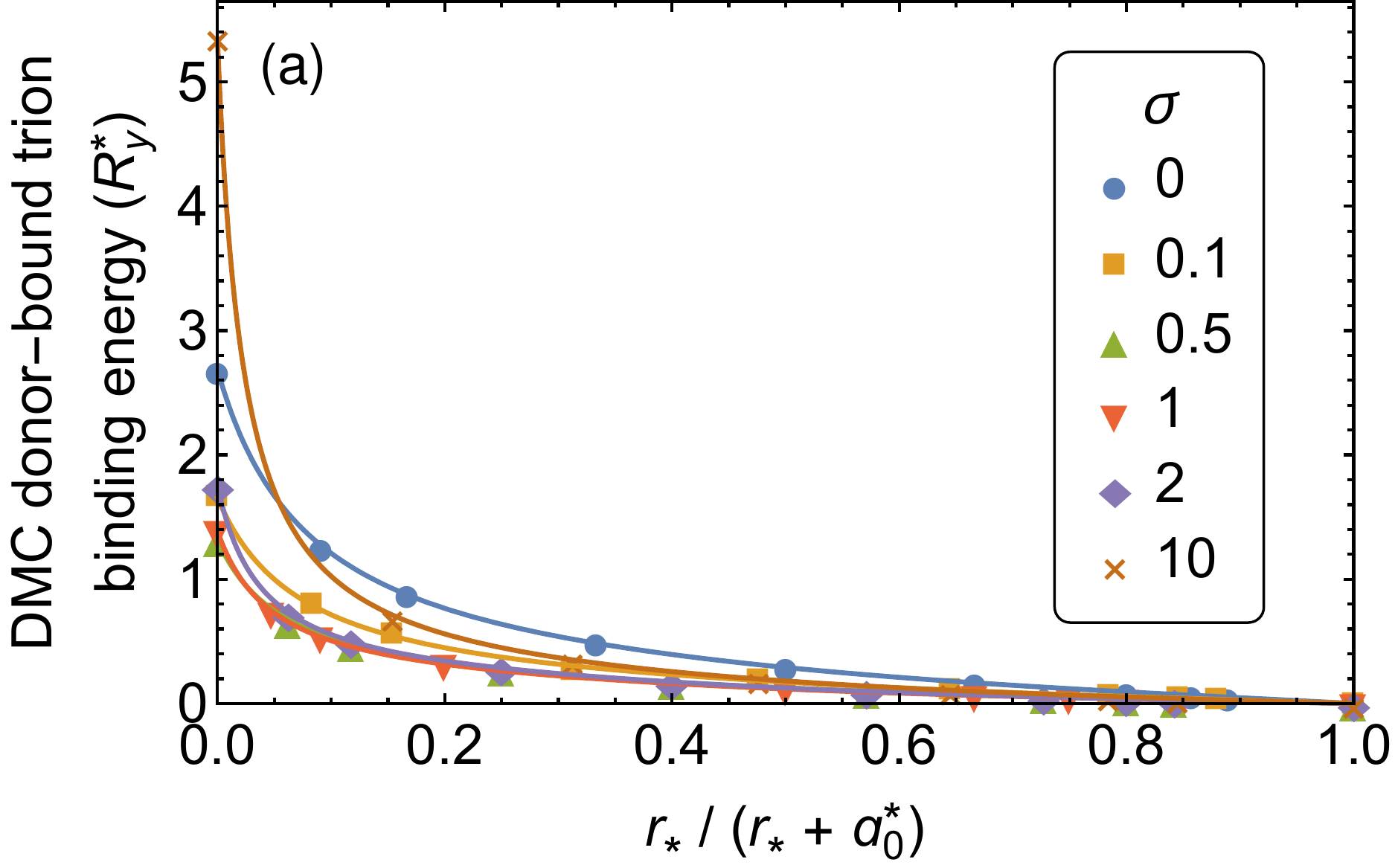}
\\ \includegraphics[clip,width=0.45\textwidth]{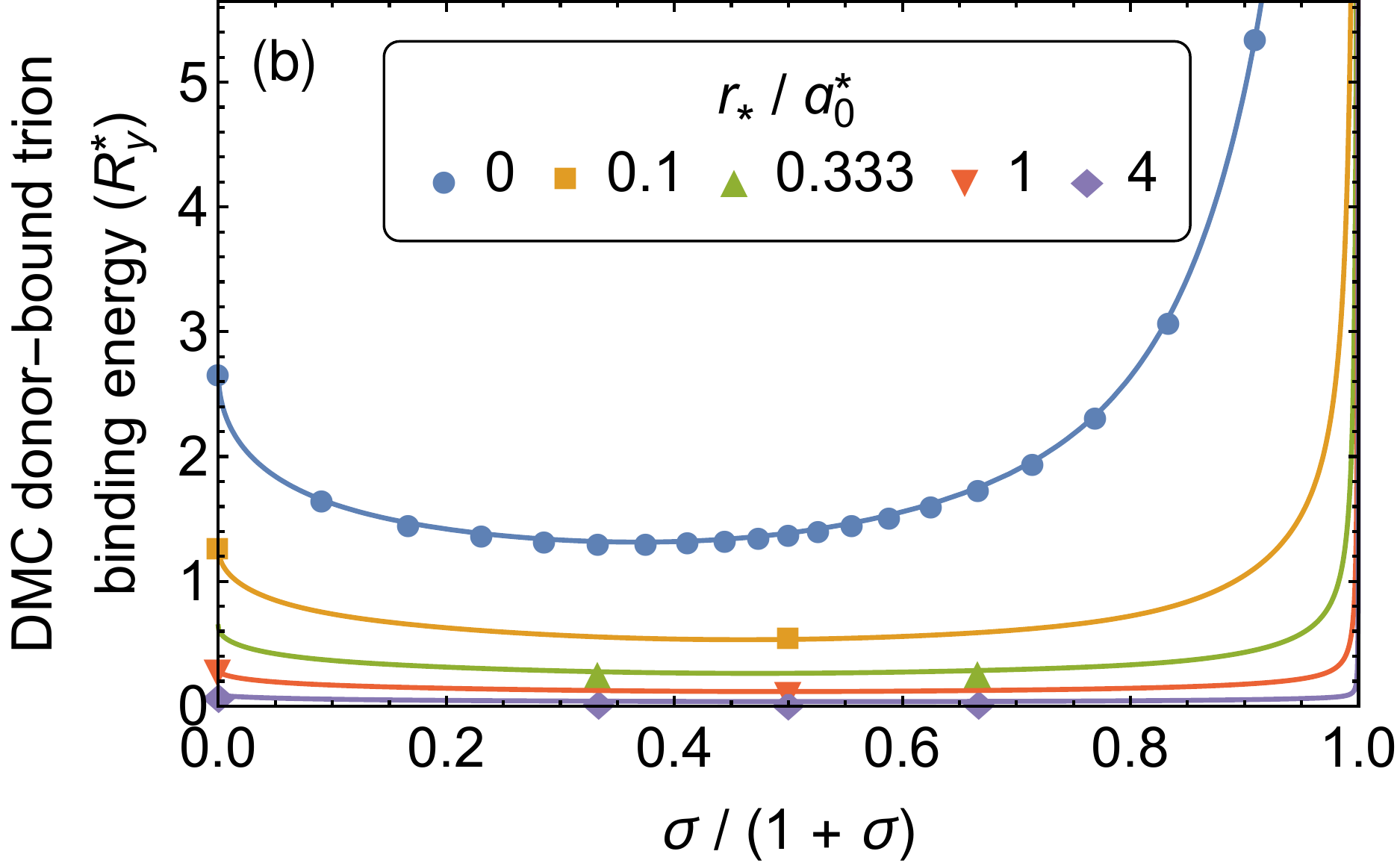}
\\ \includegraphics[clip,width=0.45\textwidth]{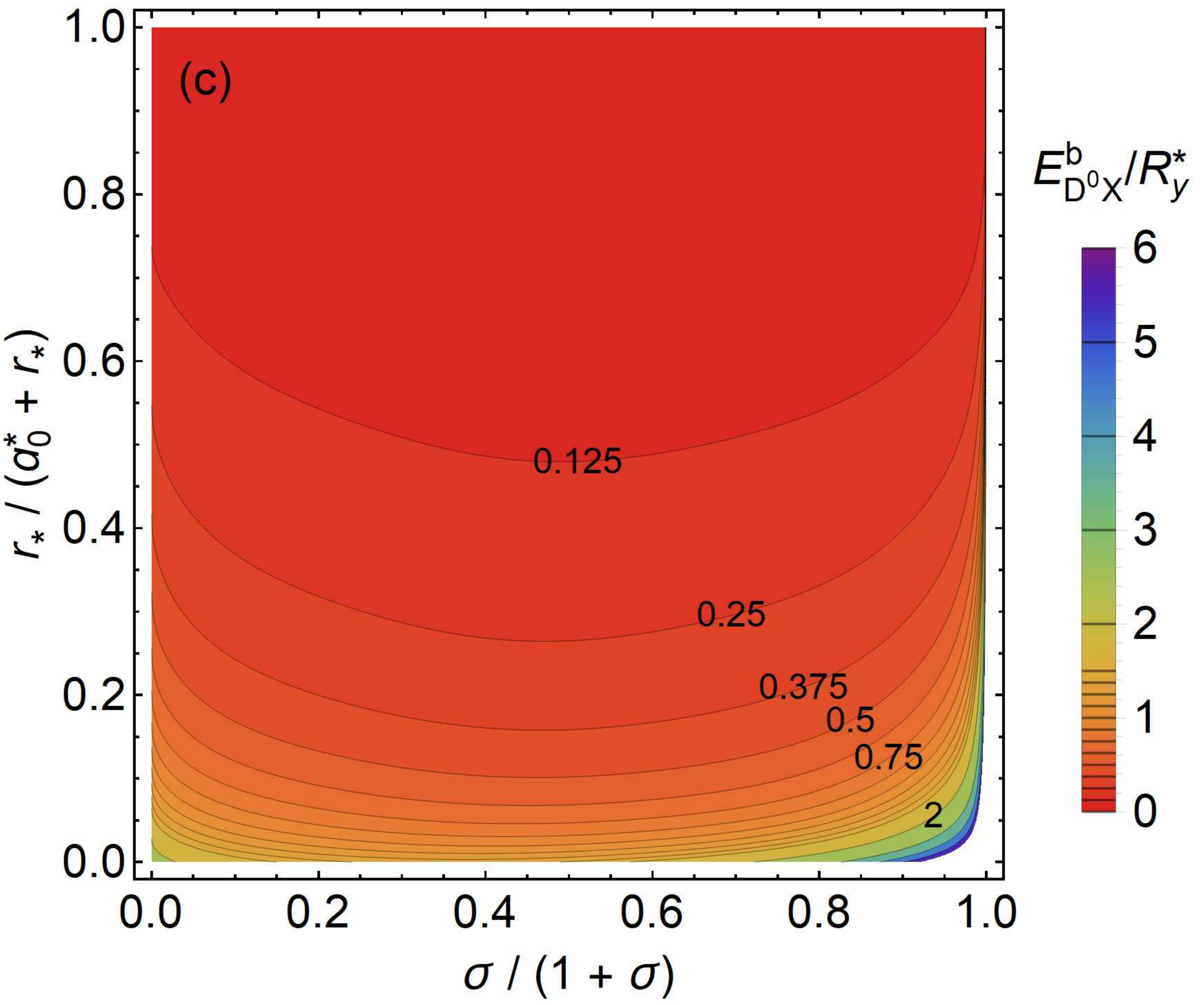}
\end{center}
\caption{(Color online) (a) DMC binding energies of donor-bound trions
  with distinguishable particles against rescaled susceptibility
  $r_*/(a_0^*+r_*)$. (b) DMC binding energies of donor-bound trions
  with distinguishable particles against rescaled mass ratio
  $\sigma/(1+\sigma)$. (c) DMC binding energies of donor-bound trions
  with distinguishable particles against rescaled susceptibility and
  rescaled mass ratio.
  \label{fig:be_dxm}}
\end{figure}

Figure \ref{fig:be_dxm} presents the binding energies of donor-bound
trions.  We have devised the following 30-parameter fitting formula:
\begin{equation}
  \frac{E^{\text{b}}_{\text{D}^0 \text{X}}}{R_{\rm y}^{\ast}}
  =\frac{(1 - y) \left[ \sum_{i, j} a_{ij} x^i y^j + b_0 (1 - y)
      \frac{x}{1 - x} - b_1 \sqrt{x} \right]}{1 + \sum_{i, j} c_{ij}
    x^i y^j}, \label{eq:EbDplusXm_fit}
\end{equation}
which includes the correct divergence as $\sigma \to \infty$ and
appropriate square-root behavior for the heavy-hole limit $\sigma\to
0$ (see App.\ \ref{app:dxm_extreme}).  The $\{a_{ij}\}$, $\{b_i\}$,
and $\{c_{ij}\}$ are fitting parameters.  The fractional error in the
fit to our DMC data is less than 3\%. The program in our Supplemental
Material can be used to evaluate
Eq.\ (\ref{eq:EbDplusXm_fit}).\cite{supplemental} Binding-energy
results in the limit of large $r_\ast$, where the interaction is of
logarithmic form, are given in Sec.\ \ref{sec:logint_results}.

Table \ref{table:be_db_ab_complexes} reports theoretical binding
energies for donor-bound trions with biexciton energies for several
real materials. The binding energy of a donor-bound trion is slightly
larger than the binding energy of a free biexciton. This leads us to
expect two lines close together in the absorption/emission spectra of
TMDCs, one corresponding to biexcitons, and another at slightly larger
energy corresponding to donor-bound trions.

Contact PDFs for donor-bound trions have been extracted from our QMC
data and are presented in Fig.\ \ref{fig:pdf_dxm} and in the
Supplemental Material.\cite{supplemental}

\begin{figure}[!htbp]
\begin{center}
\includegraphics[clip,width=0.45\textwidth]{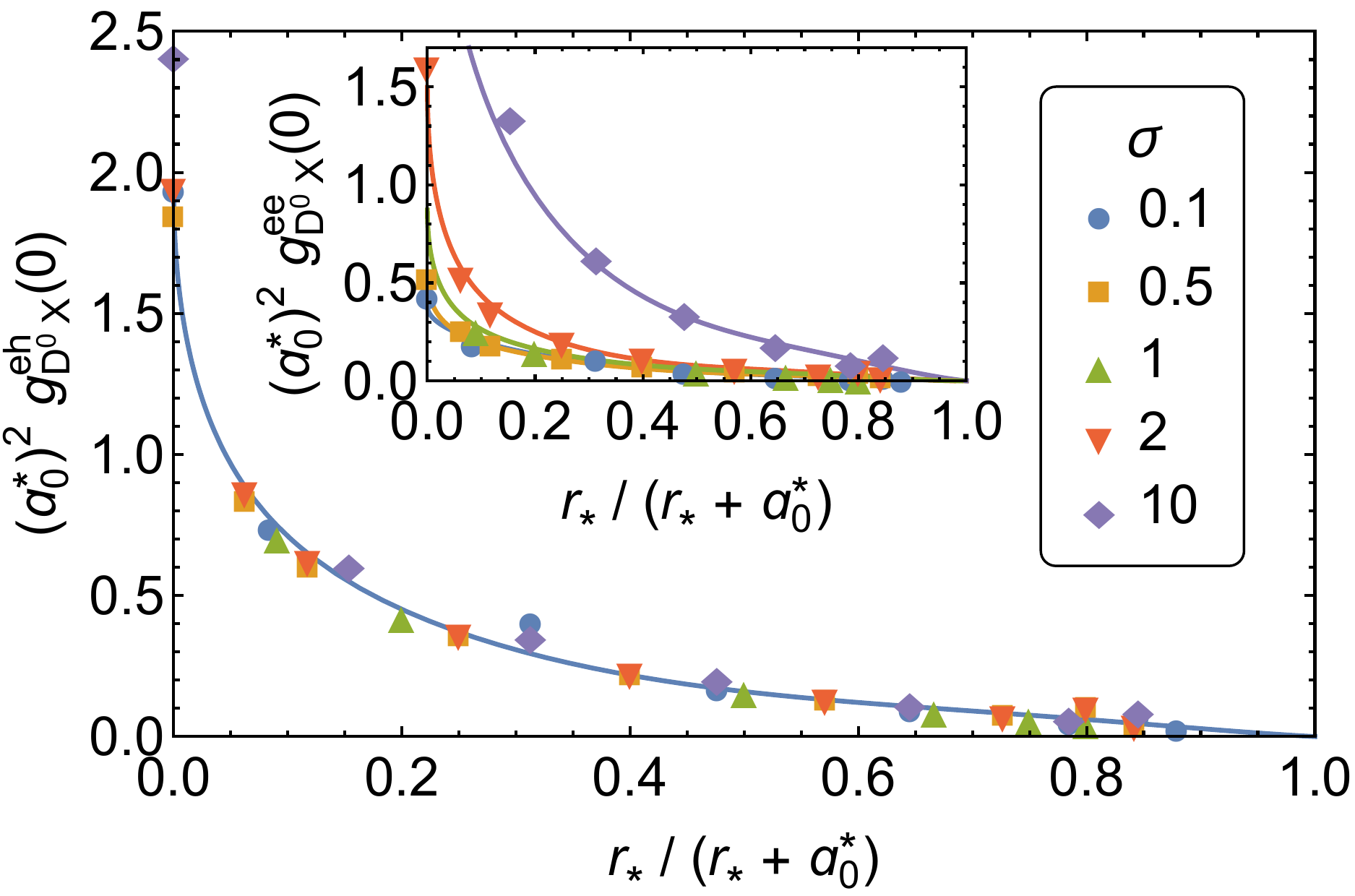}
\end{center}
\caption{(Color online) Electron-hole contact PDFs of a donor-bound
  trion complex. The inset shows electron-electron contact PDFs.  The
  solid lines were obtained using the fitting function reported in the
  Supplemental Material.\cite{supplemental}
  \label{fig:pdf_dxm}}
\end{figure}

\subsection{Donor/acceptor-bound biexcitons \label{sec:dbbiex_results}}

\begin{figure}[!htbp]
\begin{center}
\includegraphics[clip,width=0.45\textwidth]{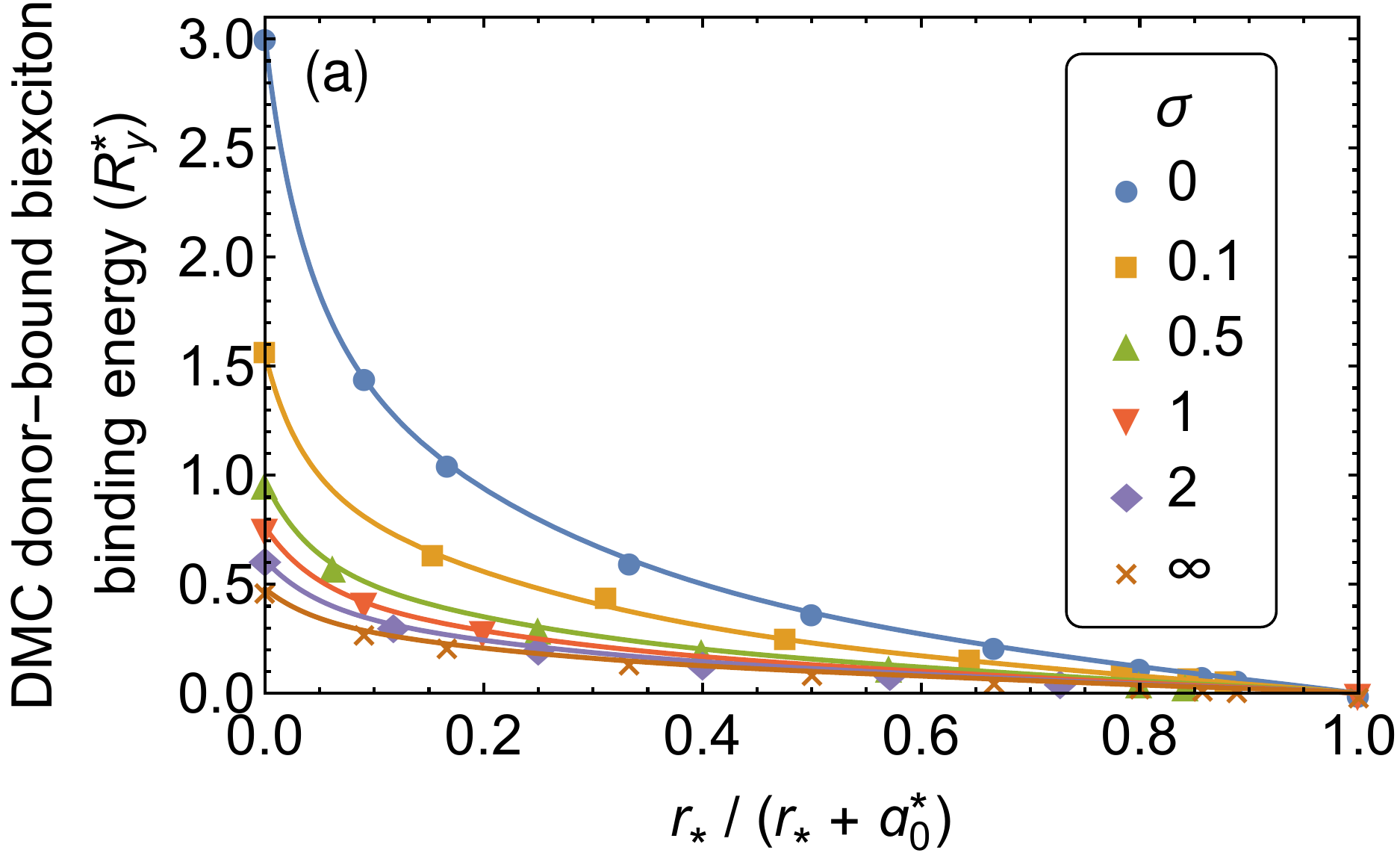}
\\ \includegraphics[clip,width=0.45\textwidth]{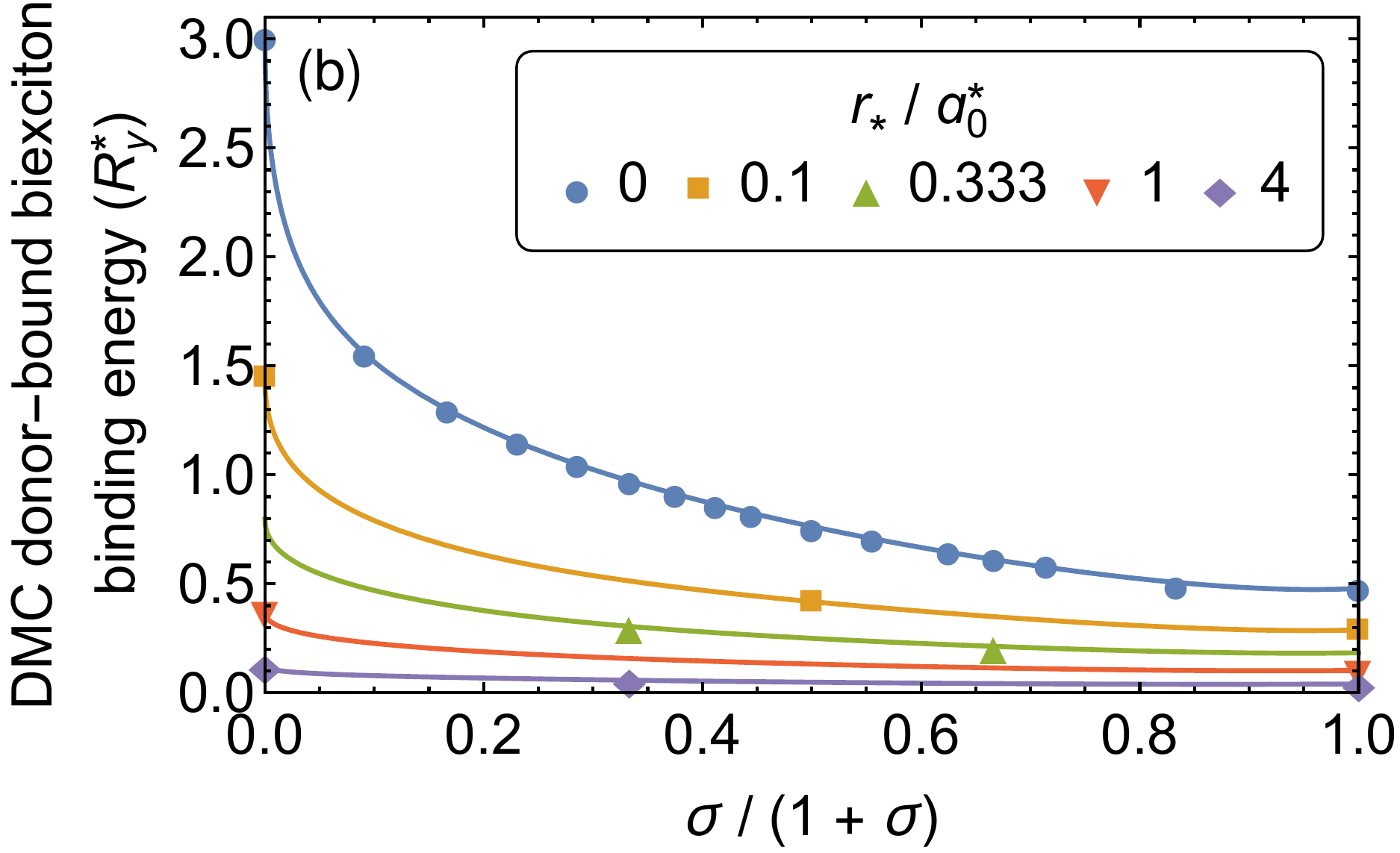}
\\ \includegraphics[clip,width=0.45\textwidth]{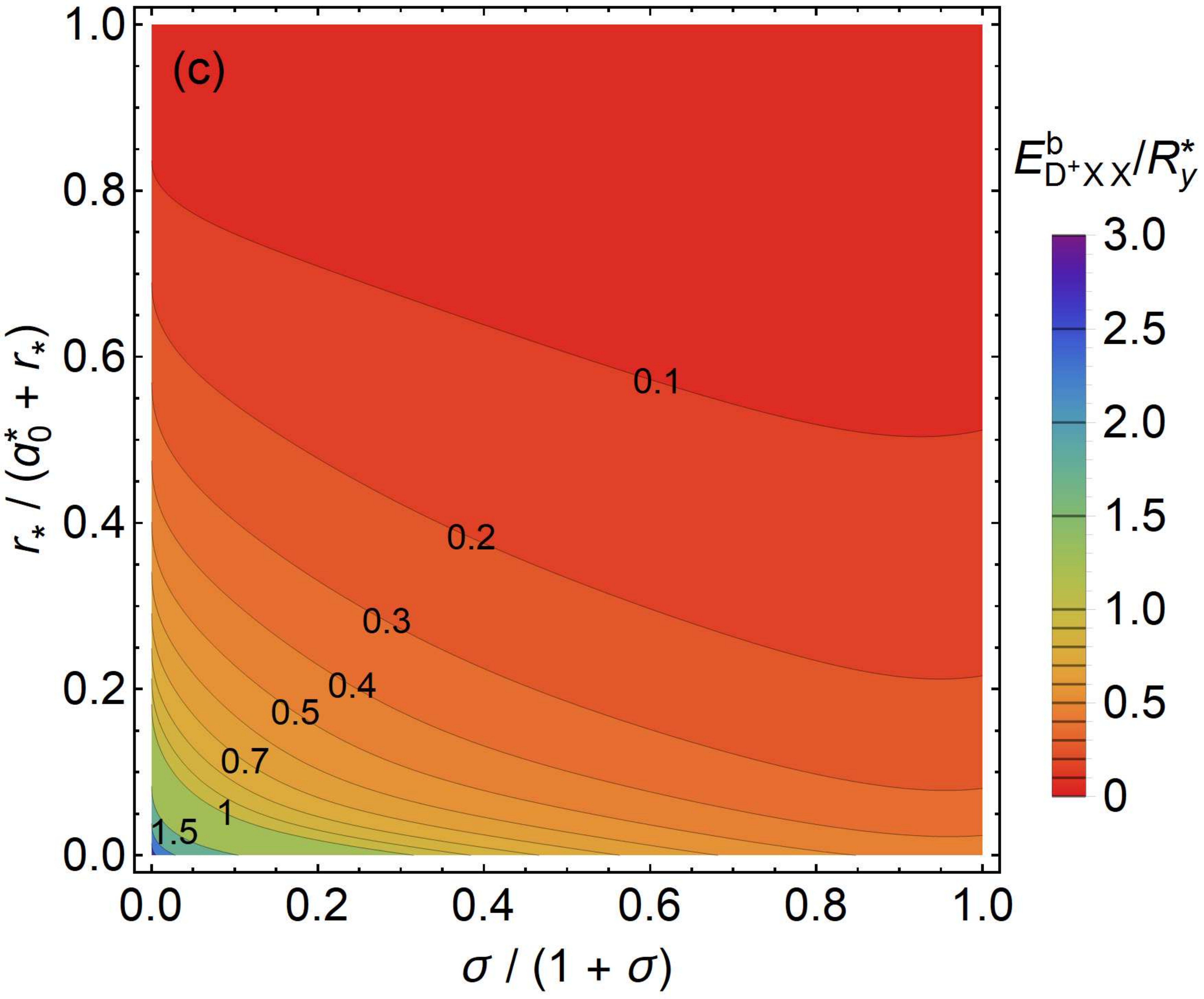}
\end{center}
\caption{(Color online) (a) DMC binding energies of donor-bound
  biexcitons with distinguishable particles against rescaled
  susceptibility $r_*/(a_0^*+r_*)$. (b) DMC binding energies of
  donor-bound biexcitons with distinguishable particles against
  rescaled mass ratio $\sigma/(1+\sigma)$. (c) DMC binding energies of
  donor-bound biexcitons with distinguishable particles against
  rescaled susceptibility and rescaled mass ratio. \label{fig:be_dxx}}
\end{figure}

Donor-bound biexciton binding energies have also been calculated and
are presented in Fig.\ \ref{fig:be_dxx}.  A 38-parameter fitting
formula similar to that of a donor-bound trion
[Eq.\ (\ref{eq:EbDplusXm_fit})] was used:
\begin{widetext}
\begin{equation}
\frac{E^{\text{b}}_{\text{D}^+ \text{XX}}}{R_{\rm y}^{\ast}} =
\frac{(1 - y) \left[ \sum_{i, j} a_{ij} x^i y^j - b_0 \sqrt{x} + b_1
    x^{3 / 2} \right]}{1 + \sum_{i, j} c_{ij} x^i
  y^j}, \label{eq:EbDplusXX_fit}
\end{equation}
\end{widetext}
where $x=\sigma/(1+\sigma)$ and $y=r_\ast/(a_0^\ast+r_\ast)$, while
the $\{a_{ij}\}$, $\{b_i\}$, and $\{c_{ij}\}$ are fitting parameters.
This gives a fractional error of less than 3\% everywhere when
fitted to our DMC data.  Equation (\ref{eq:EbDplusXX_fit}) can be
evaluated using the program supplied as Supplemental
Material.\cite{supplemental} We summarize our theoretical predictions
for the binding energies of donor/acceptor-bound biexcitons in various
TMDCs in Table \ref{table:be_db_ab_complexes}.  Binding-energy results
in the limit of large $r_\ast$, where the interaction is of
logarithmic form, are given in Sec.\ \ref{sec:logint_results}.

\begin{figure}[!htbp]
\begin{center}
\includegraphics[clip,width=0.45\textwidth]{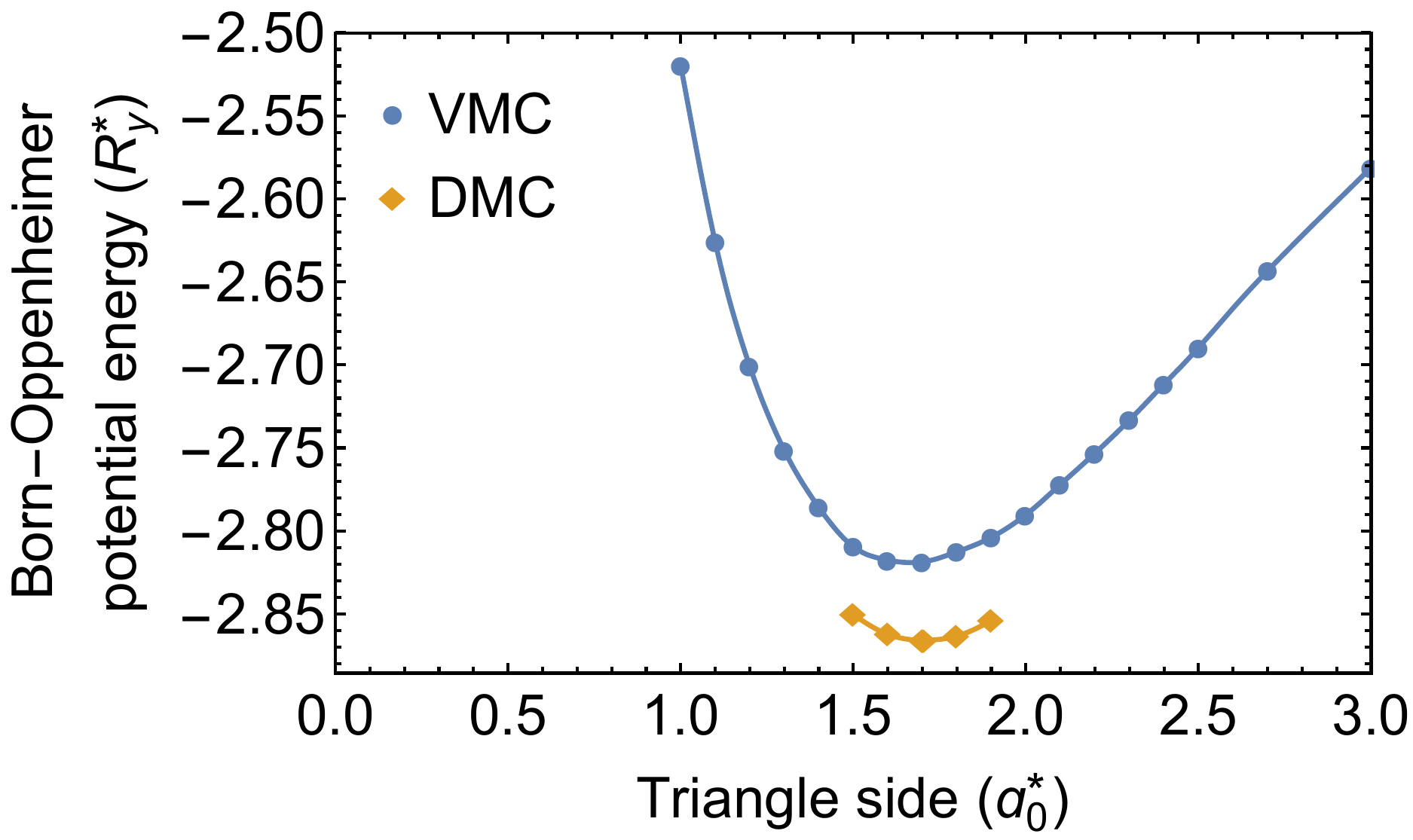}
\end{center}
\caption{(Color online) Born-Oppenheimer potential energy of a complex
  of three positive, fixed ions and two electrons, with the positive
  ions placed at the corners of an equilateral triangle. Example for
  $r_{\ast} / a_0^{\ast} = 1$. \label{fig:dddee-side}}
\end{figure}

\begin{figure}[!htbp]
\begin{center}
\includegraphics[clip,width=0.45\textwidth]{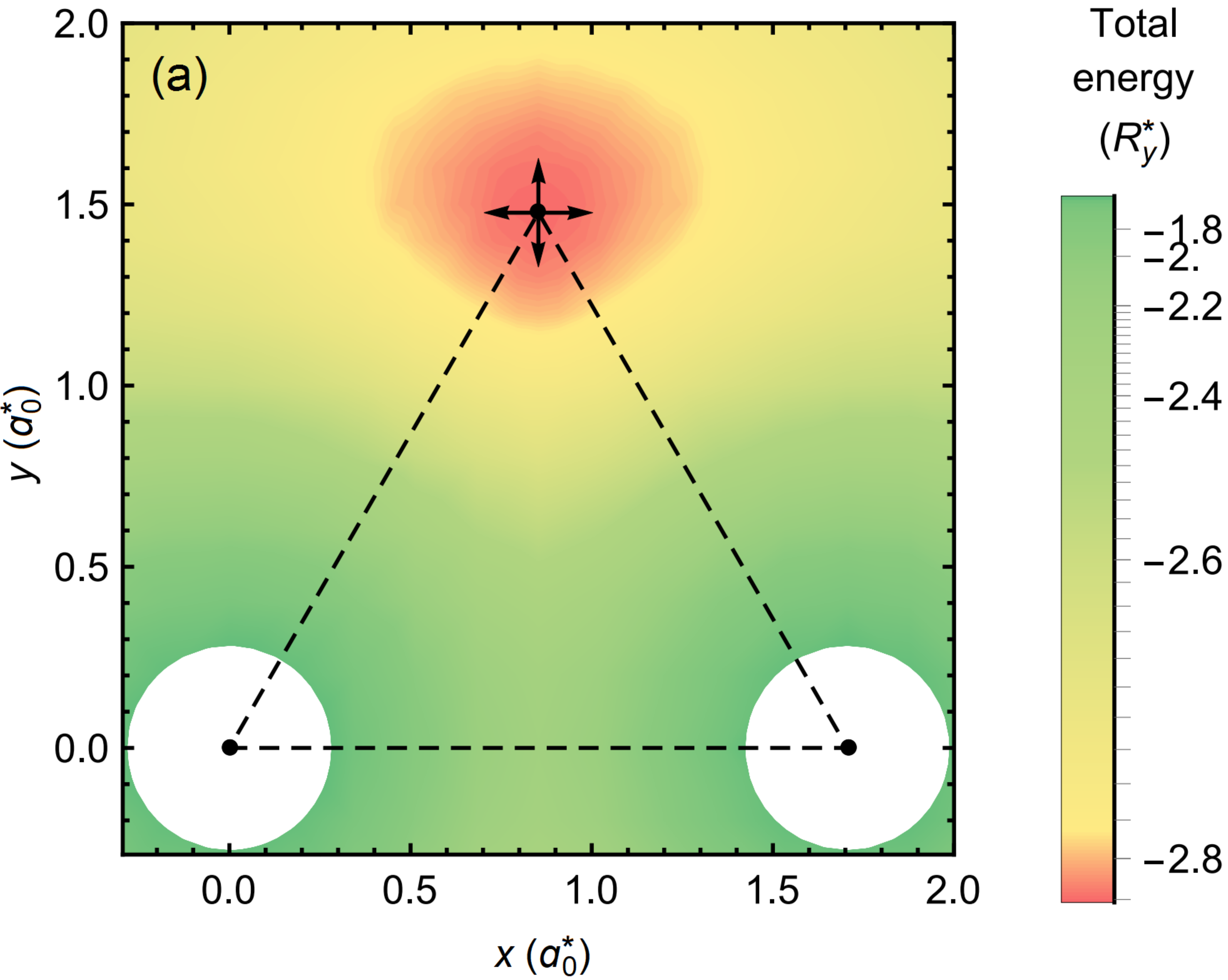}
\\ \includegraphics[clip,width=0.45\textwidth]{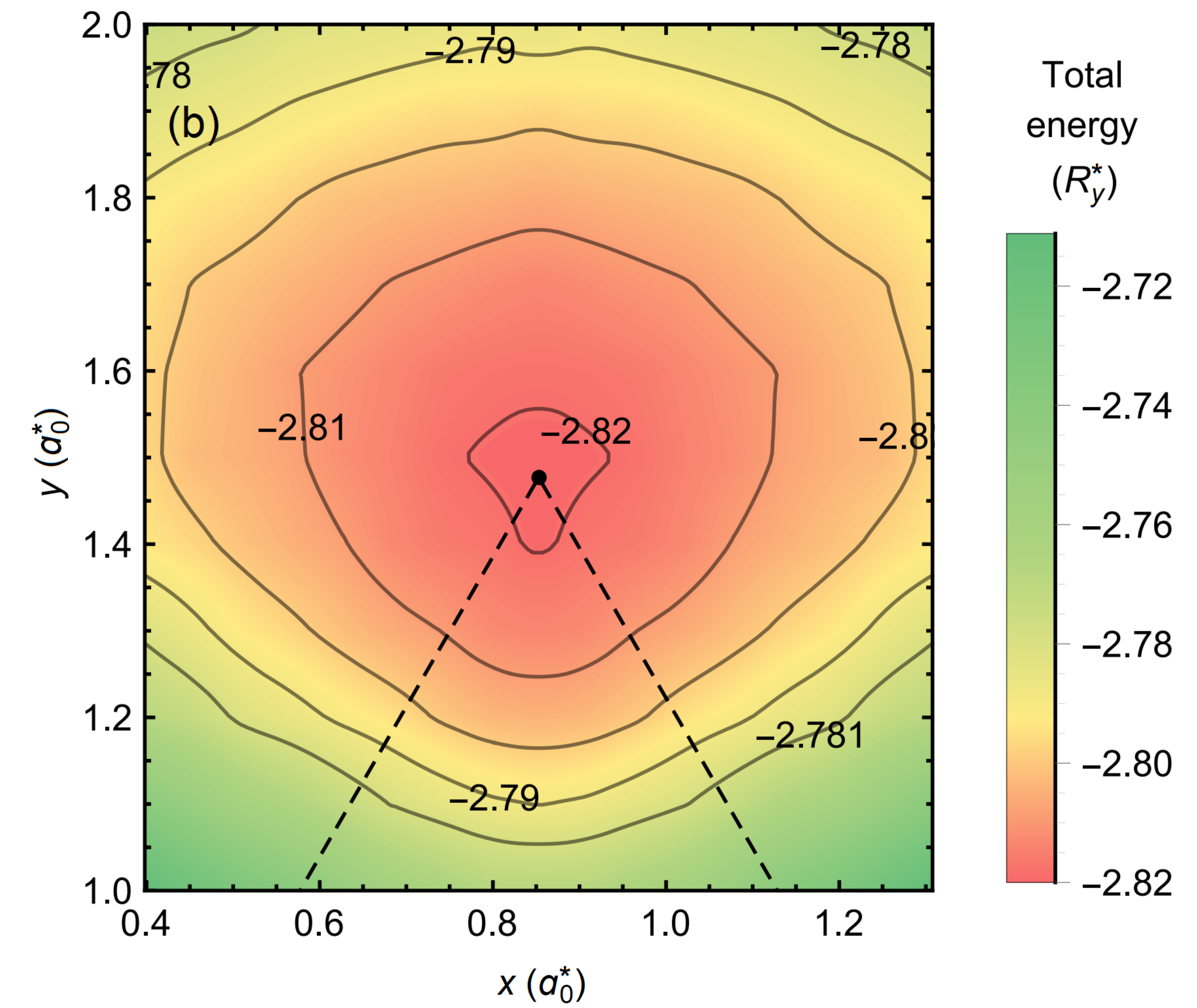}
\end{center}
\caption{(Color online) (a) Born-Oppenheimer potential energy of a
  complex of three positive, fixed ions and two electrons. We fix two
  of the ions and change the position of the third one. Example for
  $r_{\ast} / a_0^{\ast} = 1$. (b) Vertex of the triangle of fixed,
  positive charges in greater detail.
\label{fig:dddee-triangle}}
\end{figure}

The behavior of a donor-bound biexciton in the limit of heavy
electrons is discussed in App.\ \ref{app:dbxx_extreme}. In the limit
of heavy holes ($\sigma \to 0$), this complex consists of three fixed
positive particles and two light electrons and thus the question
arises of how the three fixed, positive charges are positioned with
respect to each other. The most natural position that three positive
particles would assume is an equilateral triangle. To check if this
assumption is correct we first determined how the Born-Oppenheimer
potential energy changes if we distribute the three positive charges
in the corners of equilateral triangle and then vary the triangle
side. Figure \ref{fig:dddee-side} shows the case of $r_{\ast} /
a_0^{\ast} = 1$ as an example. After finding the side length that
minimizes the Born-Oppenheimer potential energy, we changed the
position of one of the positive particles (fixing the remaining two)
and again observed the effect on the Born-Oppenheimer potential
energy. Figure \ref{fig:dddee-triangle} presents the results, which
clearly show that the equilateral triangle is a local minimum of the
Born-Oppenheimer potential-energy surface.

Closely related to donor-bound biexcitons are five-carrier complexes
known as charged biexcitons or \textit{quintons} (XX$^-$, i.e.,
e$^-$e$^-$e$^-$h$^+$h$^+$). In molybdenum and tungsten dichalcogenides
these consist of two distinguishable holes with opposite spin and
valley indices, and three distinguishable electrons that differ in
either their spin or their valley indices: see
Fig.\ \ref{fig:classification}(a).  The binding energy of a quinton is
defined as the energy required to split it into a free exciton and a
free trion.\footnote{The most energetically competitive products of
  the dissociation of a quinton are either a trion plus an exciton, or
  a biexciton plus a free charge carrier; the energy difference
  between the two possible sets of products is given by the difference
  of the trion and biexciton binding energies.  For all the 2D
  semiconductors considered here, the trion binding energy is larger
  than the biexciton binding energy,\cite{Szyniszewski_2017} so that
  an exciton and a trion are the most energetically competitive
  species into which a quinton may dissociate.}  Other possible large
complexes are donor-bound double-negative excitons (D$^-$X, i.e.,
D$^+$e$^-$e$^-$e$^-$h$^+$), donor-bound quintons (D$^0$XX, i.e.,
D$^+$e$^-$e$^-$e$^-$h$^+$h$^+$), and even donor-bound double-negative
biexcitons (D$^-$XX, i.e., D$^+$e$^-$e$^-$e$^-$e$^-$h$^+$h$^+$).  For
molybdenum and tungsten dichalcogenides there are no further
possibilities: we have exhausted the possible neutral or singly
charged complexes that can be constructed from up to four
distinguishable electrons, up to two distinguishable holes, and zero
or one donor ions. Any larger charge-carrier complexes in molybdenum
or tungsten dichalcogenides inevitably either include
indistinguishable particles or involve the much larger energies
required to excite holes in the lower spin-split valence bands.  In
Table \ref{table:big_complexes} we present our DMC binding-energy
results for quintons and the other large complexes.  Donor-bound
double-negative biexcitons appear to be unstable to dissociation into
free excitons plus donor-bound double-negative excitons, and hence are
not included in Table \ref{table:big_complexes}.  As with donor-bound
biexcitons, the energies required to remove excitons from the larger
complexes such as quintons are relatively close to the experimentally
observed biexciton binding energies.

\begin{table}[!htbp]
\caption{Binding energies of larger charge-carrier complexes in
  different TMDCs.  Binding energies are presented for: quintons
  (XX$^-$), which dissociate into excitons (X) and negative trions
  (X$^-$); donor-bound double-negative excitons (D$^-$X), which
  dissociate into excitons (X) and negative donor ions (D$^-$); and
  donor-bound quintons (D$^0$XX), which dissociate into excitons (X)
  and donor-bound trions (D$^0$X)\@. The binding energies were
  evaluated using the effective mass and in-plane permittivity
  parameters reported in bold in Table \ref{table:be_mass_rst}. The
  fitting functions of Eqs.\ (\ref{eq:EbXm_fit}) and
  (\ref{eq:EbDplusXm_fit}) were used to evaluate the energies of
  negative donor ions and donor-bound
  trions. \label{table:big_complexes}}
\begin{center}
\begin{tabular}{lccc}
\hline \hline

 & \multicolumn{3}{c}{Binding energy (meV)} \\

\raisebox{1.5ex}[0pt]{TMDC} & XX$^-$ & D$^-$X & D$^0$XX \\

\hline

MoS$_2$  & $58.6(6)$ & $84.4(4)$ & $61.6(6)$ \\

MoSe$_2$ & $57.0(4)$ & $57.9(2)$ & $56.9(9)$ \\

MoTe$_2$ & $33.8(3)$ & & \\

WS$_2$   & $57.4(3)$ & $59.2(4)$ & $58.2(6)$ \\

WSe$_2$  & $52.5(7)$ & $51.3(4)$ & $51(1)$ \\

WTe$_2$  & $47.5(3)$ & & \\

\hline \hline
\end{tabular}
\end{center}
\end{table}

Figure \ref{fig:pdf_dxx} presents the contact PDFs of a donor-bound
biexciton, which are also reported in the Supplemental
Material.\cite{supplemental}

\begin{figure}[!htbp]
\begin{center}
\includegraphics[clip,width=0.45\textwidth]{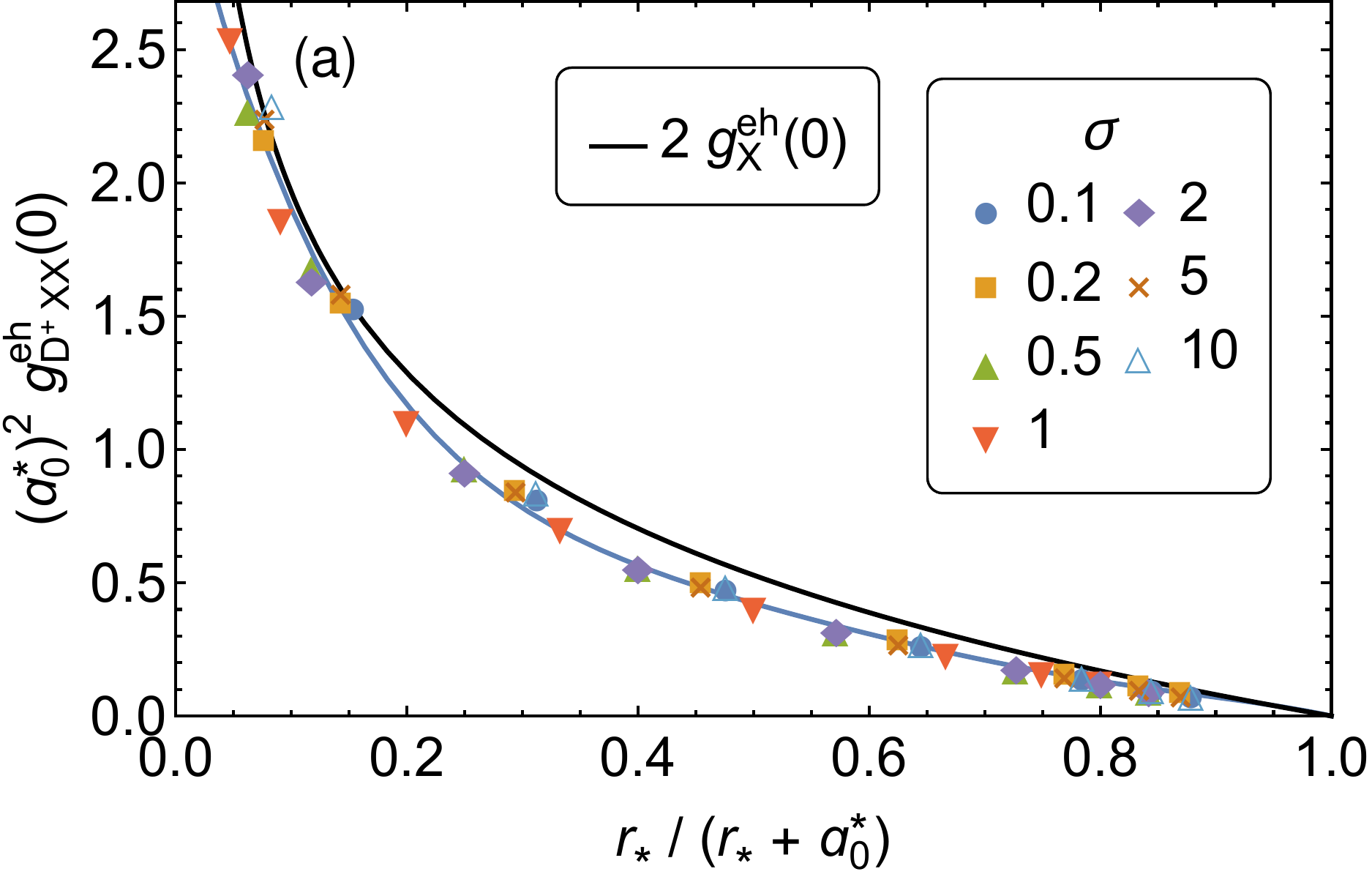}
\\ \includegraphics[clip,width=0.45\textwidth]{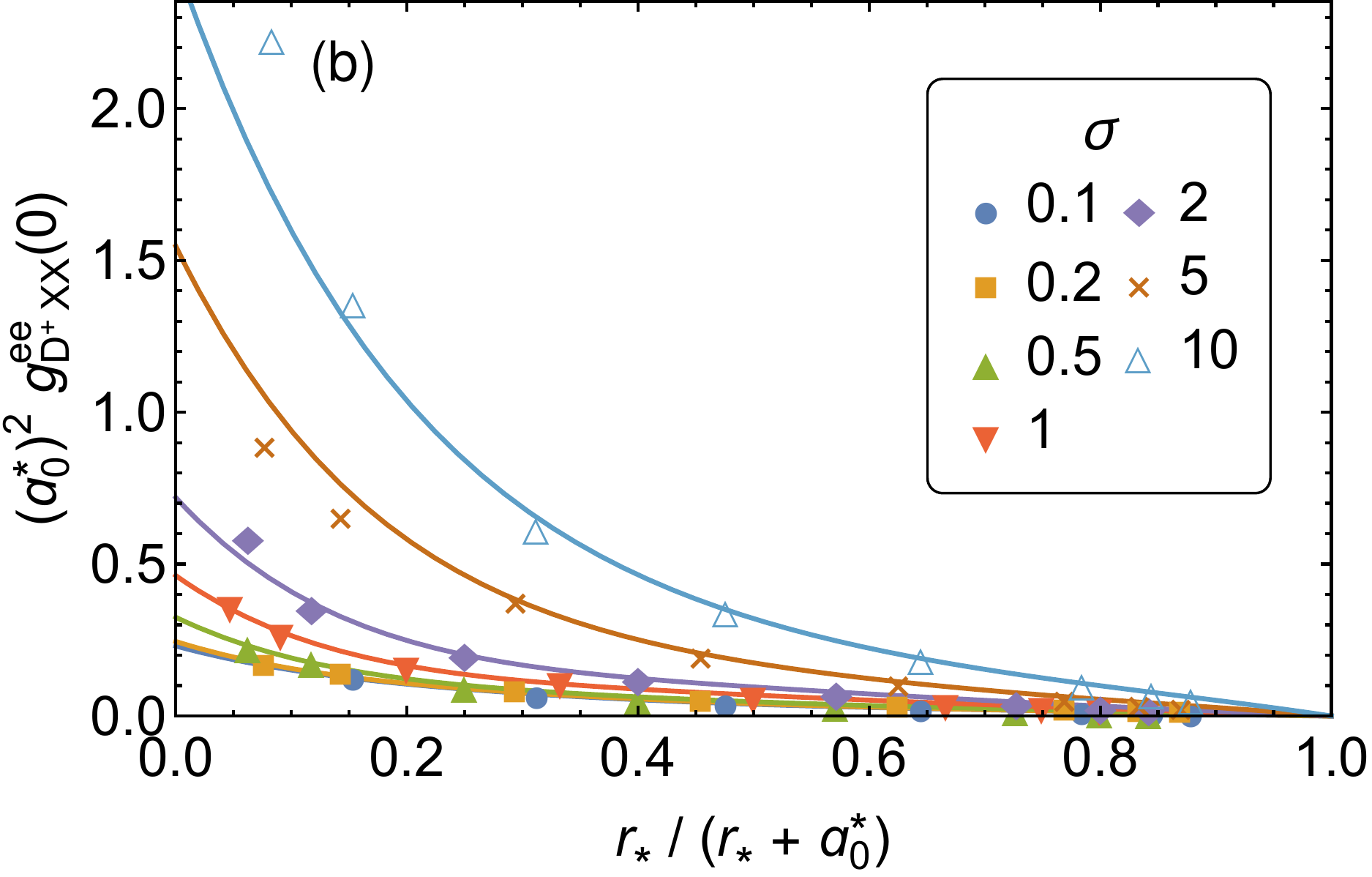}
\\ \includegraphics[clip,width=0.45\textwidth]{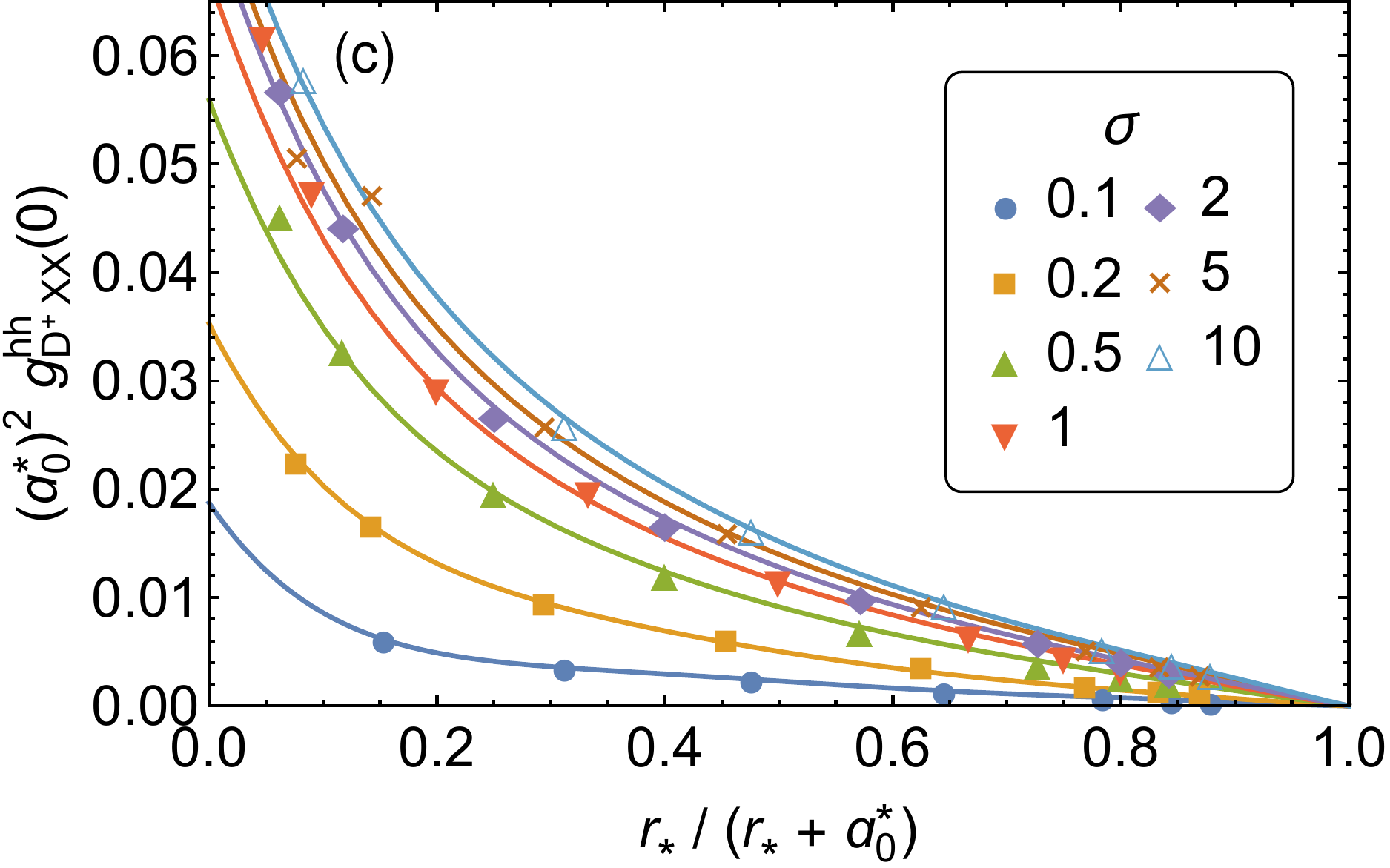}
\end{center}
\caption{(Color online) (a) Electron-hole contact PDFs of a
  donor-bound biexciton. For comparison, the black line indicates
  twice the exciton contact PDF\@. (b) Electron-electron and (c)
  hole-hole contact PDFs of a donor-bound biexciton.  The solid lines
  were obtained using the fitting function reported in the
  Supplemental Material.\cite{supplemental}
  \label{fig:pdf_dxx}}
\end{figure}

\subsection{Complexes with the logarithmic interaction \label{sec:logint_results}}

\begin{figure}[!htbp]
\begin{center}
\includegraphics[clip,width=0.45\textwidth]{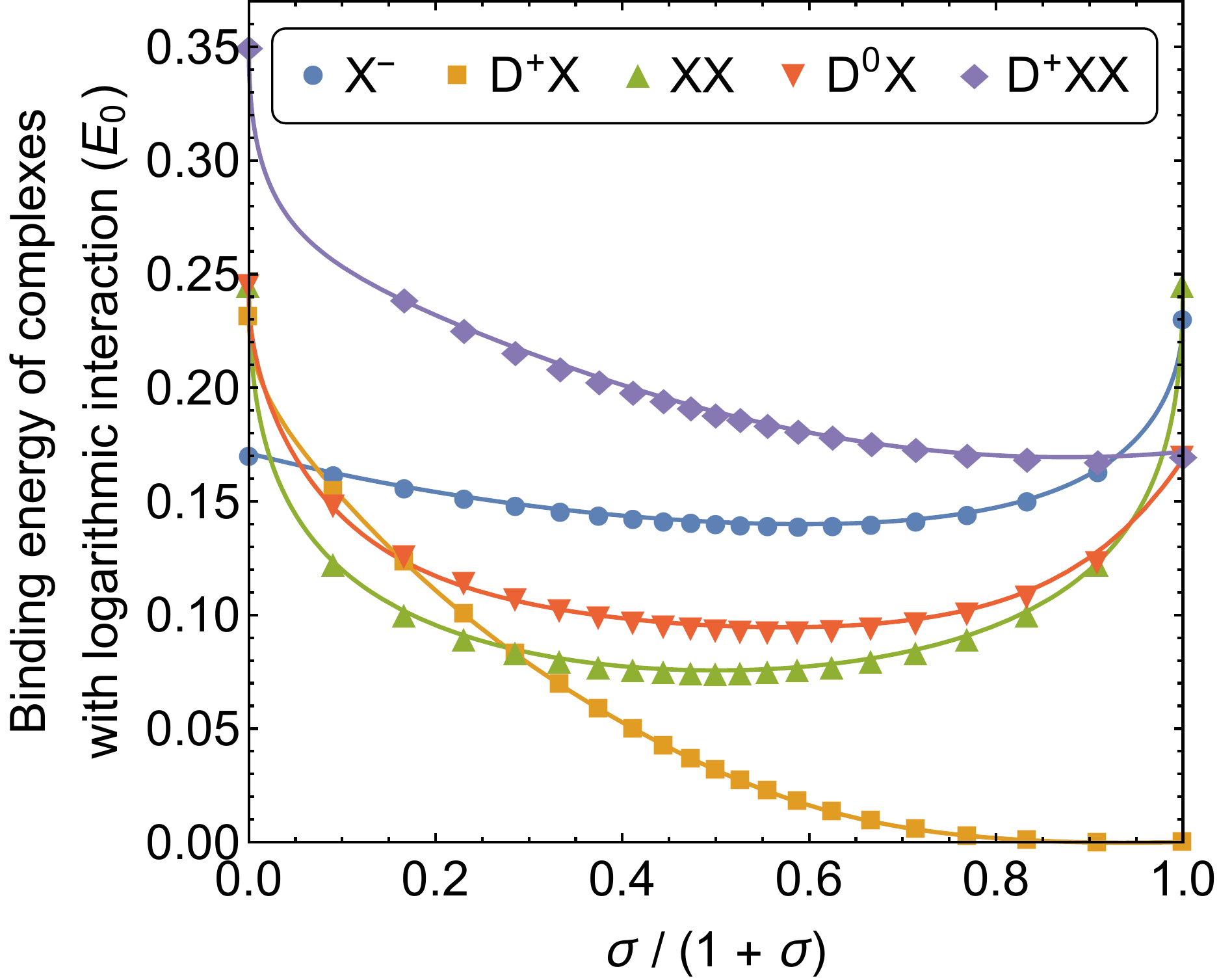}
\end{center}
\caption{(Color online) DMC binding energies of negative trions
  (X$^-$), biexcitons (XX), donor-bound excitons (D$^+$X), donor-bound
  trions (D$^0$X), and donor-bound biexcitons (D$^+$XX)\@. Particles
  in the complexes interact via the logarithmic interaction.  The
  X$^-$ data were presented in Ref.\ \onlinecite{Ganchev_2015}, and
  are shown here for completeness.
  \label{fig:be_ntrion_log}}
\end{figure}

We have also studied complexes of distinguishable particles
interacting with the purely logarithmic form of
Eq.\ (\ref{eq:log_int}). The binding energies are presented in
Fig.\ \ref{fig:be_ntrion_log}.  The lines shown in
Fig.\ \ref{fig:be_ntrion_log} were obtained using
Eqs.\ (\ref{eq:EX_fit}), (\ref{eq:EbXX_fit}), (\ref{eq:EbXm_fit}),
(\ref{eq:EbDplusX_fit}), (\ref{eq:EbDplusXm_fit}), and
(\ref{eq:EbDplusXX_fit}).  To convert from excitonic units to
logarithmic units we multiply the fitting function by $R_{\rm
  y}^\ast/E_0=r_\ast/(2a_0^\ast)=y/(2-2y)$ and take the limit that
$r_\ast\to \infty$, i.e., that $y\to 1$.  For complexes that have been
studied previously, our results are in good agreement with earlier
exact calculations.\cite{Ganchev_2015}

\section{Conclusions \label{sec:conclusions}}

In summary, we have discussed the different types of biexciton and
trion that can be observed in molybdenum and tungsten
dichalcogenides. Furthermore, we have presented statistically exact
DMC binding-energy data for biexcitons, trions, donor/acceptor-bound
trions, and donor/acceptor-bound biexcitons in 2D semiconductors,
including an analysis of extreme mass ratios.  We have shown that
biexcitons with indistinguishable charge carriers are unstable at
experimentally relevant electron-hole mass ratios.  Our calculations
have used the effective interaction between charge carriers arising
from screening effects in such materials.  We have also presented
contact PDF data that allow the investigation of additional contact
interaction energies between charge carriers in 2D semiconductors
within first-order perturbation theory. Our work provides a complete
reference for the interpretation of spectral lines in photoabsorption
and photoluminescence experiments on monolayer TMDCs in terms of a
model of charge carriers moving within the effective mass
approximation.

A broad range of theoretical works on 2D biexciton binding energies
show excellent quantitative agreement with each other, but an
enormous, threefold disagreement with experiment. By contrast, for
trions there is good agreement between theory and experiment. We have
considered and discounted various possible deficiencies in the
theoretical models of charge-carrier complexes.  We believe that the
most likely explanation for the disagreement with experiment is a
misinterpretation or misclassification of experimental optical
spectra.  In particular, we note that the energies require to remove
excitons from donor-bound biexcitons are similar to the binding
energies of experimentally observed biexcitons, suggesting that larger
charge-carrier complexes could be responsible for the observed peak
ascribed to biexcitons.

\appendix

\section{Character and product tables for the $D_{3h}''$ symmetry group}

Character and product tables for the $D_{3h}''$ symmetry group of
TMDCs are given in Tables \ref{table:d3htt} and
\ref{table:prod_d3htt}.

\begingroup \squeezetable
\begin{table}[!htbp]
\caption{Character table for the irreps of the extended point group
  $D_{3h}''$. The superscript ($\pm$) denotes the transformation under
  the $\sigma_h$ operation and $'$ denotes representations with
  nontrivial transformation under translation. \label{table:d3htt}}
\begin{center}
\begin{tabular}{lccccccccccccc}
\hline \hline

$D_{3h}''$ & $E$ & $2t$ & $2C_3$ & $2tC_3$ & $2t^2C_3$ & $9t\sigma_v$
& $\sigma_h$ & $2t\sigma_h$ & $2S_3$ & $2tS_3$ & $2t^2S_3$ & $9tC_2'$
\\ \hline

$A^+_1$ & $1$ & $1$ & $1$ & $1$ & $1$ & $1$ & $1$ & $1$ & $1$ & $1$ &
$1$ & $1$ \\

$A^+_2$ & $1$ & $1$ & $1$ & $1$ & $1$ & $-1$ & $1$ & $1$ & $1$ & $1$ &
$1$ & $-1$ \\

$A^-_1$ & $1$ & $1$ & $1$ & $1$ & $1$ & $-1$ & $-1$ & $-1$ & $-1$ &
$-1$ & $-1$ & $1$ \\

$A^-_2$ & $1$ & $1$ & $1$ & $1$ & $1$ & $1$ & $-1$ & $-1$ & $-1$ &
$-1$ & $-1$ & $-1$ \\

$E^+$ & $2$ & $2$ & $-1$ & $-1$ & $-1$ & $0$ & $2$ & $2$ & $-1$ & $-1$
& $-1$ & $0$ \\

$E^-$ & $2$ & $2$ & $-1$ & $-1$ & $-1$ & $0$ & $-2$ & $-2$ & $1$ & $1$
& $1$ & $0$ \\ \hline

$E'^+_1$ & $2$ & $-1$ & $-1$ & $2$ & $-1$ & $0$ & $2$ & $-1$ & $-1$ &
$2$ & $-1$ & $0$ \\

$E'^-_1$ & $2$ & $-1$ & $-1$ & $2$ & $-1$ & $0$ & $-2$ & $1$ & $1$ &
$-2$ & $1$ & $0$ \\

$E'^+_2$ & $2$ & $-1$ & $2$ & $-1$ & $-1$ & $0$ & $2$ & $-1$ & $2$ &
$-1$ & $-1$ & $0$ \\

$E'^-_2$ & $2$ & $-1$ & $2$ & $-1$ & $-1$ & $0$ & $-2$ & $1$ & $-2$ &
$1$ & $1$ & $0$ \\

$E'^+_3$ & $2$ & $-1$ & $-1$ & $-1$ & $2$ & $0$ & $2$ & $-1$ & $-1$ &
$-1$ & $2$ & $0$ \\

$E'^-_3$ & $2$ & $-1$ & $-1$ & $-1$ & $2$ & $0$ & $-2$ & $1$ & $1$ &
$1$ & $-2$ & $0$ \\

\hline \hline
\end{tabular}
\end{center}
\end{table}
\endgroup

\begin{table*}[!htbp]
\caption{Product table of the extended point group $D_{3h}''$. The
  relevant classification of the irreps according to
  $C_s=\{I,\sigma_h\}$, denoted by the superscript $\pm$ in Table
  \ref{table:d3htt}, is included for a given product by using
  $+\otimes +=+, +\otimes - =-$ and $-\otimes - =+$ and noting that
  all irreps in a given direct sum have the same $C_s$
  classification. \label{table:prod_d3htt}}
\begin{center}
\begin{tabular}{lcccccccc}
\hline\hline

$\otimes$ & $A_1$ & $A_2$ & $E$ &$E'_1$ & $E'_2$ & $E'_3$ \\

\hline

$A_1$ & $A_1$ & $A_2$ & $E$ & $E'_1$ & $E'_2$ & $E'_3$ \\

$A_2$ & $A_2$ & $A_1$ & $E$ & $E_1'$ & $E_2'$ & $E_3'$ \\

$E$ & $E$ & $E$ & $A_1\oplus A_2\oplus E$ & $E_2'\oplus E_3'$ &
$E_1'\oplus E_3'$ & $E_1'\oplus E_2'$ \\

$E'_1$ & $E'_1$ & $E'_1$ & $E_2'\oplus E_3'$ & $A_1\oplus A_2\oplus
E'_1$ & $E\oplus E'_3$ & $E\oplus E'_2$ \\

$E'_2$ & $E'_2$ & $E'_2$ & $E_1'\oplus E_3'$ & $E\oplus E_3'$ &
$A_1\oplus A_2\oplus E_2'$ & $E\oplus E_1'$ \\

$E'_3$ & $E'_3$ & $E'_3$ & $E'_1\oplus E'_2$ & $E\oplus E_2'$ &
$E\oplus E_1'$ & $A_1\oplus A_2\oplus E_3'$ \\

\hline\hline
\end{tabular}
\end{center}
\end{table*}

\section{Molecular analogies and behavior of binding energies at extreme
mass ratios}

\subsection{Biexcitons \label{app:bo_biex}}

In the limit that the hole mass is large, a biexciton resembles a 2D
H$_2$ molecule, and we may use the Born-Oppenheimer
approximation.\cite{Spink_2015} The leading-order biexciton total
energy is given by the minimum of the Born-Oppenheimer potential
energy $U(r)$, where $r$ is the exciton-exciton separation, plus the
harmonic zero-point energy of the exciton-exciton vibrations.
Higher-order corrections to the energy arise from vibrational
anharmonicity.

Consider a biexciton in which the charge carriers interact via the
Keldysh interaction.  Let ${\cal U}(r/a_0)$ be the Born-Oppenheimer
potential energy in Rydberg units for the case that the electron mass
$m_{\rm e}$ is finite but the hole mass $m_{\rm h}$ is infinite.
Then, at finite electron and hole masses, the Born-Oppenheimer
potential is $U(r)={\cal U}(r/a_0)R_{\rm y}$ where $R_{\rm y}=m_{\rm
  e}e^4/[2 (4 \pi \epsilon)^2 \hbar^2]=(m_{\rm e}/\mu)R_{\rm y}^\ast$
is the Rydberg and $a_0=4 \pi \epsilon \hbar^2/(m_{\rm
  e}e^2)=(\mu/m_{\rm e})a_0^\ast$ is the Bohr radius.  Note that
${\cal U}(r/a_0)$ does not depend on the electron or hole mass.

Near the minimum of the potential $r_{\rm min}$ we may write
\begin{eqnarray} U(r) & \approx & U_{\rm min} +
  \frac{1}{2} U^{\prime \prime}_{\rm min}(r-r_{\rm min})^2 \nonumber
  \\ & \equiv & U_{\rm min} + \frac{1}{2} \frac{m_{\rm h}+m_{\rm
      e}}{2} \omega^2 (r-r_{\rm min})^2, \end{eqnarray} where $(m_{\rm
  e}+m_{\rm h})/2$ is the reduced mass of the two excitons, $U_{\rm
  min}={\cal U}_{\rm min} R_{\rm y}$ is the minimum of potential,
$U^{\prime \prime}_{\rm min}={\cal U}^{\prime \prime}_{\rm min} R_{\rm
  y}/a_0^2$ is the second derivative of the potential at the minimum,
and
\begin{equation} \omega=\sqrt{\frac{2U^{\prime
  \prime}_{\rm min}}{m_{\rm h}+m_{\rm e}}}=\sqrt{\frac{2R_{\rm y}{\cal
        U}^{\prime \prime}_{\rm min}}{a_0^2(m_{\rm h}+m_{\rm
        e})}}. \end{equation}

The resulting ground-state energy in the harmonic approximation is
\begin{eqnarray} E & \approx & U_{\rm min} + \hbar \omega/2
  \nonumber \\ & \approx & \left[ {\cal U}_{\rm min} +
    \sqrt{\frac{{\cal U}^{\prime \prime}_{\rm min} m_{\rm e}}{m_{\rm
          h}}} \right] R_{\rm
    y}^\ast, \label{eq:heavy_biex_BO} \end{eqnarray} where we have
used $m_{\rm h} \gg m_{\rm e}$ in the last step.  This suggests that a
suitable fitting function for the binding energy of a biexciton with
small $\sigma \equiv m_{\rm e}/m_{\rm h}$ is a polynomial in powers of
$\sqrt{\sigma}$.  Similar conclusions hold for the case where the
interaction between the charge carriers is logarithmic.

In the limit of heavy holes, the total energies of biexcitons with
distinguishable and indistinguishable holes are identical, because
exchange effects become negligible as the heavy holes localize.  Hence
a biexciton with indistinguishable holes must be bound when the hole
mass is sufficiently large.  Likewise, a biexciton with
indistinguishable electrons has the same total energy as a biexciton
with distinguishable electrons in the limit that the electron mass is
large.

\subsection{Negative trions \label{app:trion_extreme}}

In the limit of heavy holes ($\sigma\to0$), a negative trion resembles
a 2D H$^-$ ion. The leading-order correction to the energy of an
infinite-hole-mass negative trion is therefore due to the reduced-mass
and mass-polarization perturbative corrections encountered in atomic
physics, each of which gives a contribution to the energy that is
linear in the electron-hole mass ratio $\sigma$.

In the limit of heavy electrons ($\sigma\to\infty$), a negative trion
resembles a charge-conjugated 2D H$^+_2$ ion, and hence one can use
the Born-Oppenheimer and harmonic approximations, as was done in
App.\ \ref{app:bo_biex}. The binding energy near the extreme mass
limit varies as the square-root of the mass ratio $\sigma$.

\subsection{Donor-bound excitons \label{app:dx_extreme}}

A donor-bound exciton in the limit of heavy holes is a charge
conjugate of a negative trion with heavy electrons, and therefore will
have a binding energy that varies as the square root of the mass ratio
$\sigma$.

In a donor-bound exciton with heavy electrons, the positive donor ion
and the heavy electron overlap, so the light hole is
unbound. Therefore the binding energy in this limit is zero.

\subsection{Donor-bound trions \label{app:dxm_extreme}}

The Born-Oppenheimer potential energy curve of a donor-bound trion
with a heavy hole is the same as that of a biexciton, but this time
the reduced mass is simply equal to the exciton mass.  The binding
energy varies as the square root of the mass ratio $\sigma$.

Now consider a donor-bound trion with two heavy electrons and a light
hole.  If the hole is very much lighter than the electrons then the
hole will be extremely delocalized and will see the positive donor ion
and two electrons (${\rm D^-}$) as a fixed, negative point charge; the
system therefore resembles an acceptor atom in which the hole is bound
to a fixed, negative point charge.  Hence $E_{\rm D^0X} \approx E_{\rm
  D^-}+E_{\rm A^0}$ in this limit, where $E_{\rm A^0}$ is the energy
of an acceptor atom.  In addition, if the electron mass is very much
larger than the hole mass, the exciton ground-state energy is $E_{\rm
  X} \approx E_{\rm A^0}$.  The binding energy of a donor-bound trion
in the limit that the hole is much lighter than the electron mass is
therefore $E_{\rm D^0X}^{\rm b}=E_{\rm D^0}+E_{\rm X}-E_{\rm D^0X}
\approx E_{\rm D^0}-E_{\rm D^-}$, which is the electron affinity of a
donor atom. Note that the electron affinity of a donor atom is equal
to the binding energy of a negative trion in the limit of large hole
mass.

The exciton Rydberg goes to zero in the limit that the hole mass goes
to zero; hence the binding energy of a donor-bound trion in excitonic
units goes to infinity as the hole-to-electron mass ratio goes to zero
($\sigma \to \infty$).

\subsection{Donor-bound biexcitons \label{app:dbxx_extreme}}

A donor-bound biexciton with two heavy holes resembles a trihydrogen
cation (H$_3^+$).  This molecular ion is an important component of the
interstellar medium,\cite{Martin_1961} and it is known that the
protons in H$_3^+$ form an equilateral triangle.  In
Sec.\ \ref{sec:dbbiex_results} we verify that 2D donor-bound
biexcitons with heavy holes also adopt an equilateral triangular
structure, and we calculate the bond length by minimizing the
Born-Oppenheimer potential energy.

Consider a donor-bound biexciton with two heavy electrons and two
light holes.  The binding energy of a donor-bound biexciton in the
limit that the holes (h$^+_{\rm light}$) are much lighter than the
electrons (e$^-_{\rm heavy}$) is
\begin{eqnarray}
  E^{\rm{b}}_{\rm{D}^+ \rm{XX}} & = & E_{
      \rm{D}^+ \rm{e}_{\rm{heavy}}^- \rm{e}_{\rm{heavy}}^-
      \rm{h}_{\rm{light}}^+}-E_{ \rm{D}^+ \rm{e}_{\rm{heavy}}^-
      \rm{e}_{\rm{heavy}}^- \rm{h}_{\rm{light}}^+
      \rm{h}_{\rm{light}}^+} \nonumber\\ & \approx & E_{\rm A^0}-E_{\rm A^+},
\end{eqnarray}
which is the hole affinity of an acceptor atom (in the limit of large
electron mass, D$^+$e$^-$e$^-$ acts like a fixed negative point
charge).  Note that the hole affinity of an acceptor atom is equal to the
binding energy of a positive trion in the limit of large electron
mass.

\begin{acknowledgments}
We acknowledge support from the European Research Council synergy
grant \textit{Hetero2D}, the United Kingdom's Engineering and Physical
Sciences Research Council (EPSRC), and the European Graphene Flagship
project.  M.S.\ was funded by the EPSRC doctoral training centre
``NoWNANO'' (grant no.\ EP/G03737X/1) and R.J.H.\ was funded by the
EPSRC doctoral training centre ``Graphene NOWNANO'' (grant
no.\ EP/L01548X/1).  Computer resources were provided by Lancaster
University's High-End Computing Cluster. This work made use of the
facilities of N8 HPC provided and funded by the N8 consortium and
EPSRC (grant no.\ EP/K000225/1).  R.M.\ is grateful for financial
support from MEXT-KAKENHI grant 17H05478, support by FLAGSHIP2020,
MEXT for computational resources, and project nos.\ hp170269 and
hp170220 at the K-computer. We acknowledge useful discussions with
I.\ Aleiner, A.\ Tartakovski, M.\ Potemski, and T.\ Heinz.
\end{acknowledgments}

\bibliography{2D_biex}

\end{document}